\documentclass[sigconf, nonacm, table]{acmart}
\pdfoutput=1

\usepackage{graphicx} 
\usepackage{multirow}
\usepackage{xspace}
\usepackage{booktabs,subcaption,amsfonts,dcolumn}
\usepackage[noend,linesnumbered,algoruled]{algorithm2e}
\usepackage{setspace}
\usepackage{amsmath}
\usepackage{xcolor}
\usepackage{hhline}
\usepackage{mathtools}
\usepackage{enumitem}
\usepackage[title]{appendix}
\usepackage{etoolbox}

\newtoggle{full}
\toggletrue{full}

\newcommand\vldbdoi{XX.XX/XXX.XX}
\newcommand\vldbpages{XXX-XXX}
\newcommand\vldbvolume{14}
\newcommand\vldbissue{7}
\newcommand\vldbyear{2025}
\newcommand\vldbauthors{\authors}
\newcommand\vldbtitle{\shorttitle} 
\newcommand\vldbpagestyle{empty} 

\newcommand{\ignore}[1]{}
\newcommand{\sys}{\textsc{Auto-Prep}\xspace}
\newcommand{\sysa}{\textsc{AP}\xspace}

\newcommand{\nit}[1]{\textit{#1}}

\newcommand{\code}[1]{{\tt #1}}
\newcommand{\codeq}[1]{{\tt ``#1''}}
\definecolor{mypurple}{HTML}{9f3dcc}
\definecolor{mygreen}{HTML}{13d4aa}
\newtheorem{df}{Definition}
\newtheorem{ex}{Example}
\newtheorem{pr}{Proposition}
\newtheorem{lemm}{Lemma}
\newtheorem{lem}{Theorem}
\newtheorem{prf}{Proof}

\newenvironment{example}
  {\begin{ex} \nopagebreak \begin{rm}}
  {\end{rm}\hfill$\Box$\end{ex}}

\newenvironment{definition}
  {\begin{df} \nopagebreak \begin{rm}}
  {\end{rm}\hfill$\Box$\end{df}}

\newenvironment{proposition}
  {\begin{pr} \nopagebreak \begin{rm}}
  {\end{rm}\hfill$\Box$\end{pr}}

\newenvironment{THEOREM}
  {\begin{lem} \nopagebreak \begin{rm}}
  {\end{rm}\end{lem}}

\newcommand{\yeye}[1]{\textcolor{blue}{[{\bf Yeye: }#1]}}
\newcommand{\eugenie}[1]{\textcolor{teal}{[{\bf Eugenie: }#1]}}

\DeclareMathOperator*{\argmax}{arg\,max}
\DeclareMathOperator*{\argmin}{arg\,min}

\definecolor{green-col}{HTML}{d2fce6}
\definecolor{yellow-col}{HTML}{fcefd2}
\definecolor{purple-col}{HTML}{f2d2fc}

\begin{document}

\title{Auto-Prep: Holistic Prediction of Data Preparation Steps for Self-Service Business Intelligence}

\author{Eugenie Y. Lai}
\affiliation{%
  \institution{MIT}
}
\email{eylai@mit.edu}

\author{Yeye He}
\affiliation{%
  \institution{Microsoft Research}
}
\email{yeyehe@microsoft.com}

\author{Surajit Chaudhuri}
\affiliation{%
  \institution{Microsoft Research}
}
\email{surajitc@microsoft.com}







\begin{abstract}
Business Intelligence (BI) plays a critical role in empowering modern enterprises to make informed data-driven decisions, and has grown into a billion-dollar business. Self-service BI tools like Power BI and Tableau have democratized the ``dashboarding'' phase of BI, by offering user-friendly, drag-and-drop interfaces that are tailored to non-technical enterprise users. However, despite these advances, we observe that the ``data preparation'' phase of BI continues to be a key pain point for BI users today.

In this work, we systematically study around 2K real BI projects harvested from public sources, focusing on the data-preparation phase of the BI workflows. We observe that users often have to program both (1) data transformation steps and (2) table joins steps, before their raw data can be ready for dashboarding and analysis.
A careful study of the BI workflows reveals that transformation and join steps are often intertwined in the same BI project, such that considering both holistically is crucial to accurately predict these steps. 
Leveraging this observation, we develop an \sys system to holistically predict transformations and joins, using a principled graph-based algorithm inspired by Steiner-tree, with provable quality guarantees. Extensive evaluations using real BI projects suggest that \sys can correctly predict over 70\% transformation and join steps, significantly more accurate than existing algorithms as well as language-models such as GPT-4. 

\end{abstract}

\maketitle

\pagestyle{\vldbpagestyle}
\begingroup\small\noindent\raggedright\textbf{PVLDB Reference Format:}\\
\vldbauthors. \vldbtitle. PVLDB, \vldbvolume(\vldbissue): \vldbpages, \vldbyear.\\
\href{https://doi.org/\vldbdoi}{doi:\vldbdoi}
\endgroup


\section{Introduction}

Business Intelligence (BI) plays a crucial role in modern enterprises, by empowering non-technical enterprise users to make informed data-driven decisions, and has grown into a billion-dollar business.

Over the past decades, ``self-service'' BI tools like Power BI and Tableau have democratized BI~\cite{ssbi-1, ssbi-2}, by providing intuitive drag-and-drop interfaces, to enable even non-technical users to create engaging BI dashboards for interactive data analysis.

\begin{figure}
    \centering
    \includegraphics[width=0.97 \linewidth]{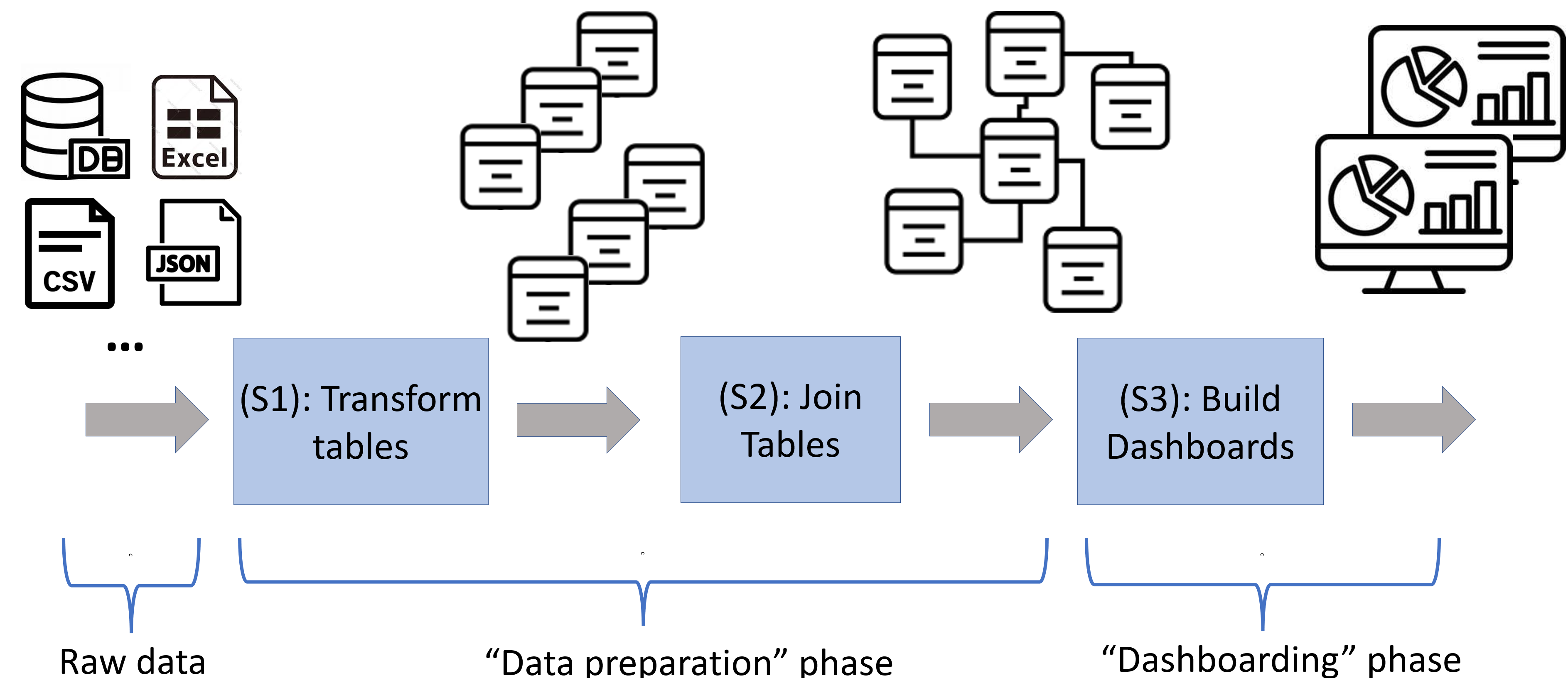} 
    \vspace{-2mm}
    \caption{Typical workflows of end-to-end BI. (S1) Raw data is first transformed; (S2) the transformed tables are then joined to create a BI model; (S3) finally, users use the BI model to create dashboards. We in this work focus on (S1) ``transform'' and (S2) ``join'', in the ``data preparation'' phase of BI.}
    \vspace{-5mm}
    \label{fig:bi-steps}
\end{figure}

\textbf{Data preparation remains a pain point in BI.}
Despite the advances in the ``dash-boarding phase'' of BI, which is a strength of today's BI tools, depicted as the final step (S3) in Figure~\ref{fig:bi-steps}, we observe that non-technical users continue to struggle with the ``data preparation'' phase, which in a typical BI workflow often needs to happen first before dashboards
can be built and analysis can be performed, like depicted as step
(S1) and (S2) in the figure.


Specifically, starting from raw data (which can be database tables, or flat files like CSV and Excel, etc.), users typically need to go through two steps in the data preparation phase, which are \underline{(S1) Transform}: or convert data files into proper tables, by relationalizing data~\cite{barowy2015flashrelate, li2023autotables} and standardizing values~\cite{tde, flashfill}; and then \underline{(S2) Join}: or link transformed tables via logical join relationships~\cite{join-pbi, join-tableau}, similar to how primary keys / foreign keys (PKs/FKs) are specified in databases, so that inter-linked analysis (e.g., cross-filtering) between multiple tables becomes possible on dashboards~\cite{why-join-pbi, why-join-tableau}.

Both of these data preparation steps, (S1) transform and (S2) join,  are common in practice. In a sample of 1837 real BI projects, we found $89\%$ have multiple tables that would require joins, and $43\%$ require transformations. We use an example simplified from real BI projects, to illustrate the two types of data-preparation steps.

\begin{example} \label{ex:case1}
    Consider our BI expert Alex who collected Table~\ref{tab:example-fertility}-\ref{tab:example-country}, and now wants to build BI dashboards to visualize how birth rates and economic statistics changed in different countries over time.

    For a BI expert and from a dimensional-modeling perspective~\cite{kimball2011data,  chaudhuri2011overview, negash2008business}, it would be obvious that \texttt{Fertility} and \texttt{Economics} are what is known as ``\emph{fact tables}'' that contain key numerical measures of interest (e.g., fertility rates and economic readings), while \codeq{Date} and \codeq{Country} are ``\emph{dimension tables}'', whose primary keys (\codeq{Year} and \codeq{Country}) can join with corresponding foreign-key attributes in the fact tables, so that users can slice/dice to perform analysis over these dimensional attributes (e.g., to answer questions like ``{in which decade and which continent is fertility the highest}?'')

    However, as is often the case in real-world BI, these raw input tables are  not yet ready for analysis in their current forms, because:
    \begin{itemize}[noitemsep,topsep=0pt,leftmargin=*]
    \item[] (1) The year values (``2010'', ``2011'', etc.) are presented as column headers in the \codeq{Fertility} table (Table~\ref{tab:example-fertility}), in a cross-tabulated format, which makes it hard to join these years with the \codeq{Year} column in the dimension-table \codeq{Date} (Table~\ref{tab:example-date}) for analysis;
    \item[] (2) Even though both  \codeq{Fertility} (Table~\ref{tab:example-fertility}) and \codeq{Economics} (Table~\ref{tab:example-econ}) need to join with the dimension-table \codeq{Country} (Table~\ref{tab:example-country}) for analysis, countries in the \codeq{Country} table are not shaped in the vertical direction to make such joins possible;
    \item[] (3) The table \codeq{Economics} (Table~\ref{tab:example-econ}) has a \codeq{Line-ID} column using the country-code-then-year format (``CHL-CY2010''), which needs to be transformed (e.g., into ``2010''), before it can equi-join with the \codeq{Year} column in \codeq{Date}  (Table~\ref{tab:example-date});
    \item[] (4) Finally, the fact table \codeq{Economics}  (Table~\ref{tab:example-econ}) contains different types of economic readings (GPD, CPI, Payroll, etc.), which however are stored as key-value pairs in the column \codeq{Metric} and \codeq{Value}, that should move into separate columns for analysis;
    \end{itemize}

    \begin{table}[t!]
    \iftoggle{full}
    {
    }
    {
        \vspace{-18mm}
    }
    \begin{minipage}{.42\linewidth}
    \begin{center}
    \renewcommand{\arraystretch}{1.1}
    \setlength{\tabcolsep}{1.4pt}
    \caption{Fertility (fact)}
    \label{tab:example-fertility}
    \vspace{-4mm}
    \resizebox{0.85\linewidth}{!}{
    \begin{tabular}{| l | l | l | l |}
    \cline{1-4}
    \hhline{|*4-}
    \rowcolor[HTML]{FFFFFF}
     {\textbf{Country}} & {\cellcolor[HTML]{FFFFFF}\textbf{2010}} & {\cellcolor[HTML]{FFFFFF}\textbf{2011}} &
     {\cellcolor[HTML]{FFFFFF}\textbf{2012}} \\ \cline{1-4}
     Poland & 1.38 & 1.3 & 1.3 \\ \hline
     Chile & 1.86 & 1.84 & 1.83 \\ \hline
     Morocco & 2.58 & 2.65 & 2.71 \\ \hline
     Turkey & 2.1 & 2.08 & 2.06 \\
     \cline{1-4}
    \end{tabular}
    }
    
    \caption{Date (dim)}
    \label{tab:example-date} 
    \vspace{-4mm}
    \resizebox{0.6\linewidth}{!}{
    \begin{tabular}{| l | l | l |} 
    \cline{1-3}
    \hhline{|*2-}
    \rowcolor[HTML]{FFFFFF}
     {\cellcolor[HTML]{FFFFFF}\textbf{Year}} & {\cellcolor[HTML]{FFFFFF}\textbf{IsLeap}} & {\cellcolor[HTML]{FFFFFF}\textbf{Days}} \\ \cline{1-3}
     2010 & No & 365 \\ \hline
     2011 & No & 365 \\ \hline
     2012 & Yes & 366 \\ 
     \cline{1-3}
    \end{tabular}
    }

    \end{center}
    \end{minipage}
    \begin{minipage}{.57\linewidth}
    \begin{center}
    \renewcommand{\arraystretch}{1.1}
    \setlength{\tabcolsep}{1.6pt}
    \caption{Economics (fact)}
    \label{tab:example-econ}  
    \vspace{-4mm}
    \resizebox{0.8\linewidth}{!}{
    \begin{tabular}{| l | l | l | l |}
    \cline{1-4}
    \hhline{|*4-}
    \rowcolor[HTML]{FFFFFF}
      {\textbf{Line-ID}} &
      {\textbf{Code}} &{\cellcolor[HTML]{FFFFFF}\textbf{Metric}} &{\cellcolor[HTML]{FFFFFF}\textbf{Value}} \\ \cline{1-4} 
      CHL-CY2010 & CHL & GDP & 12756 \\ \hline
      CHL-CY2011 & CHL & GDP & 14637\\ \hline
      CHL-CY2012 & CHL & GDP & 15397 \\ \hline
      CHL-CY2010 & CHL & CPI & 100 \\ \hline
      CHL-CY2011 & CHL & CPI & 103.3 \\ \hline
      CHL-CY2012 & CHL & CPI & 106.4 \\ \hline
      CHL-CY2010 & CHL & Payroll & 55.16 \\ \hline
      CHL-CY2011 & CHL & Payroll & 57.01 \\ \hline
      CHL-CY2012 & CHL & Payroll & 57.43 \\  
     \cline{1-4} 
    \end{tabular}
    }
    \end{center}
    \end{minipage}
    
    \caption{Country (dim)}
    \label{tab:example-country}
    \vspace{-4mm}
    \resizebox{0.7\linewidth}{!}{
    \begin{tabular}{| l | l | l | l | l |}
    \cline{1-5}
    \hhline{|*5-}
    \cellcolor[HTML]{FFFFFF} \textbf{Code} & \cellcolor[HTML]{FFFFFF} \textbf{POL} & \cellcolor[HTML]{FFFFFF} \textbf{CHL} & 
    \cellcolor[HTML]{FFFFFF} \textbf{TUR} & \cellcolor[HTML]{FFFFFF} \textbf{MAR} \\ \cline{1-5}
        \hline 
        \cellcolor[HTML]{FFFFFF} Country & Poland & Chile & Turkey & Morocco \\ \hline 
        \cellcolor[HTML]{FFFFFF} Continent & Europe & S. America & Europe & Africa \\ \hline 
        \cellcolor[HTML]{FFFFFF} Developed & Yes & No & Yes & No \\ \cline{1-5}
    \end{tabular}
    }
    \vspace{-4mm}
    \end{table}

    As a seasoned BI expert, Alex identified these necessary steps, often using clues based on what needs to ``join'' to decide what transformations may be needed, before any analysis can be performed. She then went on to program these steps, using her favorite tools (which can be Python Pandas, R, etc.). Specifically, she would:
        \begin{itemize}[noitemsep,topsep=0pt,leftmargin=*]
        \item[] (1) Invoke the \codeq{Unpivot} transformation on \codeq{Fertility} in Table~\ref{tab:example-fertility}, so that the years become a new column  in the transformed version of \codeq{Fertility} in Table~\ref{tab:example-fertility-after} (marked in \colorbox{yellow-col}{yellow}), which becomes joinable with the \codeq{Year} column in the dimension table \codeq{Date}, also marked in \colorbox{yellow-col}{yellow} in Table~\ref{tab:example-date-after};
        \item[] (2) Invoke the \codeq{Transpose} transformation on \codeq{Country} in Table~\ref{tab:example-country},  to produce a transformed version of \codeq{Country} in Table~\ref{tab:example-country-after}, which becomes joinable with both the transformed \codeq{Fertility} in Table~\ref{tab:example-fertility-after}, through the \codeq{Country} column (marked in \colorbox{green-col}{green}), and also \codeq{Economics} in Table~\ref{tab:example-econ-after}, through the \codeq{Code} column (marked in \colorbox{purple-col}{purple});
        \item[] (3) Invoke a string transformation \codeq{Split} on the \codeq{Line-ID} column of \codeq{Economics} in Table~\ref{tab:example-econ}, which splits the field using delimiter ``-'',  takes the second component, and then uses \codeq{Substring} to extract the last 4 characters to produce a new \codeq{Year} column in  Table~\ref{tab:example-econ-after} (marked in \colorbox{yellow-col}{yellow}), which becomes joinable with the \codeq{Year} column of \codeq{Date} in Table~\ref{tab:example-date-after} (also marked in \colorbox{yellow-col}{yellow}).
        \item[] (4) Invoke the \codeq{Pivot} transformation on \codeq{Economics} in Table~\ref{tab:example-econ},  to produce Table~\ref{tab:example-econ-after}, which now has different metrics (GDP, CPI, etc.) as separate columns for analysis;
        \end{itemize}

    After Alex programmed the transformations above, she would have completed the transformation stage of BI (stage S1 in Figure~\ref{fig:bi-steps}). Alex can then join/link the transformed tables in Table~\ref{tab:example-fertility-after}-\ref{tab:example-country-after}, which are effectively PK/FK joins as indicated by their colors, to complete the join stage (S2 in Figure~\ref{fig:bi-steps}). These logical join relationships create what is known as a ``BI model''~\cite{join-pbi, join-tableau}, from which dashboards can then be easily built using drag-and-drop in tools like Tableau and Power BI  (reaching stage S3 of Figure~\ref{fig:bi-steps}).  Note that in this process, data formats across tables have been automatically standardized, so that tables can be linked and enriched using other tables. 
    \end{example}
\begin{table}[t!]
\iftoggle{full}
{
}
{
    \vspace{-18mm}
}
\begin{minipage}{.45\linewidth}
\begin{center}
\renewcommand{\arraystretch}{1.1}
\setlength{\tabcolsep}{1.4pt}
\caption{Fertility (fact)}
\label{tab:example-fertility-after}
\vspace{-4mm}
\resizebox{0.75\linewidth}{!}{
\begin{tabular}{| l | l | l |}
\cline{1-3}
\hhline{|*3-}
\rowcolor[HTML]{FFFFFF}
 {\cellcolor[HTML]{13d4aa}\textbf{Country}} & {\cellcolor[HTML]{FFC107}\textbf{Year}} & {\cellcolor[HTML]{FFFFFF}\textbf{Fertility}} \\ \cline{1-3}
 \cellcolor[HTML]{d2fce6} Chile & \cellcolor[HTML]{fcefd2} 2010 & 1.86 \\ \hline
 \cellcolor[HTML]{d2fce6} Chile & \cellcolor[HTML]{fcefd2} 2011 & 1.84 \\ \hline
 \cellcolor[HTML]{d2fce6} Chile & \cellcolor[HTML]{fcefd2} 2012 & 1.83 \\ \hline
 \cellcolor[HTML]{d2fce6} Turkey & \cellcolor[HTML]{fcefd2} 2010 & 2.30 \\ \hline
 \cellcolor[HTML]{d2fce6} Turkey & \cellcolor[HTML]{fcefd2} 2011 & 2.21 \\ \hline
 \cellcolor[HTML]{d2fce6} Turkey & \cellcolor[HTML]{fcefd2} 2012 & 2.34 \\ 
 \cline{1-3}
\end{tabular}
}
\end{center}

\end{minipage}
\begin{minipage}{.53\linewidth}
\begin{center}
\renewcommand{\arraystretch}{1.1}
\setlength{\tabcolsep}{1.4pt}
\caption{Economics (fact)}
\label{tab:example-econ-after}
\vspace{-4mm}
\resizebox{0.85\linewidth}{!}{
\begin{tabular}{| l | l | l | l | l |}
\cline{1-5}
\hhline{|*5-}
\rowcolor[HTML]{FFFFFF}
 {\cellcolor[HTML]{FFC107}\textbf{Year}} & {\cellcolor[HTML]{d451fc}\textbf{Code}} & {\cellcolor[HTML]{FFFFFF}\textbf{GDP}} & {\cellcolor[HTML]{FFFFFF}\textbf{CPI}} & 
 {\cellcolor[HTML]{FFFFFF}\textbf{Payroll}} \\ \cline{1-5}
 \cellcolor[HTML]{fcefd2} 2010 & \cellcolor[HTML]{f2d2fc} CHL & 12756 & 100 & 55.16 \\ \hline
 \cellcolor[HTML]{fcefd2} 2011 & \cellcolor[HTML]{f2d2fc} CHL & 14637 & 103.3 & 57.01 \\ \hline
 \cellcolor[HTML]{fcefd2} 2012 & \cellcolor[HTML]{f2d2fc} CHL & 15397 & 106.4 & 57.43 \\ \hline
 \cellcolor[HTML]{fcefd2} 2010 & \cellcolor[HTML]{f2d2fc} TUR & 71022 & 100 & 93.29 \\ \hline
 \cellcolor[HTML]{fcefd2} 2011 & \cellcolor[HTML]{f2d2fc} TUR & 78102 & 102.3 & 98.10 \\ \hline
 \cellcolor[HTML]{fcefd2} 2012 & \cellcolor[HTML]{f2d2fc} TUR & 82019 & 104.1 & 99.38 \\ 
 \cline{1-5}
\end{tabular}
}

\end{center}
\end{minipage}

\begin{minipage}{.34\linewidth}
\begin{center}
\renewcommand{\arraystretch}{1.1}
\setlength{\tabcolsep}{1.4pt}
\caption{Date (dim)}
\label{tab:example-date-after}
\vspace{-4mm}
\resizebox{0.75\linewidth}{!}{
\begin{tabular}{| l | l | l |}
\cline{1-3}
\hhline{|*3-}
\rowcolor[HTML]{FFFFFF}
 {\cellcolor[HTML]{FFC107}\textbf{Year}} & {\cellcolor[HTML]{FFFFFF}\textbf{IsLeap}} &  {\cellcolor[HTML]{FFFFFF}\textbf{Days}} \\ \cline{1-3}
 \cellcolor[HTML]{fcefd2} 2010 & No & 365 \\ \hline
 \cellcolor[HTML]{fcefd2} 2011 & No & 365 \\ \hline
 \cellcolor[HTML]{fcefd2} 2012 & Yes & 366 \\ 
 \cline{1-3}
\end{tabular}
}
\end{center}
\end{minipage}
\vspace{-3mm}
\begin{minipage}{.64\linewidth}
\begin{center}
\renewcommand{\arraystretch}{1.1}
\setlength{\tabcolsep}{1.4pt}
\caption{Country (dim)}
\label{tab:example-country-after}  
\vspace{-4mm}
\resizebox{0.8\linewidth}{!}{
\begin{tabular}{| l | l | l | l |}
\cline{1-4}
\hhline{|*3-}
\rowcolor[HTML]{FFFFFF}
  {\cellcolor[HTML]{d451fc}\textbf{Code}} & {\cellcolor[HTML]{13d4aa}\textbf{Country}} &
  {\cellcolor[HTML]{FFFFFF}\textbf{Continent}} &
  {\cellcolor[HTML]{FFFFFF}\textbf{Developed}}\\ \cline{1-4} 
  \cellcolor[HTML]{f2d2fc} POL & \cellcolor[HTML]{d2fce6} Poland & Europe & Yes \\ \hline
  \cellcolor[HTML]{f2d2fc} CHL & \cellcolor[HTML]{d2fce6} Chile & S. America & No \\ \hline
  \cellcolor[HTML]{f2d2fc} TUR & \cellcolor[HTML]{d2fce6} Turkey & Europe & Yes \\ \hline
  \cellcolor[HTML]{f2d2fc} MAR & \cellcolor[HTML]{d2fce6} Morocco & Africa & No \\ 
 \cline{1-4} 
\end{tabular}
}
\end{center}

\end{minipage}
\vspace{-2mm}
\end{table}

While our BI professional Alex has the expertise to program all these transformation and join steps, it is clearly challenging for non-technical users in self-service BI-tools like Tableau and Power BI, who typically are neither DBAs nor BI professionals. This challenge is evidenced by large numbers of user questions on Power BI and Tableau forum~\cite{pbi-forum, tableau-forum}, where users are often stuck and raise various questions about the data preparation process (e.g., how to transform a particular table~\cite{forum-q-1, forum-q-2, forum-q-3, forum-q-4}, and how best to join two related tables~\cite{forum-join-q-1, forum-join-q-2, forum-join-q-3, forum-join-q-4}, etc.).  



\textbf{Prior art: predictions for transform and join only}.
We are not the first to recognize the need to predict transformations or joins --- while there is extensive prior research, existing methods predominantly predict either only transformations, or only joins, as two \textit{separate and standalone} problems. For example, prior work on join predictions, such as PK/FK joins~\cite{fk-ml, lin2023autobi, hpi, chen2014fast, zhang2010multi}, all assumes that input tables are already properly transformed, and therefore does not consider transformations that need to happen before joins.

Similarly, while there is existing research on predicting transformations~\cite{yan2020auto-suggest, li2023autotables, lee2021lux, chen2022rigel, barowy2015flashrelate}, existing approaches all operate on \textit{individual tables} that predict one table at a time, without taking into account signals across tables, and how transformations interact with joinability in a global sense. 

When testing existing join-only and transform-only prediction methods on real BI projects, we find that they perform poorly on complex multi-table BI projects, because transformation and join are not considered in a holistic manner.

\textbf{\sys: ``holistic'' prediction across tables and steps}. 
Our insights from inspecting real BI projects reveal that, interestingly, (1) Considering join and transform jointly can allow the two to ``help each other'', in finding correct joins and transforms; (2) Considering transforms across multiple tables in the same BI project also allows these tables to ``help each other'' in finding the most likely transforms. 
We argue that by modeling end-to-end data preparation (both transform and join) holistically and across all tables, and by focusing on the set of most common and important transforms in the context of BI that we will precisely define using a DSL,
predictions for both transform and join can be made more accurately, as we illustrate in the following example.

\begin{example}
\label{ex:holistic}
We revisit Table~\ref{tab:example-fertility}-\ref{tab:example-country} in Example~\ref{ex:case1} again. Looking at \codeq{Fertility} (Table~\ref{tab:example-fertility}) alone, we may not be sure if an \codeq{Unpivot} is necessary to transform the column headers ``2010/2011/2012'' into column values. However, inspecting other related tables in the same BI project, such as \codeq{Date} (Table~\ref{tab:example-date}) with a column \codeq{Year} containing value ``2010/2011/2012'', should give us clues that ``2010/2011/2012'' are data values, and an \codeq{Unpivot} is needed for Table~\ref{tab:example-fertility} (to produce Table~\ref{tab:example-fertility-after}), so that the two tables can join.

Similarly, looking at the \codeq{Country} table (Table~\ref{tab:example-country}) in isolation, we may not be sure if \codeq{Transpose} or  \codeq{Unpivot} would be required, but the presence of the \codeq{Country} column in Table~\ref{tab:example-fertility} and the \codeq{Code} column in Table~\ref{tab:example-econ} gives strong indication that Table~\ref{tab:example-country} is oriented in the row direction, and a \codeq{Transpose} is needed to turn its rows into columns (producing Table~\ref{tab:example-country-after}), which can then become joinable with both Table~\ref{tab:example-fertility} and  Table~\ref{tab:example-econ} for analysis.

As a final example, in the \codeq{Economics} table (Table~\ref{tab:example-econ}), the need to perform \codeq{Split} and \codeq{Substring}  to extract year values like ``2010/2011'', is obvious only in the presence of the \codeq{Year} column in \codeq{Date} (Table~\ref{tab:example-date}), for the two to join in downstream analysis.
\end{example}

As we can see from these examples, holistic reasoning across transform/join, and for all tables in the same project, helps us accurately predict both transforms and joins, which is a unique opportunity overlooked by existing methods that tend to consider transform and join as separate problems. 

While the intuition for holistic prediction is clear, the technical challenges lie in principled modeling of disparate types of transformation steps (Unpivot, Transpose, Pivot, Split, Substring, etc.) that are most common in BI transformations based on our analysis of real BI projects, together with join steps. In this work, we develop \sys that builds on a new graph-based formulation, which seamlessly integrates diverse classes of transforms and joins in a unified search graph, with probabilistic interpretations of the transform/join steps as weighted edges on the graph. We develop new graph algorithms inspired by Steiner-tree, to solve the resulting optimization problem both efficiently, and with provable quality guarantees. 

We evaluate \sys using real BI projects harvested from public sources, treating user-programmed transforms and joins as ground truth. Our results suggest that \sys can accurately predict over $70\%$ of the data preparation steps, substantially better than the existing methods. 



\begin{table*}[t]
\iftoggle{full}
{
}
{
    \vspace{-18mm}
}
\caption{DSL of transformation operators in \sys.}
\label{tab:dsl}
\vspace{-4mm}
\scalebox{0.8}{
\begin{tabular}{|l|l|l|l|}
\hline
Category & Operator & Operator parameters & Operator description (example in parenthesis) \\
\hline
 & Unpivot~\cite{pandas-melt} &  start\_column, end\_column & Collapse homogeneous columns into one column (e.g., unpivot Table~\ref{tab:example-fertility} to produce Table~\ref{tab:example-fertility-after}) \\
 Table-reshaping & Pivot~\cite{pandas-pivot} &  pivot\_column, value\_column & Lift values in a column into column headers (e.g., pivot Table~\ref{tab:example-econ} to produce Table~\ref{tab:example-econ-after}) \\
 & Transpose~\cite{pandas-transpose} &  none &  Convert rows into columns and columns into rows (e.g., transpose Table~\ref{tab:example-country} to produce Table~\ref{tab:example-country-after})  \\

\hline
 & Split~\cite{pandas-split} &  delimiter, select\_pos &  Split a string using delimiters (e.g., split to extract years from Table~\ref{tab:example-econ} to produce Table~\ref{tab:example-econ-after}) \\
String-transform & Concatenate~\cite{pandas-concat} &  array of strings or columns & Concatenate two strings and produce an output string \\
 & Substring~\cite{pandas-substring} &  column, start, length & Extract substrings by positions  (e.g., substring to extract years from Table~\ref{tab:example-econ}  to produce Table~\ref{tab:example-econ-after})  \\ \hline
  & No-op &  none  & No transformation is required for a table  (e.g., Table~\ref{tab:example-date} requires no transformation for the BI project) \\ \hline
\end{tabular}
}
\vspace{-4mm}
\end{table*}

\section{Related Work} 
\label{sec:related}

\textbf{Business intelligence (BI).} 
Business intelligence (BI) plays a crucial role in modern enterprises, with leading vendors such as Tableau~\cite{tableau} and Power-BI~\cite{power-bi} offering ``self-service'' BI tools that are gaining popularity, especially among non-technical users~\cite{ssbi-1, ssbi-2}, by enabling them to build visualization dashboards with simple drag-and-drop~\cite{mackinlay2007show}. BI is closely related to research topics such as data warehousing~\cite{kimball2011data, chaudhuri2011overview}, OLAP~\cite{olap-1, olap-2}, and ETL~\cite{etl-1, etl-2}.

\textbf{Join-only prediction methods.}
Join is a core operation in data management, with a large body of work studying join predictions, which, however, do not consider the need for transformations. 

For example, the canonical join prediction problem studies the detection of PK/FK joins in classical database settings, with many influential methods developed in the literature, such as MC-FK~\cite{zhang2010multi}, Fast-FK~\cite{chen2014fast}, HoPF~\cite{hpi},
ML-FK~\cite{fk-ml}, BI-Join~\cite{lin2023autobi}, etc. All of these methods consider a classical database scenario, where tables loaded into databases are already properly transformed/cleaned a priori (e.g., by ETL engineers), where additional transformations are not necessary. However, this is insufficient for the real BI workflows we study, where users often have to start from raw data tables that need to be transformed first, before they are ready for analysis.


\textbf{Transformation-only prediction methods.}
There is also extensive research on transformation predictions, which, however, do not consider joins that need to happen afterwards. Furthermore, existing transformation-prediction methods make predictions \textit{one table at a time}, overlooking the rich interactions between tables that can provide signals for what transformations may be needed. 
For instance, transformations are predicted based on user-provided examples~\cite{tde, flashfill, barowy2015flashrelate, foofah, chen2022rigel, tde-excel}, inferred data patterns~\cite{trifacta-transform-by-pattern, jin2020auto},  characteristics of input tables~\cite{yan2020auto-suggest, mackinlay2007show, lee2021lux}, or relationships between input/output tables~\cite{auto-pipeline, AutoPandas, scythe, synthe-sql}, but most focusing on one single table at a time. We show that on complex BI projects, such methods tend to produce inaccurate predictions.

In this work, we demonstrate how existing transform and join predictions can be seamlessly integrated into a graph-based \sys framework, and reasoned holistically in a principled manner, for accurate predictions at a global level.

\iftoggle{full}
{
    Although most of the existing transformation-prediction methods do not consider joins, a few prior studies~\cite{warren2006multi, zhu2017auto, dargahi2024dtt} consider a limited class of string transforms for joins, between two tables only. We compare with these methods and show that they do not cover the diverse transformations we aim to handle in Table~\ref{tab:dsl}, and do not handle multiple tables holistically. 
    
    
}
{

}

\section{Preliminaries} 
\label{sec:preliminary}


\subsection{Real-world BI projects}
\label{sec:real-data}
To study real-world BI workflows end-to-end, we use 1837 real BI projects created using the popular Power BI tool (in the .pbix format), crawled from public sources. Our inspection suggests that these BI projects cover diverse application domains, including financial reporting, inventory management, sales analytics, etc. 

For each crawled BI project, we programmatically extract the following: (1) a set of raw input tables, (2) user-programmed data transformation steps on each table (the ground-truth transformations we want to predict), and (3) user-specified join relationships on the transformed tables (the ground-truth joins we want to predict). Example~\ref{ex:case1} shows a simplified version of such a BI project, as well as what we extract from the BI project. 
Detailed statistics of the BI projects we collected will be discussed in Section~\ref{sec:experiments}.


\subsection{Transformations in BI} 
    In this work, we consider 6 common transformation operators (listed in Table~\ref{tab:dsl}) that are essential for BI analysis, which cover two broad categories: (1) table-reshaping transformations, and (2) string transformations.
    These transformations are both common and crucial to enable multi-table end-to-end BI analysis, as illustrated in Example~\ref{ex:case1}. 


\textbf{Table-reshaping transformations.} Reshaping transformations such as \code{Unpivot}~\cite{pandas-melt}, \code{Transpose}~\cite{pandas-transpose}, \code{Pivot}~\cite{pandas-pivot}, change the basic shape/structure of an input table, which are frequently used to convert raw data into a suitable relational form for analysis. Variants of these reshaping transformations have been studied in prior work (e.g.,~\cite{barowy2015flashrelate, li2023autotables, foofah, yan2020auto-suggest}), and we will give a short overview of these transformations below.

\underline{Unpivot.} The unpivot operator collapses a set of selected columns (specified by a \codeq{selected\_columns} parameter, shown in the first line of Table~\ref{tab:dsl}) into one single column, while keeping the remaining columns unchanged.  For instance, in Example~\ref{ex:case1}, the column headers \codeq{2010/2011/2012} in Table~\ref{tab:example-fertility} need be to unpivoted, in order to produce Table~\ref{tab:example-fertility-after}, which becomes joinable with Table~\ref{tab:example-date-after}, and is more amenable to analysis (e.g., for range-filters or aggregation).

Unpivot is a common operator that is available in Python Pandas~\cite{pandas-melt} and R~\cite{r-melt} for programmers and data scientists, as well as in Power BI~\cite{pbi-unpivot} and Tableau~\cite{tableau-unpivot} for less technical BI users.

\underline{Pivot.} The pivot operator is the inverse operator of unpivot, which lifts the values in a column into column headers. In Example~\ref{ex:case1}, values in the \codeq{Metric} column of Table~\ref{tab:example-econ} (\codeq{GDP/CPI/Payroll}) need to be pivoted into separate columns for relational analysis, in accordance with principles such as the First Normal Form~\cite{1nf}. Like unpivot, this is a common operator in Python Pandas~\cite{pandas-pivot}, R~\cite{r-pivot}, Power BI~\cite{pbi-pivot}, and Tableau~\cite{tableau-pivot}.

\underline{Transpose.} The transpose operator turns rows to columns and vice versa. In Example~\ref{ex:case1}, transpose is necessary to convert Table~\ref{tab:example-country} to Table~\ref{tab:example-country-after}, so that joins become possible for cross-table analysis. 

\textbf{String transformations.} Unlike reshaping transformations, string transformations operate on values in the same row and do not alter the shape of the input table. We give an overview of these operators (Table~\ref{tab:dsl}) in the interest of space and refer the reader to~\cite{tde, flashfill, barowy2015flashrelate, foofah, chen2022rigel} for more details.

\underline{Split.} The split operator uses a given \codeq{delimiter} parameter to separate an input string into an array of segments, from which one segment is selected using a \codeq{select\_pos} parameter, to produce an output string. 

\underline{Concatenate.} The concatenate operator is the reverse of split, which assembles multiple strings into one output string. 

\underline{Substring.} The substring operator selects a sub-part of a string column, based on a \codeq{start\_pos} and \codeq{length} parameter. 

String transformation operators often need to be composed, to produce a desired output~\cite{tde, flashfill}. For instance, in Example~\ref{ex:case1}, the \codeq{Line-ID} column in Table~\ref{tab:example-econ} is first split using \code{delimiter = ``-''}, to produce an array of segments, \code{\{[CY2010, CHL], [CY2011, CHL], ...\}}, from which the first segment is selected (\code{select\_pos = ``1''}) to produce \code{\{CY2010, CY2011, CY2012\}}. Then a substring operator, with parameters \code{start\_pos = 2} and \code{length = 4}, is used to produce values \code{\{2010, 2011, 2012\}}, which become the \codeq{Year} column in Table~\ref{tab:example-econ-after} that is joinable with Table~\ref{tab:example-fertility-after} and Table~\ref{tab:example-date-after}.




\underline{No-op.} Finally, many tables in BI projects may require no  transformations (e.g., Table~\ref{tab:example-date}). It is crucial that our  predictions do not ``over-trigger'', and should correctly predict ``no-op'' on such tables. 

    \textbf{Additional transformations.} We focus on the transformation in Table~\ref{tab:dsl}, which are key operators to enable cross-table BI in real BI projects. We also note that our approach is general and extensible, where new transformation operators can easily ``plug-in''.
\iftoggle{full}
{

    Note that there are transformations that are more ``cosmetic'' in nature in the context of BI, which usually do not affect users' ability to perform cross-table analysis. Based on our analysis of real BI projects, common transformations in this category include operators such as: change column types, remove/reorder columns, and row-to-row transformations to beautify column values (e.g., adding a dollar sign to the \codeq{GDP} column of Table~\ref{tab:example-econ-after}\ignore{, or inserting thousands of separators into the numbers, etc.}). 
    We in this work do not attempt to predict such transformations, because
    (1) they tend to be cosmetic in nature that do not typically affect cross-table BI analysis (i.e., tables can still be joined for cross-table analysis), and (2) these transformations are often handled  well by existing techniques already (e.g., by input/output examples using program-synthesis~\cite{tde, flashfill}).
    \iftoggle{full}
    {
        \footnote{Certain transformations, such as beautifying a column by inserting dollar-sign, requires additional user input (e.g., input/output examples), as otherwise it is hard to know the exact format that users would prefer (e.g., with a dollar sign or not). Such transformations are handled well by existing techniques (e.g., by-example transformations), and are out of scope for this work.}
    }
}
{
We give more discussions on additional transformations and extensibility in our technical report~\cite{full}.
}



\subsection{Join relationships in BI}
Real BI projects often come with multiple tables (89\% of BI projects we study have more than one table). 
With multiple tables in one BI project, it is crucial to create accurate join relationships across tables, to enable cross-table analysis. Popular BI tools all offer intuitive GUI tools to help users define join relationships between tables, as documented in tutorials for Power-BI~\cite{join-pbi} and Tableau~\cite{join-tableau}. 

From a database perspective, these join relationships in BI are often PK/FK joins, whose detection has been extensively studied~\cite{zhang2010multi, chen2014fast, hpi, fk-ml, lin2023autobi}. However, existing PK/FK methods are designed for the classical database setting where input tables are assumed to be properly transformed already, which is insufficient for the end-to-end BI scenario we study. 

\section{Problem Definition} \label{sec:problem-def}


\ignore{
\begin{table}
\begin{center}
    \caption{Summary of symbols and notations}
    \vspace{-4mm}
    \label{tab:symbols}
    \setlength\tabcolsep{2 pt}
    \small \begin{tabular}{l l}
    \toprule
    Symbol              & Description \\
    \midrule
    $\mathcal{T}$, $T_i$ & set of input tables, a table in $\mathcal{T}$\\
    $C(T_i)$, $c_{ix}$ & set of column names of $T_i$, a column name in $C(T_i)$\\
    $\nit{Dom}(c_{ix})$ & domain of column $c_{ix}$\\
    $\mathbf{O}$, $O_j$ & set of operators, an operator in $\mathbf{O}$\\
    $P(O_j)$, $p_k$ & parameter space of $O_j$, one way to parameterize $O_j$\\
    $O_{jk}$ & transformation using $O_j$ parameterized by $p_k$\\
    $O_{jk}(T_i)$ & transformed table $T_i$ using $O_j$ parameterized by $p_k$\\
    $J(T_i, T_j)$ & table join between $T_i$ and $T_j$ \\
    $M(\mathcal{T})$ & BI model of $\mathcal{T}$, defined as a set of table joins \\
    $\mathcal{S}$ & a set of transformation sequences \\
    $G, G(T_i)$ & search graph, search tree for $T_i$ \\
    $V_t, E_t, E_{\text{join}}$ & transformation nodes, transformation edges, join edges\\
    \bottomrule
    \end{tabular}
    \end{center}
\end{table}
}

We now define our high-level \emph{Most-Probable BI-Prep (MPBP)} problem, as finding the most likely data preparation (join and transform) steps, for a given set of input tables, such that the joint probability of these (join and transform) steps is maximized. We will state the high-level MPBP problem below first, before making it a concrete graph optimization problem (in Section~\ref{sec:steiner-tree}).





\begin{definition}[Most-Probable BI-Prep (MPBP)].
\label{def:highlevel-problem}
    Given a set of $n$ input tables $\mathcal{T} = \{T_i | i \in [n]\}$, and a space of transformation operators $\mathbf{O} = \{\code{no}\text{-}\code{op}, $ $ \code{transpose}, $  $\code{unpivot},$ $ \code{pivot}, $ $ \code{split}, \code{concatenate}, $ $\code{substring}, \ldots\}$, where each operator $O \in \mathbf{O}$ can be parameterized as $O(P)$, using parameter $P$ drawn from a space of parameters $\mathcal{P}$.  
    For each input table $T_i \in \mathcal{T}$, let $S_i = (O_1(P_1), O_2(P_2), \ldots)$ be a candidate sequence of appropriately parameterized transformations on $T_i$, and let $p(S_i | T_i)$ be the probability of transformation $S_i$ given the input table $T_i$. Denote by $S_i(T_i)$ the transformed version of $T_i$, $\mathcal{S} = \{S_i | i \in [n]\}$ the set of all transformations for each $T_i \in \mathcal{T}$, and by $\mathcal{S(T)} = \{S_i(T_i) | i \in [n]\}$ the set of all transformed tables. Finally, let  $J(\mathcal{S(T)})$ be the candidate joins found on the transformed $\mathcal{S(T)}$. 
    
    The \emph{Most-Probable BI-Prep (MPBP)} problem is to find the optimal set of transformations $\mathcal{S}^*$,  such that the overall transformation probability, written as $p(\mathcal{S}|\mathcal{T}) = \prod_{i \in [n]}{p(S_i|T_i)}$, and the join probability of $\mathcal{S(T)}$, written as $p(J(\mathcal{S(T)}))$, are jointly maximized:
\begin{align}
\label{eqn:highlevel-obj-function}
    \mathcal{S}^* & = \argmax_{ \mathcal{S}}~~p(\mathcal{S}|\mathcal{T}) \cdot  p(J(\mathcal{S}(\mathcal{T}))) \\
    \label{eqn:highlevel-obj-function-expand}
   & = \argmax_{ \mathcal{S}} \left( \prod_{S_i \in \mathcal{S}}{p(S_i|T_i)} \right) \cdot  p(J(\mathcal{S}(\mathcal{T}))) 
\vspace{-2mm}
\end{align}
\end{definition}
\vspace{-2mm}
Note that both the term $p(\mathcal{S}|\mathcal{T})$, to estimate the probability of transformations, and the term $p(J(\mathcal{S(T)}))$, to estimate the probability of joins, have been studied separately in the literature (e.g.,~\cite{li2023autotables, foofah, yan2020auto-suggest} for transformations, and ~\cite{chen2014fast, hpi, fk-ml, lin2023autobi} for joins). We  will describe how we enhance and integrate these estimations in our holistic graph formulation in Section~\ref{sec:offline-training}. 

The main challenge of MPBP lies in the need to \emph{jointly optimize} both transforms and joins across all tables, so that the predicted transformations $\mathcal{S}$ not only have high transformation probabilities, but also the transformed tables $\mathcal{S}(\mathcal{T})$ should have high join probabilities. This is important, because greedily finding the most likely transform/join in isolation often leads to sub-optimal solutions, as we illustrate in the following example.


\begin{example}
\label{ex:def1}
We revisit Example~\ref{ex:case1}, and use $T_1$ to $T_4$ to denote Table~\ref{tab:example-fertility} to Table~\ref{tab:example-country} for short. Recall that the desired transformations for $T_1$ to $T_4$, denoted as $S_1$ to $S_4$, are: 

\begin{small}
$S_1 = (\code{Unpivot}(\text{``2010'', ``2012''}))$

$S_2 = (\code{no}\text{-}\code{op})$ 

$S_3 = (\code{Pivot}(\text{``Metric''},\text{``Value''}),$ $\code{Substring}(\code{Split} $ $ (\text{``Line-ID''},   \text{``-''})[1]), 2, 4)$

$S_4 = (\code{Transpose}())$
\end{small}

Let $\mathcal{S}^{*} = \{S_1, S_2, S_3, S_4\}$ be a candidate solution, and let the transformation probability of $p(S_1|T_1)$, $p(S_2|T_2)$, $p(S_3|T_3)$, $p(S_4|T_4)$ be $0.8$, $0.8$, $0.8$, and $0.4$, respectively. 
The overall transformation probability $p(\mathcal{S^{*}|T})$ is then $(0.8)^3 \cdot 0.4$. Furthermore, on the transformed tables $\mathcal{S^*(T)}$, let the join probability of the resulting joins  (color-coded in Table~\ref{tab:example-fertility-after} to Table~\ref{tab:example-country-after}), be $p(J(\mathcal{S^*(T)})) = 0.9$. The overall objective function in Equation~\eqref{eqn:highlevel-obj-function} is then $(0.8)^3 \cdot 0.4 \cdot 0.9 = 0.18$. 

Now suppose there is an alternative transformation for $T_4$, $S_4' = (\code{no}\text{-}\code{op})$ (or not transposing $T_4$), which has a slightly higher probability $p(S_4'|T_4) = 0.6$. If we were to optimize transformations alone, independent of the joins, we may find $\mathcal{S}^- = \{S_1, S_2, S_3, S_4'\}$ to have better transformation probability, $p(\mathcal{S^{-}|T}) = 0.8^3 \cdot 0.6$, which is greater than $p(\mathcal{S^{*}|T}) = (0.8)^3 \cdot 0.4$. 

However, when we bring the ``goodness'' of joins into the picture, we find the join probability $p(J(\mathcal{S^-(T)})) = 0.5$ to be considerably inferior to $p(J(\mathcal{S^*(T)})) = 0.9$, because not transforming $T_4$ (no-op) leads to worse overall joins, when the original $T_4$ cannot join with any other table. 
Overall, because $p(J(\mathcal{S^*(T)})) \cdot p(\mathcal{S^{*}|T}) = (0.8)^3 \cdot 0.4 \cdot 0.9 = 0.18$ is better than  $p(J(\mathcal{S^-(T)})) \cdot p(\mathcal{S^{-}|T}) = (0.8)^3 \cdot 0.6 \cdot 0.5 = 0.15$, the MPBP formulation in Definition~\ref{def:highlevel-problem} would select $\mathcal{S}^*$, which are the desired transformations in Example~\ref{ex:case1}.
\end{example}

Since we are searching transform/join candidates across all $n$ tables simultaneously, it is easy to see that the search space of MPBP can become intractable quickly. This is because for each input table, with 7 operators, and hundreds of possible parameters for each operator, only considering one-step transformations already yields a search space of at least $(7 \cdot 100)^n$, which is before we even consider multi-step transformations, and different join paths. 

In the following, we describe how to solve MPBP using principled graph optimizations, both efficiently and with quality guarantees.

\section{\sys: Overview}

\begin{figure}[t!]
    \iftoggle{full}
    {
    }
    {
        \vspace{-8mm}
    }
    \centering
    \includegraphics[width=0.98 \linewidth]{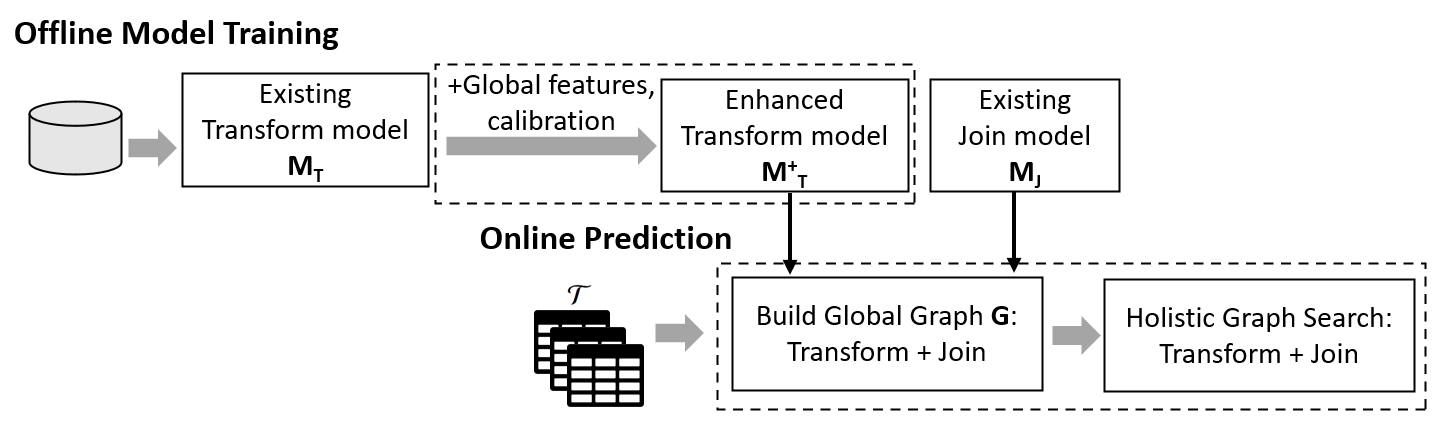}
    \vspace{-4mm}
    \caption{Architecture overview of \sys. 
    }
    \vspace{-3mm}
    \label{fig:sys-overview}
\end{figure}

In this section, we give an overview of how we solve MPBP using \sys.
The architecture of our system is shown in Figure~\ref{fig:sys-overview}, which at a high level has (1) an offline model training step, and (2) an online prediction step using graph-based optimizations. 



In the \underline{offline model training} stage, shown in the top half of Figure~\ref{fig:sys-overview}, we leverage existing join prediction models~\cite{lin2023autobi, fk-ml},  denoted as $M_J$ in the figure, as well as existing transformation prediction models $M_T$ designed for single tables (e.g. Pivot~\cite{yan2020auto-suggest}, Unpivot~\cite{yan2020auto-suggest}, String-Transform~\cite{zhu2017auto}). Using all input tables holistically, we enhance the transformation classifiers using global-level features, which we denote as $M_T^+$, 
to better estimate $p(\mathcal{S}|\mathcal{T})$ in Equation~\eqref{eqn:highlevel-obj-function}. 
We will highlight the key intuition in the offline part in Section~\ref{sec:offline-training} . 

The \underline{online prediction} phase, shown in the lower half of Figure~\ref{fig:sys-overview}, is what we consider the core contributions of \sys, where we holistically consider all transformation and join candidates across all tables defined in MPBP (Definition~\ref{def:highlevel-problem}). 
Specifically, for a given set of input tables $\mathcal{T}$, we construct a global search graph $G$, where we model each (original or transformed) table as a vertex, and all transformation or join candidates as edges, using probabilities from $M_T^+$ and $M_J$ models as edge-weights. We show that MPBP can then be equivalently solved on the graph, using a new graph algorithm we develop. We describe the core algorithm in Section~\ref{sec:steiner-tree} in detail.

It should be noted that our \sys framework is general and extensible, as we can extend and ``plug-in'' new transformation operators (together with their model $M_T$), while still utilizing the same graph algorithm, which is a salient feature of \sys. 





\section{Offline Model Training} \label{sec:offline-training}
We now describe the offline part of \sys (the upper half of Figure~\ref{fig:sys-overview}). Note that while it is an integral part of our system (and we create useful enhancements to existing transformation models), we do not consider these to be core contributions of this work. We will therefore only highlight the key intuition here for completeness,  leaving more details to
\iftoggle{full}
{Appendix~\ref{ap:local-features}.
}
{the technical report~\cite{full}.
}

\subsection{Transformation models $\mathbf{M_T}$}
\label{sec:transform_model}

Recall that in MPBP of Definition~\ref{def:highlevel-problem},   we need to estimate the transformation probability
$p(\mathcal{S}|\mathcal{T}) = \prod_{i \in [n]}{p(S_i|T_i)}$ in Equation~\eqref{eqn:highlevel-obj-function}. Given that each $S_i$ is a sequence of transformations $S_i = \{O_1, O_2, O_3,$ $ \ldots \}$\footnote{We write $O_1(P_1), O_2(P_2)$ in the original notation of  Definition~\ref{def:highlevel-problem}, simply as $O_1, O_2$, omitting the parameters $P_1, P_2$ for simplicity,   when the context is clear.}, the probability $p(S_i|T_i)$ can in turn be estimated as:
\begin{equation}
p(S_i|T_i) = p(O_1 | T_i) \cdot p(O_2 | O_1(T_i)) \cdot p(O_3 | O_2(O_1((T_i)))  \ldots
\end{equation}
where each $p(O|T)$ denotes the estimated  probability of  transformation $O$ given table $T$. 

Note that the problem of estimating $p(O|T)$ has been studied in program synthesis, where transformations need to be predicted based on a table $T$, often using machine learning classification models~\cite{yan2020auto-suggest, zhu2017auto}, which we write as $M_T$. 

In \sys, we build on top of these existing models $M_T$, and use the same $p(O|T)$ abstraction to estimate the probability of each transformation $O$. Additionally, we  perform enhancements to $M_T$, and create $M_T^+$, after observing that we operate on \emph{a set} of tables $\mathcal{T}$ in the BI setting (as opposed to one individual table), which presents opportunities for leveraging global signals from all tables in $\mathcal{T}$. 
\iftoggle{full}
{
    We illustrate the intuition using an example below.


    
    
    
    
    \begin{example}
        We revisit Example~\ref{ex:holistic}, and focus on the \codeq{Country} table $T_4$ (Table~\ref{tab:example-country}), where the desired transformation $S_4 = (\code{Transpose}())$. 
        
    
        Recall that for the \codeq{Country} table, looking at this table alone it may not be clear whether an \codeq{Transpose} is required. However, the strong overlap between the first row of Table~\ref{tab:example-country} and the \codeq{Country} column in Table~\ref{tab:example-fertility}, together with the overlap between the second row of Table~\ref{tab:example-country} and the \codeq{Code} column in Table~\ref{tab:example-econ}, provides global-level signals from other tables, that Table~\ref{tab:example-country} is not oriented correctly, and a \codeq{Transpose} is required.
    \end{example}
    
    We therefore develop a number of global-level features leveraging all tables in $\mathcal{T}$, which enhance $M_T$ by making the models globally-aware. Some of the key global features include:
    (1) \underline{Column-header-overlap} that captures the column-header overlap between table $T_i$ and other tables $T_j \in \mathcal{T} \setminus \{T_i\}$ in the same BI project; (2) \underline{Value-domain-overlap} that captures how much column value domain overlap between table $T_i$ and other tables $T_j \in \mathcal{T} \setminus \{T_i\}$; and (3) \underline{Headers-value-overlap} that captures the overlap between column-header of $T_i$ and column values of $T_j \in \mathcal{T} \setminus \{T_i\}$. All these global-level features provide signals for whether column-headers are oriented correctly and whether reshaping operators like transpose may be needed. Additional details of the features can be found in Appendix~\ref{ap:local-features}.
}
{
    We leave details of these $M_T^+$ models to our technical report~\cite{full} in the interest of space, since we do not consider them our core contributions in \sys.
}

We use boosted decision trees~\cite{Chen2016xgboost} to train our transformation models $M_T^+$, whose outputs are calibrated to true probabilities, or $p(O|T) \in [0, 1]$, using probability calibration techniques in machine learning~\cite{sklearn-calibration}. This makes sure that $p(O|T)$ from different transformation operators are true probabilities, that are comparable on the same scale.

\subsection{Join models $\mathbf{M_J}$}
\label{sec:join_model}

Recall that for a candidate set of transformations $\mathcal{S}$, the objective function of MPBP in Equation~\eqref{eqn:highlevel-obj-function} has another term $p(J(\mathcal{S(T)}))$, which estimates the ``goodness'' (probability) of all joins, for a set of transformed tables $\mathcal{S(T)}$ that are induced by $\mathcal{S}$.

Since join prediction has been extensively studied in the literature~\cite{zhang2010multi, chen2014fast, hpi, fk-ml, lin2023autobi}, we use a standard abstraction for join probabilities as follows. Let $M_J$ be a join model (e.g., a regression model), that predicts the join probability between two columns $(C, C')$, as $M_J(C, C') \in [0,1]$. We then use $\tilde{p}(T, T')$ to denote the join probability between any two given tables $(T, T')$, defined as:
\begin{equation}
\label{eqn:join-model-unnormalized}
\tilde{p}(T, T') = \max_{C \in T, C' \in T'}{M_J(C, C')}
\end{equation}
which estimates the join probability between $(T, T')$, as the maximum join probability between any two columns $C \in T, C' \in T'$.

We use probability calibration~\cite{sklearn-calibration, platt1999probabilistic} to ensure that $\tilde{p}(T, T') \in [0, 1]$, and are true probabilities, where $1$ indicates full confidence that $T, T'$ should join, $0$ indicates full confidence that they should not join, while $0.5$ is right on the decision boundary. Given that  $\tilde{p}(T, T')$ is true probability,  we use 
$\tilde{p}(T, T') > 0.5$ as the cutoff for whether two tables should join, similar to prior work~\cite{lin2023autobi}.

As a final normalization step, since we want to compare the goodness of joins using between tables predicted to join (i.e., any table-pairs with $\tilde{p}(T, T') < 0.5$ would not join and therefore should not contribute to the joinability score), we use a normalized version of $\tilde{p}(T, T')$, written as $p(T, T')$:
\begin{equation}
\label{eqn:join-model}
p(T, T') = \max( \tilde{p}(T, T'), 0.5)
\end{equation}
which is lower-bounded at $0.5$ (the decision-boundary for joins), to normalize the effect of non-joinable table-pairs\footnote{The reason to lower-bound $p(T, T')$ at 0.5 in Equation~\eqref{eqn:join-model}, is to ensure that ``bad'' (non-joinable) candidates have equalized effects. For example, given two join candidates, $\tilde{p}(T, T') = 0.2$ and $\tilde{p}(T, T'') = 0.1$, the former should not contribute more to the overall objective-function than the latter (even though the score of the former is better), as neither will be selected to join in the final solution and, therefore, should be treated equally. Note that in the normalized $p(T, T')$, both will have the identical score $0.5$, normalizing the score difference of such non-joinable candidates.}. 
We will see how $p(T, T')$ is used as edge-weights in our graph formulation next.  

\section{Online Global Graph Search} \label{sec:steiner-tree}
We are now ready to describe the core online phase of \sys, which is our graph-based formulation and its graph algorithms. 


\subsection{Graph representations} \label{sec:graph-rep}
In this section, we will introduce how we represent the entire search space, including transformation steps and join candidates across all tables, using a graph in a unified manner.

\textbf{Transformation trees}. We will start by representing all possible transformations that originate from a single input table, $T \in \mathcal{T}$, using a tree we call \emph{transformation-tree}, denoted by $G(T)$.

\begin{definition}
\label{def:tree}
[Transformation tree $G(T)$]. Let $T \in \mathcal{T}$ be an input table, $O \in \mathcal{O}$ be a predicted transformation on $T$, 
with probability $p(O|T)$. Let $O(T)$ be a transformed version of $T$ after applying $O$, and $O'(O(T))$ in turn be a transformed version of $O(T)$ after applying $O'$, etc., up to $m$-level deep. 

We construct a weighted tree $G(T)=(V(T), E(T))$ to represent all possible transformations from $T$ as follows. We first represent the input table $T$ as a vertex $v(T)$, which is the root of $G(T)$.

We represent each of $T$'s transformed descendants $O(T)$,  $\forall O\in \mathcal{O}$, also as a vertex, written as $v(O(T))$. We connect $v(O(T))$ and $v(T)$ with an edge $e(O)$ to 
represent the transformation $O$ on $T$, where the edge weight $w(e(O))$ is the transformation probability $P(O|T)$. This forms the first level of nodes in $G(T)$, or $m=1$.

Similarly and recursively, each transformed descendant of  $O(T)$ using transformation $O'\in \mathcal{O}$, written as $O'(O(T))$, is also represented as a vertex $v(O'(O(T)))$, which is connected to $v(O(T))$ 
with an edge, 
whose edge weight is set similarly as $P(O'|O(T))$. This forms additional levels of nodes in $G(T)$, for $m>1$.

The resulting $G(T) = (V(T), E(T))$ is our \emph{transformation tree} for one input table $T$, up to $m$ levels (we use $m=2$ in this work).
\end{definition}

\begin{figure}

    \centering
    \includegraphics[width=0.7 \linewidth]{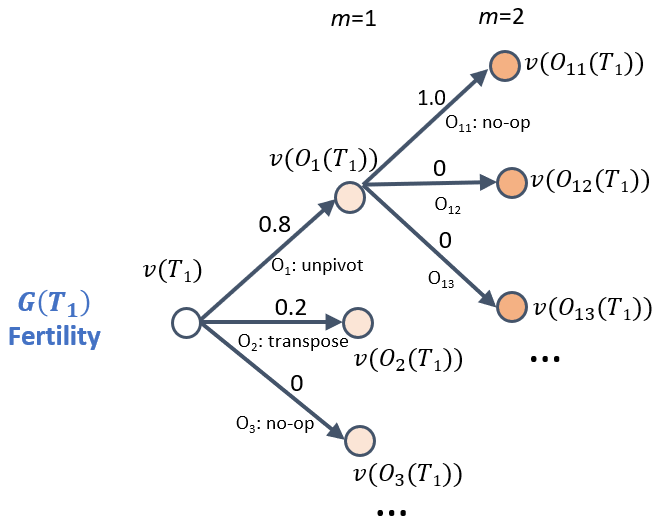}
    \vspace{-4mm}
    \caption{Transformation tree $G(T_1)$ to represent all possible transformations on the \code{Fertility} table $T_1$ (Table~\ref{tab:example-fertility}). Edge weights represent transformation probabilities (from $M_T$).}
    \vspace{-2mm}
    \label{fig:ex-tree}
\end{figure}

We illustrate transformation trees $G(T)$ with an example below. 
\begin{example} \label{ex:tree} We revisit Example~\ref{ex:def1}, and focus on the \codeq{Fertility} table in Table~\ref{tab:example-fertility}, which we refer to as $T_1$. We explain how to represent all possible candidate transformations on this table $T_1$, using 
a transformation tree  $G(T_1)$ shown in Figure~\ref{fig:ex-tree}. 

We first represent the \codeq{Fertility} table $T_1$ as a root node $v(T_1)$ in the figure. Suppose the candidate transformations on $T_1$ include   $O_1= \code{Unpivot}(\text{``2010'', ``2012''})$, $O_2=\code{Transpose}()$,~ $O_3=\code{no}\text{-}\code{op}$, etc., where $O_1$ is the desired transformation for $T_1$, to produce $O_1(T_1)$  that corresponds to the transformed table in Table~\ref{tab:example-fertility-after}.

We represent each transformation $O_i$ as an edge from $v(T_1)$, weighted by transformation probability $p(O_i|T_1)$, and each  table resulting from a transformation $O_i$ as a vertex $v(O_i(T_1))$ in $G(T_1)$. This forms the first level of nodes in the tree ($m=1$). For example, for $O_1= \code{Unpivot}$, with predicted probability $p(O_1|T_1) = 0.8$, then the edge weight between $v(T_1)$ and $v(O_1(T_1))$ is set to $p(O_1|T_1) = 0.8$, as shown in Figure~\ref{fig:ex-tree}. The same is true for $O_2$, $O_3$, etc.

Note that each transformed table $O_i(T_1)$ may be transformed again, which are represented as a second level of edges in the transformation tree ($m=2$). For example, we may predict $O_{11}, O_{12}, O_{13}$ on the transformed table $O_1(T_1)$, with probability $1.0$, $0$, and $0$, respectively, like represented in the figure.
\end{example}

\textbf{Global search graph}. We can now connect all transformation trees $G(T_i), \forall T_i \in \mathcal{T}$, to represent the global search graph $G(\mathcal{T})$ for both transformations and joins, defined as follows.

\begin{definition}
\label{def:graph}
[Global search graph $G(\mathcal{T})$].   Let $G(T_i) = (V(T_i),$ $E(T_i))$ be the transformation tree for each $T_i \in \mathcal{T}$ in Definition~\ref{def:tree}. The set of transformation trees, $\{G(T_i) | T_i \in \mathcal{T}\}$, induces a \emph{global search graph} $G(\mathcal{T}) = (V(\mathcal{T}), E(\mathcal{T}))$, where the vertex set $V(\mathcal{T}) = \bigcup_{T_i \in \mathcal{T}}{V(T_i)}$ is simply the union of vertices in $G(T_i)$, whereas the edge set $E(\mathcal{T}) = E_T(\mathcal{T}) \cup E_J(\mathcal{T})$ consists of two types edges, transformation-edges $E_T(\mathcal{T})$ and join-edges $E_J(\mathcal{T})$:

(1) \emph{Transformation-edges} $E_T(\mathcal{T})$, defined as $E_T(\mathcal{T}) = \bigcup_{T_i \in \mathcal{T}}{E(T_i)}$, is simply the union of all transformation-edges $E(T_i)$ in $G(T_i)$, where each edge $e \in E_T(\mathcal{T})$ represents a transformation, whose edge weight is its transformation probability $w(e) = p(O|T)$, based on the transformation model $M_T$ trained offline (Section~\ref{sec:transform_model});

(2) \emph{Join-edges} $E_J(\mathcal{T})$, written as $E_J(\mathcal{T}) = \{ e(v, v') | v \in V_{\text{leaf}}(T_i),$ $v' \in V_{\text{leaf}}(T_j), i \neq j\}$,  represent all candidate joins between tables, where each edge $e(v, v') \in E_J(\mathcal{T})$ stands for a candidate join between two tables represented by leaf vertices $(v, v')$\footnote{Note that \codeq{no-op} is also a predicted transformation, therefore leaf vertices can naturally represent both transformed tables, as well as un-transformed original tables.}, from two different transformation trees $G(T_i)$ and $G(T_j)$. The edge weight of each join-edge $e(v, v')$ is the normalized join probability  $w(e) = p(v, v')$ from the join model $M_J$ (Equation~\eqref{eqn:join-model} in Section~\ref{sec:join_model}).
\end{definition}


\iftoggle{full}
{
    \begin{algorithm}[h!]
    \begin{scriptsize}
    
    \SetKwComment{Comment}{/* }{ */}
    \SetKwData{transformationOnlyModel}{transformationOnlyModel}
    \SetKwData{extractFeatures}{extractFeatures}
    \SetKwData{getParentNode}{getParentNode}
    \SetKwData{transformClassifier}{transformClassifier}
    \SetKwData{joinModel}{joinModel}
    \SetKwData{TransformationSearch}{BIPrepSearch}
    \SetKw{Not}{not}
    \SetKw{Or}{or}
    \caption{Build global search graph $G(\mathcal{T})$}
    \label{alg:graph}
    \KwIn{$\mathcal{T}$ (input tables)}
    \KwOut{$G(\mathcal{T})$ (global search trees)}

    \ForEach{$T_i \in \mathcal{T}$}{ 
    
        $V(T_i) = \{ v(T_i) \} \cup \{ v(O(T_i)) \cup \{ v(O'(O(T_i))) | \forall O, O' \in \mathcal{O} \} $\;

        $E_T(T_i) = \{ (v(T_i), v(O(T_i)) | \forall O \in \mathcal{O} \}  \}$\;
        $E_T(T_i) = E_T(T_i) \cup \{ (v(O(T_i)), v(O'(O(T_i))) | \forall O', O \in \mathcal{O} \}$\;
        
    }

    $V(\mathcal{T}) = \bigcup_{T_i \in \mathcal{T}} V(T_i)$
    
    $E_T(\mathcal{T}) = \bigcup_{T_i \in \mathcal{T}} E_T(T_i)$

    \ForEach{$e =(T, O(T)) \in E_T$}{ 
    
        $w(e) = p(O|T)$ \tcp{from Section~\ref{sec:transform_model}}
    
    }

    $E_J(\mathcal{T}) = \{ (v, v') | v \in V_{\text{leaf}}(T_i), v' \in V_{\text{leaf}}(T_j)\}$
    
    \ForEach{$e =(v, v') \in E_J$}{ 
    
        $w(e) = p(v, v')$  \tcp{from Equation~\ref{eqn:join-model} in Section~\ref{sec:join_model}}
    
    }

    return $G(\mathcal{T}) = (V(\mathcal{T}), E_T(\mathcal{T}) \cup E_J(\mathcal{T}) )$;
    \end{scriptsize}
    \end{algorithm}

    The pseudo-code for building global search graphs is shown in Algorithm~\ref{alg:graph}.
    We illustrate the global search graph using an example.
}
{
    The pseudo-code for building global search graphs can be found in~\cite{full}.
    We illustrate the global search graph using an example.
}


\begin{figure}[t!]
\iftoggle{full}
{
}
{
}
    \centering
    \includegraphics[width=0.9 \linewidth]{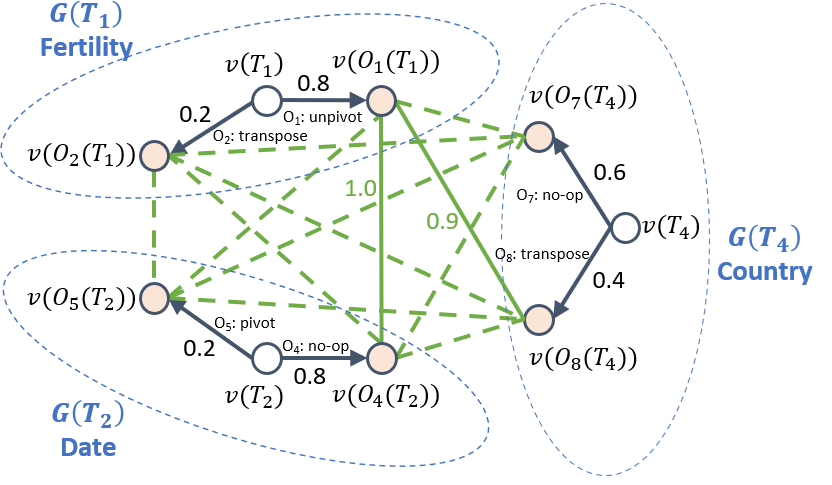}
    \vspace{-6mm}
    \caption{Global search graph for Example~\ref{ex:search-graph}. Each dashed-circle corresponds to a transformation tree $G(T_i)$, whose black edges represent possible transformations for table $T_i$. Join candidates are represented as green edges between transformation trees, where joinable tables are represented by solid edges (with $w(e) > 0.5$), and non-joinable ``placeholder edges'' are represented as dashed green edges (with $w(e) = 0.5$).}
    \vspace{-2mm}
    \label{fig:ex-graph}
\end{figure}

\begin{example} 
\label{ex:search-graph}
    We continue with Example~\ref{ex:tree}, and use
    Figure~\ref{fig:ex-graph} to represent the global search graph $G(\mathcal{T})$ for the tables in Example~\ref{ex:def1}. For simplicity, we show only the first level of transformations for each transformation tree $G(T_1), G(T_2), G(T_4)$, marked in dashed-circles. We omit table $T_3$ from Example~\ref{ex:def1} here (for $T_3$ requires two-level transformations that is too big to show on the figure). 

    Note that $G(T_1)$ in $G(\mathcal{T})$ directly corresponds to the transformation tree for $T_1$ in Figure~\ref{fig:ex-tree} (Example~\ref{ex:tree}), where $O_1$ is the desired \codeq{Unpivot} transformation for $T_1$ with probability $p(O_1|T_1) = 0.8$, $O_2$ is the second-ranked \codeq{Transpose} with probability $p(O_2|T_1) = 0.2$, etc. We omit additional transformations like $O_3$, and the second level of transformations like $O_{11}, O_{12}$  from Figure~\ref{fig:ex-tree}, to  simplify this illustration and avoid clutter. 
    
    We construct $G(T_2)$ and $G(T_4)$ similarly, which are shown in their respective circles. Note that all black edges in the graph are now  the transformation-edges $E_T(\mathcal{T})$ in Definition~\ref{def:graph}.

Finally, we construct join-edges $E_J(\mathcal{T})$ in $G(\mathcal{T})$, between any two vertices $(v, v')$  at the leaf-level of two different transformation trees $G(T_i)$ and $G(T_j)$,  represented as green edges. The edge-weight of each $(v, v')$ corresponds to the joinability of the two tables $(v, v')$, computed by invoking the join models $M_J$  (Section~\ref{sec:join_model}). 

In the figure, we use solid green edges to represent join candidates $e = (v, v')$ that are predicted as joinable, whose edge weights  $w(e) = p(v, v') \geq 0.5$. For example, $(v(O_1(T_1)), v(O_4(T_2)))$ and $(v(O_1(T_1)), v(O_8(T_4)))$ are clearly joinable, with edge-weights 1.0 and 0.9, respectively.  The remaining non-joinable candidates are represented as dashed green edges, which are ``placeholder join edges'' for non-joinable candidates, whose edge weights $w(e)$  are all set to the constant $0.5$ (Equation~\ref{eqn:join-model}).\footnote{This normalizes scores for all non-joining edges (since none of them will be selected in the final solution), to equalize their effect on the objective function 
(Section~\ref{sec:join_model}).}
\end{example}

\ignore{
\begin{figure}
    \centering
    \includegraphics[width=0.55 \linewidth]{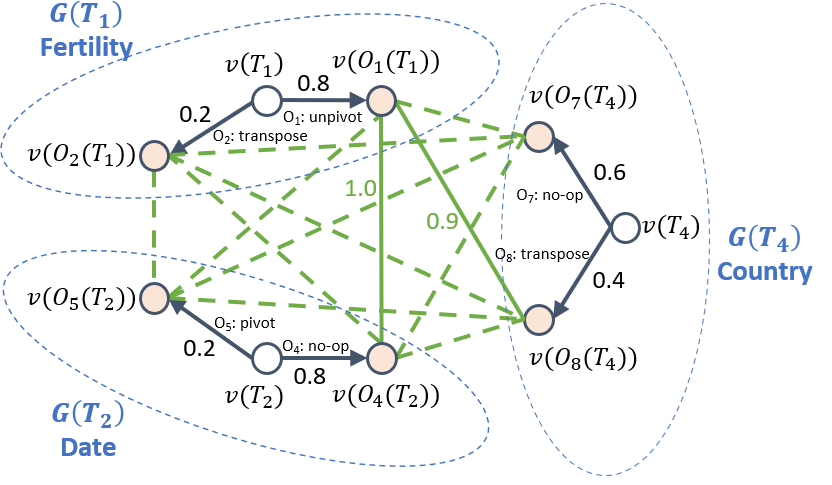}
    \vspace{-4mm}
    \caption{Example search graph consisted of one-level trees, with transformation-edges $E_T$ (gray), join-edges $E_J$ (green).}
    \vspace{-4mm}
    \label{fig:ex-graph-unweighted}
\end{figure}

\begin{example}
    Figure~\ref{fig:ex-graph-unweighted} shows an example $G$ that consists of trees $\{G(T_1), G(T_2)\}$.
\end{example}
}

\subsection{Solve MPBP using global search graph} \label{sec:simple}

\begin{figure*}[t]
\iftoggle{full}
{
}
{
    \vspace{-20mm}
}
    \centering
    \begin{minipage}{0.45\textwidth}
        \centering
    \includegraphics[width=0.9\linewidth]{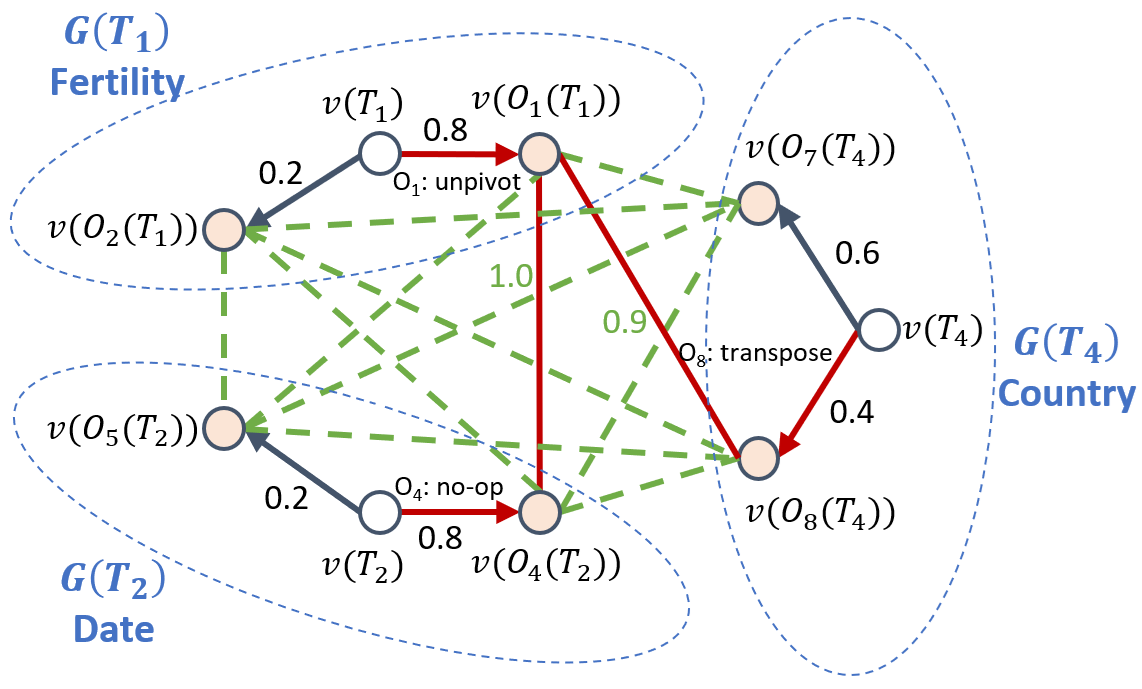}
    \vspace{-4mm}
    \caption{An optimal solution for the search graph in Figure~\ref{fig:ex-graph}, that corresponds to the solution $\mathcal{S}^*$ in Example~\ref{ex:def1}.}
    \vspace{-2mm}
    \label{fig:ex-solution-optimal}
    \end{minipage}\hfill
    \begin{minipage}{0.45\textwidth}
        \centering
    \includegraphics[width=0.9\linewidth]{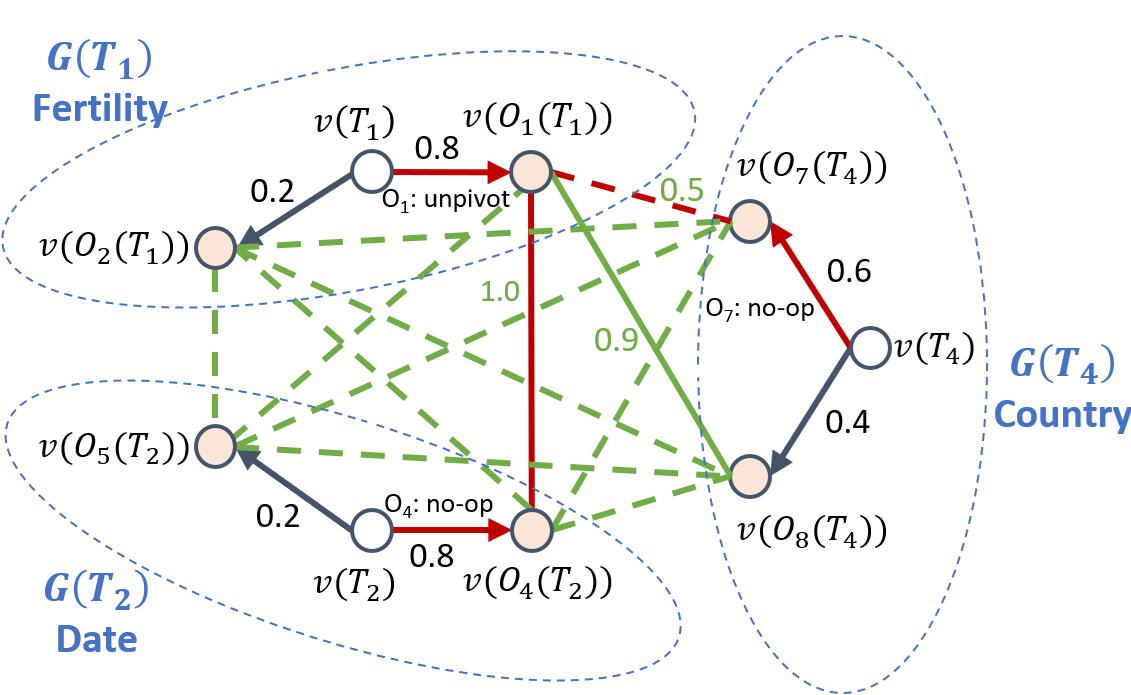}
    \vspace{-4mm}
    \caption{An inferior solution for the search graph in Figure~\ref{fig:ex-graph}, that corresponds to the solution $\mathcal{S}^-$ in Example~\ref{ex:def1}.}
    \vspace{-2mm}
    \label{fig:ex-solution-inferior}
    \end{minipage}
\end{figure*}



Using the global search graph $G(\mathcal{T})$, we now describe how the MPBP problem in Definition~\ref{def:highlevel-problem} can be solved using the new graph algorithms we design in this work.

Recall that in MPBP, given input tables $\mathcal{T}$, we need to select both (1) transformations $\mathcal{S}$, and (2) joins between the transformed tables $\mathcal{S}(\mathcal{T})$, denoted by $J(\mathcal{S}(\mathcal{T}))$, so that  both 
the ``goodness'' (probability) of the transformations  $p(\mathcal{S}|\mathcal{T})$, and the ``goodness'' (probability) of the joins $p(J(\mathcal{S}(\mathcal{T})))$, are jointly maximized in the objective function in Equation~\eqref{eqn:highlevel-obj-function}.

\textbf{Valid solutions to MPBP on search graph $\mathbf{G(\mathcal{T})}$}. We first show how a valid solution to MPBP would look like on $G(\mathcal{T})$.

First, the transformations selected in $\mathcal{S}$ for MPBP can naturally be represented as transformation-edges on  ${G(\mathcal{T})}$. Since we are only performing exactly one sequence of transformations for each table $T_i$, then in each transformation tree $G(T_i)$, the selected transformations will form a unique ``path'', from the root of $G(T_i)$, denoted by $v(T_i)$  representing the original input table $T_i$, to a unique leaf-vertex in $G(T_i)$, denoted by $v(L_i) \in V_{\text{leaf}}(G(T_i))$, representing a transformed table $L_i$ that originates from $T_i$. The unique path induced by $v(T_i)$ in $G(T_i)$, written as ${path}(v(T_i) \rightarrow v(L_i))$, naturally corresponds to the sequence of transformations selected for $T_i$. The union of such paths across all $G(T_i)$, is then the overall selected transformations, $\mathcal{S} = \{ {path}(v(T_i) \rightarrow v(L_i)) | i \in [n]\}$


Similarly, the joins selected in $J(\mathcal{S}(\mathcal{T}))$ can be represented as join-edges in the search graph $G(\mathcal{T})$. Because the selected joins $J(\mathcal{S}(\mathcal{T}))$ need to be based on the selected transformations $\mathcal{S}$, the join-edges need to be between leaf vertices representing transformed tables $\{v(L_i) | i \in [n] \}$ on the search graph $G(\mathcal{T})$. 

Since join relationships in BI settings often follow star/snowflake-schema~\cite{kimball2011data, lin2023autobi}, where given $n$ input tables, $(n-1)$ join-edges need to exist to ``connect'' all $n$ tables, we therefore use the $(n-1)$ most confident join-edges that can connect all $n$ tables, as our selected joins in $J(\mathcal{S}(\mathcal{T}))$.

A sub-graph in ${G(\mathcal{T})}$ with these required transformation-edges $\mathcal{S}$, and join-edges $J(\mathcal{S}(\mathcal{T}))$, is then a valid solution to MPBP, defined as follows.

\begin{definition}[Valid solutions to MPBP on search graph ${G(\mathcal{T})}$.] \label{def:valid-solution}
    Let ${G(\mathcal{T})}$ be the global search graph induced by transformation trees $\{G(T_i) $ $| T_i \in \mathcal{T} \}$ in Definition~\ref{def:graph}. Let $v(L_i)$ be a unique leaf-vertex selected in transformation tree $G(T_i)$, for all $i \in [n]$. 
    
    A \emph{valid solution to MPBP} on ${G(\mathcal{T})}$ is an edge set $E = E_T \cup E_J$, where $E_T = \{ e | e \in {path}( v(T_i) \rightarrow v(L_i) ), i \in [n]\}$ is the union of transformation-edges on the paths between root $v(T_i)$ and leaf $v(L_i),  \forall i \in [n]$, and $E_J =\{(v(L_i), v(L_j))\}$, $|E_J| = (n-1)$, are $(n-1)$ join-edges between vertices in $\mathcal{L} = \{v(L_i) | i\in [n]\}$, that form a spanning tree to connect $n$ transformed tables represented by $\mathcal{L}$. 
\end{definition}

We use an example to show valid solutions on search graphs.
\begin{example}
\label{ex:valid-graph-solution}  
Figure~\ref{fig:ex-solution-optimal} and Figure~\ref{fig:ex-solution-inferior} show two valid solutions to MPBP, on the search graph $G(\mathcal{T})$ from Example~\ref{ex:search-graph}. The selected transformation-edges and join-edges are all marked in red. 

In Figure~\ref{fig:ex-solution-optimal}, the selected transformation-edges are:  $O_1$ (unpivot) on $T_1$, $O_4$ (no-op) on $T_2$, and  $O_8$ (transpose) on $T_4$, which are the same set of desired transformations in Example~\ref{ex:def1}. Between the three transformed tables $v(O_1(T_1))$, $v(O_4(T_2))$ and $v(O_8(T_4))$,  two join-edges,  $(v(O_1(T_1)), v(O_4(T_2)))$ and $(v(O_1(T_1)), v(O_8(T_4)))$, are marked in red, indicating that they are selected to join (which would correspond  to the yellow and green join path on transformed tables in Table~\ref{tab:example-fertility-after}-Table~\ref{tab:example-country-after}, respectively). 

Intuitively, we can see that the red edges correspond to a valid solution to MPBP, because in each transformation tree $G(T_1)$, $G(T_2)$ and $G(T_4)$, we select one and exactly one transformation path, from the root of the transformation tree, to a unique transformed table (leaf vertex), which are $v(O_1(T_1))$, $v(O_4(T_2))$ and $v(O_8(T_4))$, respectively. Furthermore, we select exactly 2 join-edges to form a spanning tree that connects $v(O_1(T_1))$, $v(O_4(T_2))$ and $v(O_8(T_4))$.

In Figure~\ref{fig:ex-solution-inferior}, we show another valid solution, that selects a different set of transformations in red: $O_1$ (unpivot) for $T_1$, $O_4$ (no-op) for $T_2$, and $O_7$ (no-op) for $T_4$. The selected 2 join-edges are similarly marked
in red. This is still a valid solution, because a unique transformation path is selected in each transformation tree $G(T_i)$, and 2 join-edges are selected to span all transformed tables.
\end{example}

\textbf{Goodness of solutions on search graph $\mathbf{G(\mathcal{T})}$}. Given that there exist many valid solutions (exponential in the number of tables), to evaluate the relative merits of these solutions, we resort to the objective function of MPBP in Equation~\eqref{eqn:highlevel-obj-function}, which has two components, $p(\mathcal{S}|\mathcal{T})$ for the goodness of transformations, and $p(J(\mathcal{S}(\mathcal{T})))$ for the goodness of joins.

The first term for transformations, $p(\mathcal{S}|\mathcal{T})$, can be expanded as $\prod_{S_i \in \mathcal{S}}{p(S_i|T_i)}$, where each $S_i =(O_{i1}, O_{i2}, \ldots )$ is the sequence of transformations selected for table $T_i$. This can further be expanded in terms of individual transformations $O_{ij}$ as: 
\begin{align}
p(\mathcal{S}|\mathcal{T})  =  \prod_{S_i \in \mathcal{S}}{p(S_i|T_i)}  
 = \prod_{O_{ij} \in S_i, S_i \in \mathcal{S}}{p(O_{ij})} \label{eqn:transform-expanded}
\end{align}
On the search graph $G(\mathcal{T})$, the term in Equation~\eqref{eqn:transform-expanded} is effectively the cross-product of probabilities (edge weights) of all selected transformation $p(O_{ij})$ in $\mathcal{S}$, which  naturally measures the collective likelihood of all selected transformations in $\mathcal{S}$.


Given a valid solution $E = E_T \cup E_J$ on $G(\mathcal{T})$ in Definition~\ref{def:valid-solution}, where $E_T$ is the selected transformation-edges, it can be written as: 
\begin{align}
\label{eqn:transform-cross-product}
p(\mathcal{S}|\mathcal{T}) & =  \prod_{e \in E_T}{w(e)} 
\end{align}
Similarly, the goodness of joins in $p(J(\mathcal{S}(\mathcal{T})))$, is also the cross-product of join-probability of all selected join-edges in $E_J$:
\begin{align}
\label{eqn:join-cross-product}
    p(J(\mathcal{S}(\mathcal{T}))) = \prod_{e \in E_J}{w(e)}
\end{align}
The overall objective function  $p(\mathcal{S}|\mathcal{T}) \cdot p(J(\mathcal{S}(\mathcal{T})))$ in Equation~\eqref{eqn:highlevel-obj-function}, can now be equivalently written as the product of Equation~\eqref{eqn:transform-cross-product} and~\eqref{eqn:join-cross-product}, where the optimal edge-set $E^*$ on $G(\mathcal{T})$ is:
\begin{align}
\label{eqn:obj_edge_unnormalized}
    E^* & = \argmax_{ E = E_T \cup E_J} \prod_{e \in E_T}{w(e)}  \prod_{e \in E_J}{w(e)}
\end{align}
We can now define the search problem on graph ${G(\mathcal{T})}$, that is the graph-equivalent of MPBP as follows.

\begin{definition} [Most-Probable BI-Prep on Graph (MPBP-G)].
\label{def:problem-on-graph}
Let ${G(\mathcal{T})}$ be the global search graph induced by transformation trees $\{G(T_i) $ $| T_i \in \mathcal{T} \}$. The \emph{MPBP on Graph (MPBP-G)} problem, is to select a set of edges $E = E_T \cup E_J$ from ${G(\mathcal{T})}$, such that $E$ is a valid solution on ${G(\mathcal{T})}$ as defined in Definition~\ref{def:valid-solution},  while the objective function in Equation~\eqref{eqn:obj_edge_unnormalized} is maximized.
\end{definition}



\begin{example}
\label{ex:optimal-graph-solution}
We revisit Figure~\ref{fig:ex-solution-optimal} and Figure~\ref{fig:ex-solution-inferior} from Example~\ref{ex:valid-graph-solution} to illustrate Definition~\ref{def:problem-on-graph}. The set of edges selected in  Figure~\ref{fig:ex-solution-optimal}, denoted by $E^*$,  corresponds to the optimal solution $\mathcal{S}^*$ in Example~\ref{ex:def1}. There are 3 transformation-edges in $E^*$, $O_1$ (unpivot) for $T_1$, $O_4$ (no-op) for $T_2$, and $O_8$ (transpose) for $T_4$, with probability 0.8, 0.8, and 0.4, respectively, whose collective transformation probability is then $(0.8)^2 \cdot 0.4$ = $0.256$ (Equation~\eqref{eqn:transform-cross-product}). It can be shown that in order to connect the transformed tables $v(O_1(T_1))$,  $v(O_4(T_2))$ and  $v(O_8(T_4))$ and make it a valid solution, the best join-edges are  $(v(O_1(T_1)), v(O_4(T_2)))$ and $(v(O_1(T_1)), v(O_8(T_4)))$ like indicated in Figure~\ref{fig:ex-solution-optimal}, whose overall join probability is $1.0 \cdot 0.9 = 0.9$ (Equation~\eqref{eqn:join-cross-product}). The overall objective function in Equation~\eqref{eqn:obj_edge_unnormalized} is then $0.256 \cdot  0.9 = 0.23$.

In comparison, the set of edges selected in  Figure~\ref{fig:ex-solution-inferior}, denoted by $E^-$, has transformation-edges $O_1$, $O_4$, and $O_7$, with probability 0.8, 0.8, and 0.6, respectively. The overall transformation probability is then $(0.8)^2 \cdot 0.6$ = $0.384$. To  connect the transformed tables $v(O_1(T_1))$,  $v(O_4(T_2))$ and  $v(O_7(T_4))$ using join-edges and make it a valid solution, the best possible join-edges are $(v(O_1(T_1)), v(O_4(T_2)))$ and $(v(O_1(T_1)), v(O_7(T_4)))$, whose overall join probability is $1.0 \cdot 0.5 = 0.5$, leading to an overall score of $0.384 \cdot  0.5 = 0.19$, making this $E^-$ inferior  to the solution $E^*$ in Figure~\ref{fig:ex-solution-optimal}.

Note that if we were to only consider transformations, $E^-$ would have a better transformation probability ($0.384$) than $E^*$ ($0.256$), which however is an inferior solution when both transformations and joins are considered, like shown above. This demonstrates the importance of holistic optimization in \sys.
\end{example}

\begin{THEOREM} \label{theorem:np-complete}
    The MPBP-G problem in Definition~\ref{def:problem-on-graph} is NP-complete. 
\end{THEOREM}

We show the hardness of MPBP-G using a reduction from Exact Cover by 3-Sets (X3C)~\cite{karp2010reducibility}, a proof of which can be found in
\iftoggle{full}
{Appendix~\ref{ap:theorem1}.
}
{\cite{full}.
}

Intuitively, we can also see that finding the optimal edges $E = E_T \cup E_J$ is hard --- given $m$ possible multi-step transformations (easily in the thousands when each transformation can be parameterized differently), and $n$ input tables, for $E_T$ alone there are $m^n$ options to choose from.\footnote{Note that figures like Figure~\ref{fig:ex-graph} are simplified for illustration, which show only 1-step transformations, and without considering parameter options for the same operator $O$.} The selection of optimal join-edges $E_J$ on top of $E_T$ further compounds the complexity, and makes the problem hard.


\subsection{Solving MPBP-G leveraging Steiner-tree
}
We now describe our graph algorithm for MPBP-G, that can solve MPBP-G both efficiently and with provable quality guarantees.

Observing that the cross-products of probabilities (edge weights) in Equation~\eqref{eqn:obj_edge_unnormalized} are not amenable to graph optimization, our first step is to apply logarithmic transformation on edge weights $w(e)$:
\begin{align}
\label{eqn:log-transform}
\overline{w}(e) = - \log(w(e))
\end{align}
This allows us to rewrite the objective function in Equation~\eqref{eqn:obj_edge_unnormalized} as:
\begin{align}
    E^* & = \argmax_{ E = E_T \cup E_J} \prod_{e \in E_T}{w(e)}  \prod_{e \in E_J}{w(e)} \\
    & = \argmin_{ E = E_T \cup E_J}  \sum_{e \in E_T}{\overline{w}(e)} + \sum_{e \in E_J}{\overline{w}(e)}  \label{eqn:obj_sum_edge}
\end{align}

Using the new edge weights $\overline{w}(e)$, we can instead minimize the \textit{sum} of edge weights in the selected edge-set $E$, which is amenable to graph-based optimizations.
More specifically, our MPBP-G problem in Definition~\ref{def:problem-on-graph}, with the objective function in Equation~\eqref{eqn:obj_sum_edge}, now draws a close parallel to the \emph{Minimum Steiner Tree (MST)} problem~\cite{kou1981fast, karp2010reducibility} in graph theory.

\begin{figure}[t!]
\iftoggle{full}
{
}
{
    \vspace{-15mm}
}
    \centering
    \includegraphics[width=0.8 \linewidth]{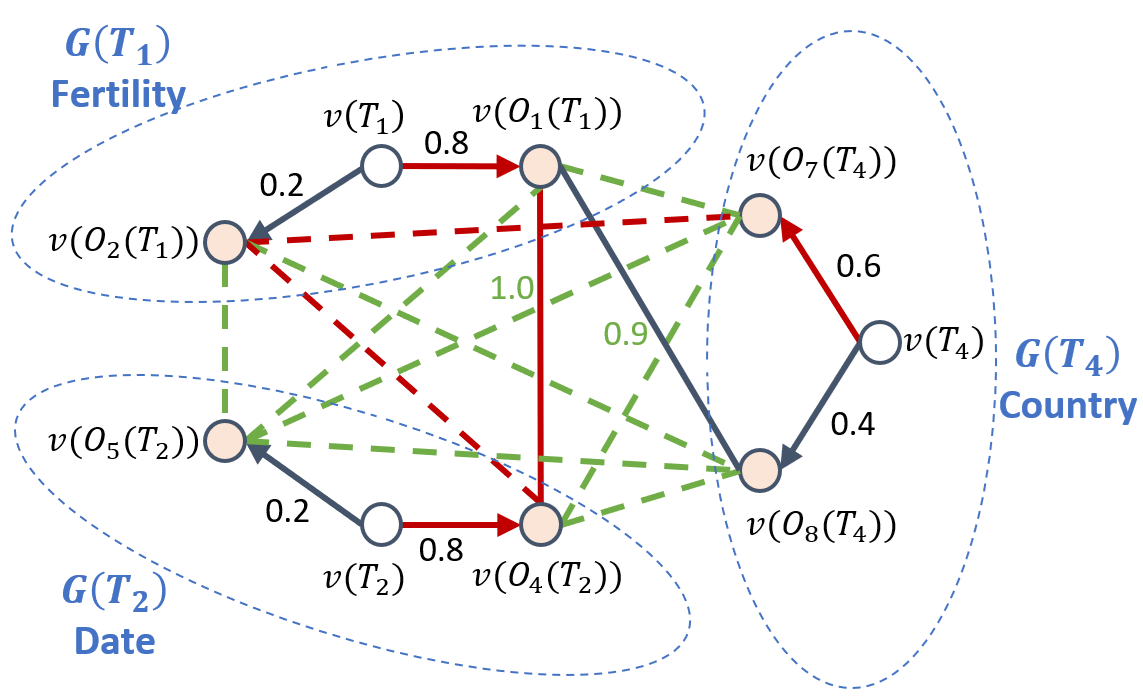}
    \vspace{-3mm}
    \caption{An invalid solution $E^x$ to MPBP-G, that is nevertheless a valid Steiner tree w.r.t terminals $R = \{ v(T_i) | i \in [n]\}$. 
    }
    \vspace{-2mm}
    \label{fig:invalid-solution}
\end{figure}

Recall that given a weighted graph $G=(V, E)$, and a subset of vertices $R \subseteq V$ known as the terminals, a \emph{valid Steiner tree} is a tree in $G$ that spans (or contains) all vertices in $R$. A \emph{minimum Steiner tree} is the valid Steiner tree with minimal total edge weights~\cite{kou1981fast, karp2010reducibility}.

In the MPBP-G problem, if we let the union of all root vertices $v(T_i)$ in transformation trees $G(T_i)$ be the terminals $R$ in the Steiner-tree problem, or $R = \{v(T_i) | i \in [n] \}$, we prove that the following connections between Steiner tree and MPBP-G hold:

\begin{proposition}
\label{prop:solution-is-steiner}
A valid solution to MPBP-G in Definition~\ref{def:valid-solution} must be a valid Steiner tree on the same graph $G(\mathcal{T})$, that spans the terminals defined by the root vertices of all transformation trees, $R = \{v(T_i) | i \in [n] \}$. 
\end{proposition}

\iftoggle{full}
{
A proof of Proposition~\ref{prop:solution-is-steiner} can be found in Appendix~\ref{ap:proof-prop-solution-is-steiner}.
}
{
A proof of Proposition~\ref{prop:solution-is-steiner} can be found in~\cite{full}.
}
Intuitively, to see why a valid solution to MPBP-G must be a valid Steiner tree for the given $R$, observe that for a valid solution to MPBP-G in Definition~\ref{def:valid-solution}, $E = E_T \cup E_J$, because $E_T$ would  connect each root $v(T_i)$  with a leaf-vertex $v(L_i)$ via a unique path, and the union of the leaf vertices $v(L_i)$, written as $\mathcal{L} = \{v(L_i) | i \in [n] \}$, would in turn be connected by $(n-1)$ edges in $E_J$ (which is the minimum number of edges to connect $\mathcal{L}$). The resulting $E = E_T \cup E_J$ therefore connects all terminals in $R$ and is a valid Steiner tree. 

Note that in Proposition~\ref{prop:solution-is-steiner}, we only state $E = E_T \cup E_J$ being a valid tree, with no statement about the minimality of such a tree. 

\begin{example}
\label{ex:steiner}
Figure~\ref{fig:ex-solution-optimal} and~\ref{fig:ex-solution-inferior} show two valid solutions to MPBP-G (red edges). Both can be verified as valid Steiner trees, as the selected edges form trees that connect all terminal vertices in $R = \{v(T_i) | i \in [n] \}$, or the root vertices of all transformation trees.
\end{example}

The reverse of Proposition~\ref{prop:solution-is-steiner}, however, is not always true, which means that not all valid Steiner trees on $G(\mathcal{T})$ are valid solutions to MPBP-G in Definition~\ref{def:valid-solution}, which we illustrate using an example.

\begin{example}
\label{ex:counter-ex}
    Figure~\ref{fig:invalid-solution} shows an example, where selected edges marked in red can be verified as a valid Steiner tree, since the red edges form a tree that connects all $R = \{v(T_1), v(T_2), v(T_4)\}$. However, it is not a valid solution to MPBP-G in Definition~\ref{def:valid-solution}, because the vertex $v(O_2(T_1))$ is incident on selected join-edges (red dashed lines in the figure), effectively used to join, but does not have corresponding transformation-edge (namely, $v(T_1) \rightarrow v(O_2(T_1))$), to create this transformed table first, which therefore is not a valid solution to MPBP-G  (Definition~\ref{def:valid-solution}).
\end{example}

What this means is that we could not hope to solve MPBP-G  by directly invoking Minimum Steiner Tree (MST) on $G(\mathcal{T})$, because solutions to MST may not be valid solutions to MPBP-G.
However, we prove the following two key properties of MPBP-G, which will allow us to construct principled algorithms to solve MPBP-G.

\begin{proposition}
\label{prop:solution-edge-count}
Any valid solution to MPBP-G in Definition~\ref{def:valid-solution} has exactly $(m \cdot n)$ transformation-edges and $(n-1)$ join edges, where $m$  is the depth of transformation trees (Definition~\ref{def:tree}), and $n$ is the number of input tables in MPBP-G.  
\end{proposition}

\iftoggle{full}
{
    A proof of Proposition~\ref{prop:solution-edge-count} can be found in Appendix~\ref{ap:proof-prop-solution-edge-cnt}.
}
{
    A proof of Proposition~\ref{prop:solution-edge-count} can be found in~\cite{full}.
}
To see why it holds, recall that valid solutions to MPBP-G requires a unique transformation-path (with exactly $m$ edges) for each transformation tree, which translates to $(m \cdot n)$ transformation-edges for $n$ transformation trees. Since a valid solution to MPBP-G also requires $(n-1)$ join edges to connect $n$ input tables, 
Proposition~\ref{prop:solution-edge-count} then follows. 

\begin{example}
\label{ex:exact-edge-cnt}
As intuitive examples for Proposition~\ref{prop:solution-edge-count}, we revisit the two valid solutions to MPBP-G in Figure~\ref{fig:ex-solution-optimal} and~\ref{fig:ex-solution-inferior}. Given $n=3$ input tables, and tree-depth set to $m=1$, both solutions have exactly $(m \cdot n) =  3$ transformation edges, and $(n-1) =  2$ join edges. 
\end{example}

\begin{proposition}
\label{prop:steiner-edge-count}
A valid Steiner tree on the search graph $G(\mathcal{T})$ of MPBP-G that connects all terminals $R = \{v(T_i) | i \in [n] \}$, has at least $(m \cdot n) + (n-1)$ edges.  Furthermore, any valid Steiner tree with exactly $(m\cdot n) + (n-1)$ edges is a valid solution to MPBP-G. 
\end{proposition}

\iftoggle{full}
{
    A proof of this can be found in Appendix~\ref{ap:proof-prop-steiner-edge-count},
}
{
    A proof of this  can be found in~\cite{full},
}
where the key idea is that for a valid Steiner tree to connect all terminals in $R = \{v(T_i) | i \in [n] \}$, at least one path of length $m$ from the root $v(T_i)$ to a leaf vertex $v(L_i)$ in each transformation tree $G(T_i)$  is required, which together with at least $(n-1)$ join edges to connect $n$ transformation trees, leads to at least $(m \cdot n) + (n-1)$ edges.

\begin{example}
Continue with Example~\ref{ex:exact-edge-cnt}, where $n=3$ and $m=1$, we can see that both trees marked in red in Figure~\ref{fig:ex-solution-optimal} and~\ref{fig:ex-solution-inferior} are valid Steiner trees with $5$ edges, while the tree in Figure~\ref{fig:invalid-solution} is also a valid Steiner tree with $6$ edges. All three cases indeed have at least $(m \cdot n) + (n-1) = (1 \cdot 3) + (3-1)=5$ edges. Furthermore, both valid Steiner trees with exactly $(m \cdot n) + (n-1) = 5$ edges in Figure~\ref{fig:ex-solution-optimal} and~\ref{fig:ex-solution-inferior}, are indeed valid solutions to MPBP-G (Example~\ref{ex:valid-graph-solution}).    
\end{example}

Using Proposition~\ref{prop:solution-is-steiner}-\ref{prop:steiner-edge-count}, we now describe our Algorithm~\ref{alg:solve-using-steiner}, which builds on top of MST that can provably solve MPBP-G.
    \begin{algorithm}
    \begin{small}
    
    \SetKwComment{Comment}{/* }{ */}
    \SetKwData{transformationOnlyModel}{transformationOnlyModel}
    \SetKwData{extractFeatures}{extractFeatures}
    \SetKwData{getParentNode}{getParentNode}
    \SetKwData{transformClassifier}{transformClassifier}
    \SetKwData{joinModel}{joinModel}
    \SetKwData{TransformationSearch}{BIPrepSearch}
    \SetKw{Not}{not}
    \SetKw{Or}{or}
    \caption{Solve MPBP-G on global search graph $G(\mathcal{T})$}
    \label{alg:solve-using-steiner}
    \KwIn{$G(\mathcal{T})$ (global search graph)}
    \KwOut{Edge set $E$ in $G(\mathcal{T})$ (solution to MPBP-G)}
    
    \ForEach{$G(T_i) \in G(\mathcal{T})$}{ 
        $v(L_i) = \argmin_{v(L_i) \in V_{leaf}(T_i)}{ \left(\sum_{e \in path(v(T_i), v(L_i))) }{\overline{w}(e)}\right) } $ \label{ln:fine-best-leaf}

        $\overline{w}(L_i) = \sum_{e \in \text{path}(v(T_i), v(L_i))) }{\overline{w}(e)}$ \label{ln:calc-best-leaf-edge-weight} 
    }


    $S_b = \bigcup_{i \in [n]}{\{e| e \in {path}(v(T_i), v(L_i))) \} }$  \label{ln:select-base-solution} 

     $S_b = S_b \bigcup \{e | e \in \text{spanning-tree}(\{v(L_i)|i\in [n]\}) \} $\label{ln:select-base-solution-add-spanning-tree-for-join} 

        
    $p = \sum_{i \in [n]}{\overline{w}(L_i)} + (n-1) (-\log(0.5))$ \label{ln:calc-penalty}


    \ForEach{$e  \in G(\mathcal{T})$ \label{ln:iterate-edge} }{ 
    
        update $e$ edge weight as $\overline{w}(e) + 2p$ \label{ln:update-penalty} 

    }
    
    $S_s \leftarrow \text{solve MST on~} G(\mathcal{T})$ \tcp{\footnotesize{using Kou's  algorithm~\cite{kou1981fast}}} \label{ln:solve_mst}
    
    $W_s =  \sum_{e \in S_s } \overline{w}(e) + 2p \cdot |S_s| $ \label{ln:calc_mst_cost}

    $W_b = p + 2p \cdot |S_b| $ \label{ln:calc-base-solution} %
    
    return the edge set $S_s$ if $W_s < W_b$, otherwise $S_b$;
    \end{small}
    \end{algorithm}
    
In Algorithm~\ref{alg:solve-using-steiner}, we start by iterating through each transformation tree $G(T_i)$, where in each $G(T_i)$, we select the leaf vertex $v(L_i)$  with the best transformation probabilities (sum of edge weights) on the path between root $v(T_i)$ and $v(L_i)$ (Line~\ref{ln:fine-best-leaf}). Note that since we are using negative log-probabilities $\overline{w}(e)$ as edge weights (Equation~\eqref{eqn:log-transform}), where smaller is better, $v(L_i)$ is selected using $\argmin$. Assign $\overline{w}(L_i)$ as the sum of the edge weights on each best path (Line~\ref{ln:calc-best-leaf-edge-weight}).  

We know at this point that there is one valid solution $S_b$ to MPBP-G, that simply selects all transformation-edges on each ${path}(v(T_i), v(L_i))$ (Line~\ref{ln:select-base-solution}), as well as any spanning-tree that connects $\{v(L_i)|i\in [n]\}$ (Line~\ref{ln:select-base-solution-add-spanning-tree-for-join}), whose objective function is at least as good as $p = \sum_{i \in [n]}{\overline{w}(L_i)} + (n-1) (-\log(0.5))$ (Line~\ref{ln:calc-penalty}), where the second term comes from the fact that the join score of $S_b$ from any spanning-tree will be no worse than $(n-1) (-\log(0.5))$ (since each non-joinable ``placeholder join edge'' has a worst-case score of $-\log(0.5)$). 

Next, we add $2p$ as an edge-weight ``penalty term'' to every edge $e \in G(\mathcal{T})$ (Line~\ref{ln:update-penalty}). We then solve the minimum Steiner tree (MST) problem on the resulting graph $G(\mathcal{T})$ using the classical Kou's algorithm~\cite{kou1981fast} to find a Steiner tree $S_s$ (Line~\ref{ln:solve_mst}), whose total edge-weight is $W_s$  (Line~\ref{ln:calc_mst_cost}). We also calculate the total edge-weight of $W_b$ for $S_b$ including the $2p$ penalty term (Line~\ref{ln:calc-base-solution}). Finally, we return the better of $S_s$ and $S_b$, based on their cost $W_s$ and $W_b$, as our solution. 

We prove that Algorithm~\ref{alg:solve-using-steiner} not only finds a valid solution to MPBP-G, but also has the following approximation guarantee. 

\begin{THEOREM}
\label{theorem:steiner-tree-solve}
Algorithm~\ref{alg:solve-using-steiner} produces a valid solution to MPBP-G in polynomial time.  Furthermore, this solution is within a $(2-(2/n))$ factor of the optimal, where $n \geq 2$ is the number of input tables.
\end{THEOREM}

\iftoggle{full}
{
    A proof of Theorem~\ref{theorem:steiner-tree-solve} can be found in Appendix~\ref{ap:proof-thm-steiner-tree-solve}, which follows from Proposition~\ref{prop:solution-is-steiner}-\ref{prop:steiner-edge-count}.
}
{
    A proof of Theorem~\ref{theorem:steiner-tree-solve} can be found in~\cite{full},  which follows from Proposition~\ref{prop:solution-is-steiner}-\ref{prop:steiner-edge-count}.
}
Intuitively, to see why the  Algorithm~\ref{alg:solve-using-steiner} must return a valid solution to MPBP-G, first notice that $S_b$ is a valid solution to MPBP-G (therefore also a valid Steiner tree per Proposition~\ref{prop:solution-is-steiner}). 
Furthermore, $S_s$ must be a valid Steiner tree by virtue of Kou's algorithm~\cite{kou1981fast}, and from Proposition~\ref{prop:steiner-edge-count}, we know that $S_s$ must have at least $(m\cdot n) + (n-1)$ edges to be a valid Steiner tree.

Suppose $S_s$ has exactly $(m\cdot n) + (n-1)$ edges in $|S_s|$, then by Proposition~\ref{prop:steiner-edge-count} it must be a valid solution to MPBP-G. Now, since both $S_s$ and $S_b$ are valid solutions to MPBP-G, the better of the two returned by Algorithm~\ref{alg:solve-using-steiner} must also be a valid solution to MPBP-G. 

Alternatively, if $S_s$ has more than $(m\cdot n) + (n-1)$ edges  in $|S_s|$, then its sum of edge-weight $W_s$ is at least $2p \cdot (m\cdot n + n)$ from the penalty weight alone, which is already greater than that of the solution $S_b$, which will have exactly $(m\cdot n) + (n-1)$ edges (Proposition~\ref{prop:solution-edge-count}), whose total edge-weight $W_b$ is therefore no greater than $2p \cdot (m\cdot n + (n-1)) + p$. This forces $S_b$ (a valid solution to MPBP-G) be picked over $S_s$. 
Together, these ensure that Algorithm~\ref{alg:solve-using-steiner} always returns a valid solution to MPBP-G. 

We prove the approximation ratio and polynomial complexity of Algorithm~\ref{alg:solve-using-steiner} \iftoggle{full}
{
    in Appendix~\ref{ap:proof-thm-steiner-tree-solve},
}
{
    in~\cite{full},
} using arguments in Kou's algorithm~\cite{kou1981fast}.

\textbf{Optimistic vs. Precise Mode of \sys.} 
\label{sec:opt-vs-precise}
So far we focused on formulating and solving \sys as a graph search problem, assuming that the  edge weights (transformation probabilities) are fixed.
In practice, since we use global features across multiple tables to estimate transformation probabilities (Section~\ref{sec:transform_model}), the probability of a transformation (edge-weight) can depend on whether certain transformations are actually selected on other tables, which can therefore evolve. 
We design two variants of \sys to handle this, 
one ``optimisitc'' and one ``precise''. 

In the ``optimistic'' mode of \sys, for each table $T$, we take an optimistic view and use the best possible feature configuration from other tables $\mathcal{T} \setminus T$ to estimate its transformation probabilities, which become \emph{fixed} edge-weights on the search graph $G(\mathcal{T})$, on which we can then directly invoke Algorithm~\ref{alg:solve-using-steiner} once to solve. 

In the ``precise'' mode of \sys, we iteratively invoke Algorithm~\ref{alg:solve-using-steiner}, where in each iteration we use the bounds from the best solutions found so far, to prune away infeasible graph edges (transformations) and update edge weights, so that Algorithm~\ref{alg:solve-using-steiner} can be invoked on the updated graph again until convergence. This tends to yield better solutions but is more expensive. 
\iftoggle{full}
{
   Algorithm~\ref{alg:precise-mode-iterative} shows the pseudo-code of the precise variant. 

   \begin{algorithm}
    \begin{small}
    
    \SetKwComment{Comment}{/* }{ */}
    \SetKwData{transformationOnlyModel}{transformationOnlyModel}
    \SetKwData{extractFeatures}{extractFeatures}
    \SetKwData{getParentNode}{getParentNode}
    \SetKwData{transformClassifier}{transformClassifier}
    \SetKwData{joinModel}{joinModel}
    \SetKwData{TransformationSearch}{BIPrepSearch}
    \SetKw{Not}{not}
    \SetKw{Or}{or}
    \caption{Solve MPBP-G on $G(\mathcal{T})$ (Precise-mode)}
    \label{alg:precise-mode-iterative}
    \KwIn{$G(\mathcal{T})$ (initial search graph)}
    \KwOut{Edge set $E$ in $G(\mathcal{T})$ (solution to MPBP-G)}

    $S_{curr} \leftarrow \text{solve MPBP-G on~} G(\mathcal{T})$ \tcp{\footnotesize{invoke Algorithm~\ref{alg:solve-using-steiner}}}

      $G_{curr}(\mathcal{T}) \leftarrow  
      G(\mathcal{T})$  
      
    \Repeat{$S_{curr} == S_{pre}$ or after $k$ iterations}{


      $G_{curr}(\mathcal{T}) \leftarrow  
      prune(G_{curr}(\mathcal{T}), S_{curr})$  \tcp{\footnotesize{prune graph}}
      
      $S_{pre} \leftarrow  
      S_{curr}$ 
      
      $S_{curr} \leftarrow \text{solve MPBP-G on~} G_{curr}(\mathcal{T})$ \tcp{\footnotesize{invoke Algorithm~\ref{alg:solve-using-steiner}}}
    }

    return $S_{curr}$;
    \end{small}
    \end{algorithm}

    As can be seen from the algorithm, we start by taking the initial $G(\mathcal{T})$, and  invoking Algorithm~\ref{alg:solve-using-steiner} to solve MPBP-G on $G(\mathcal{T})$, to get the initial solution $S_{curr}$ (which is where the Optimistic variant ends, as it returns $S_{curr}$ as the solution). 
    
    In the Precise-variant of \sys, however, we assign the current graph to $G_{curr}(\mathcal{T})$, and continue to iteratively update $G_{curr}(\mathcal{T})$, by pruning away infeasible transformation edges in $G_{curr}(\mathcal{T})$, by using the  best solution found so far $S_{curr}$ (e.g., if the best possible solution has an overall joint probability of 0.5, then any transformation-edge with predicted probability lower than 0.5 can all be pruned). Importantly, the transformation-probability (edge-weights) of the $G_{curr}(\mathcal{T})$ is updated when certain edges are pruned, because we use global features in transformation-classifiers, where the predicted probability of a transformation can be affected by the removal of certain transformations on other tables, leading to a new $G_{curr}(\mathcal{T})$.

    We then assign $S_{curr}$ into $S_{pre}$, and invoke Algorithm~\ref{alg:solve-using-steiner} again on $G_{curr}(\mathcal{T})$ to get a new $S_{curr}$. We repeat the iterative computation, until convergence ($S_{curr} == S_{pre}$), or after $k$ iterations, at which point we return $S_{curr}$ as our solution to MPBP-G.
    
}
{
    We give the pseudo-code of the precise variant in~\cite{full} in the interest of space.
}

\section{Experimental Evaluation} \label{sec:experiments}

\subsection{Evaluation Setup}

\ignore{
\begin{table}
    \centering
    
    \begin{subtable}[t]{0.5\textwidth}
    \resizebox{\linewidth}{!}{
    \begin{tabular}{*{6}{c}} 
    \toprule
    \textbf{Operator} &
        \textbf{Transpose} &
        \textbf{Unpivot} &
        \textbf{Pivot} &
        \textbf{Transform join} &
        \textbf{No-op}\\
    \midrule
        \% in data&
        6.7\% & 28.2\%   & 3.3\% & 5.1\% & 56.7\% \\
    \bottomrule
    \end{tabular}
    }
    \end{subtable}

    \begin{subtable}[t]{0.5\textwidth}
    \resizebox{\linewidth}{!}{
    \begin{tabular}{*{10}{c}} 
    \toprule
    \textbf{Num. of tables} &
        \textbf{3} &
        \textbf{4} &
        \textbf{5} &
        \textbf{6} &
        \textbf{7} &
        \textbf{8} &
        \textbf{9} &
        \textbf{10} &
        $\mathbf{\geq 11}$ \\
    \midrule
        \% in data&
        7.5\% & 8.4\%   & 11.4\% & 9.0\% & 4.7\% & 4.3\% & 3.7\% & 3.3\% & 8.0\%\\
    \bottomrule
    \end{tabular}
    }
    \end{subtable}
    \caption{Benchmark statistics.}
    \label{tab:benchmark}
    \vspace{-6mm}
\end{table}
}

\subsubsection{Benchmark} \mbox{}\\
We sampled 1837 real BI projects for evaluation, where 510 projects are set aside as test cases that are never looked at, and the rest are used as training data to calibrate our offline classification models (Section~\ref{sec:offline-training}). We extract user-programmed transformations 
and joins 
from the real BI projects as ground truth, which algorithms will need to predict correctly (Section~\ref{sec:real-data}). The distribution of transformations in our dataset is the following: unpivot: $28.2\%$; pivot: $3.3\%$; transpose: $6.7\%$; string-transform: $5.1\%$; no-op: $56.7\%$.
    Table~\ref{tab:benchmark-stats} shows the key statistics of the dataset.






\begin{table}[t]
\vspace{-6mm}
    \centering
    \caption{Benchmark statistics.}
    \vspace{-4mm}
    \label{tab:benchmark-stats}
    \resizebox{0.9\linewidth}{!}{
    \begin{tabular}{ |c|c|c|c|c| } 
    \hline
     & Average & 50\textsuperscript{th} \%tile & 90\textsuperscript{th} \%tile & 95\textsuperscript{th} \%tile \\
    \hline
    \# of rows per table & 24818.7 & 101 & 25907.2 & 103705.2 \\
    \hline
    \# of columns per table & 52.8 & 6 & 21 & 32 \\
    \hline
    \# of tables per BI project & 4.8 & 4 & 10 & 13 \\
    \hline
    \end{tabular}
    \vspace{-6mm}
    }
\end{table}

\iftoggle{full}
{
\begin{table}
    \centering
    \caption{Distribution of transform operators in BI projects.}
    \vspace{-4mm}
    \label{tab:operator-distribution}
    \resizebox{0.85\linewidth}{!}{
    \begin{tabular}{ |c|c|c|c|c|c| } 
    \hline
      & transpose & unpivot & pivot & string transform  & no-op \\
     \hline
     \% of projects&
        6.7\% & 28.2\%   & 3.3\% & 5.1\% & 56.7\% \\
    \hline
    \end{tabular}
    }
\iftoggle{full}
{
}
{
    \vspace{-6mm}
}
\end{table}
}
{
}

\subsubsection{Metrics}\mbox{}\\
In our problem, since both transformations and joins need to be predicted, we evaluate quality for both transformations and joins. 

\textbf{Transformation quality}. {We compare the predicted transformations of different methods against the ground truth (user-programmed transformations), where both the operator, parameters, and order of the transformations (Table~\ref{tab:dsl}) need to be predicted exactly correctly (we do not assign partial credit for inexact matches)}. We use the standard precision, recall, and F1-score metrics, where precision \begin{small}
$P_{\text{transform}}=\frac{\text{num-of-correct-predicted-transforms}}{\text{num-of-total-predicted-transforms}}$,
\end{small} recall 
\begin{small}
$R_{\text{transform}}=\frac{\text{num-of-correct-predicted-transforms}}{\text{num-of-total-true-transforms}}$,
\end{small}
and F1-score 
\begin{small}
$F_{\text{transform}}=\frac{2 P_{\text{transform}} R_{\text{transform}}}{P_{\text{transform}} + R_{\text{transform}}}$
\end{small}is the harmonic mean of precision and recall. 

\textbf{Join quality}. We compare the predicted joins against the ground truth joins programmed by users. 
Precision $P_{\text{join}}$, recall $R_{\text{join}}$, and F1-score $F_{\text{join}}$ are defined similarly.


\subsubsection{Methods compared}\mbox{}\\
While we propose to holistically predict both transformations and joins,  existing methods predominately predict either only transformations, or only joins (reviewed in Section~\ref{sec:related}). 
As a result, for this evaluation, we test against 3 classes of baselines:  

(1) \underline{Transformation-only methods}: We compare with the following transformation-only methods, for transformation quality. Our goal here is to show that considering only transformation (without considering joins holistically), misses the opportunity to predict accurately given the intertwined nature of the two.
\begin{itemize} [noitemsep,topsep=0pt,leftmargin=*]
    \item \textbf{Single-Operator-Predict (SOP)~\cite{yan2020auto-suggest}}. This method uses machine learning models to predict single transformation operators, based on the characteristics of input tables. This is the basis of our transformation-classifiers (Section~\ref{sec:offline-training}) used in \sys. We refer to this as Single-Operator-Predict (SOP) to differentiate with the next baseline below that considers multiple operators.
    \item \textbf{Multi-Operator-Predict (MOP)~\cite{li2023autotables}}. This is a state-of-the art method that predicts transformations from multiple operators, with a focus on reshaping tables. We refer to this as Multi-Operator-Predict (MOP).
    \item \textbf{GPT-transformation (GPT-t)}. Since language models like GPT have shown diverse capabilities, including program synthesis~\cite{brown2020language}, we use a recent version of GPT-4\footnote{GPT-4-0613, accessed from Azure OpenAI in March 2024.}~\cite{openai2023gpt4} and in-context-few-shot learning, as another baseline for transformation predictions. 
    We extensively test combinations of different prompt strategies, including few-shot examples, chain-of-thought~\cite{wei2022chain}, and dynamic RAG to retrieve most relevant examples~\cite{lewis2020retrieval}, in order to optimize the quality of the GPT-based method.
    \item \textbf{Fine-tuned-GPT-transformation (FT-GPT-t)}. In addition to invoking vanilla GPT-4, we also perform the expensive step of fine-tuning GPT-4 models, in an attempt to optimize their prediction quality in data transformation use cases. We use the OpenAI fine-tuning API~\cite{openai-ft} and the same data available to \sys to fine-tune GPT-4. We fine-tune GPT both only for transform (FT-GPT-t), and also end-to-end for both transform and join (referred to as FT-GPT). These serve as  additional strong baselines that leverage language-models for data preparation.
\end{itemize}

(2) \underline{Join-only methods}: We compare with 2 representative methods from the literature for join predictions. 
\begin{itemize}[noitemsep,topsep=0pt,leftmargin=*]
    \item \textbf{BI-join (BI-j)~\cite{lin2023autobi}}. BI-Join is a recent method designed to predict joins in BI settings, and has been shown to be competitive on a range of join benchmarks~\cite{lin2023autobi}. 
    \item \textbf{GPT-join (GPT-j)}. Since GPT is capable of understanding tables~\cite{narayan2022can, table-gpt}, we again use GPT-4 and few-shot learning, as another baseline for join predictions. We refer to this as GPT-j. 
\end{itemize}

(3) \underline{Transformation + join methods}: Given the {4} transformation-only and 2 join-only baselines mentioned above, we further compare with a third class of baselines, which sequentially invoke the 3 transformation-only methods, followed by the 2 join-only methods, for a total of $4 \times 2 = 8$ methods (which are \textbf{SOP + BI-j}, \textbf{MOP + BI-j}, \textbf{GPT-t + BI-j}, {\textbf{FT-GPT-t + BI-j}}, \textbf{SOP + GPT-j}, \textbf{MOP + GPT-j}, \textbf{GPT-t + GPT-j}), and \textbf{FT-GPT-t + GPT-j}. These methods are directly comparable to \sys (except that \sys holistically optimizes across transforms/joins using graph-algorithms, whereas these baselines simply invoke join predictions after transformations sequentially). 


We compare all baselines above, with two variants of \textit{our method}: 
\begin{itemize}[noitemsep,topsep=0pt,leftmargin=*]
    \item \textbf{\sys-Optimistic (\sysa-O)}. This is a version of our algorithm that directly invokes Algorithm~\ref{alg:solve-using-steiner} once to solve MPBP-G, based on optimistic estimates of transformation probabilities.  
    \item \textbf{\sys-Precise (\sysa-P)}. This is the precise variant of our algorithm 
    that iteratively invokes Algorithm~\ref{alg:solve-using-steiner} based on precise estimates of probabilities, which tends to give better solutions, but is more expensive than \sysa-O. We set the search depth $m$ as 2 consistently for both AP methods across all experiments.
\end{itemize}

\subsection{Experimental results} 


\begin{table}[t!]
\vspace{-6mm}
    \centering
    \caption{Transformation quality comparison.}
    \vspace{-4mm}
    \label{tab:transformation-quality}
    \resizebox{0.9\linewidth}{!}{
    \begin{tabular}{|c|cc|cccc|} 
    \hline
    \textbf{Method} &
        \textbf{\sysa-P} &
        \textbf{\sysa-O} &
        \textbf{SOP} &
        \textbf{MOP} &
        \textbf{GPT-t} &
        {\textbf{FT-GPT-t}}\\
    \hline
        $P_{\text{transform}}$ &
        \textbf{0.785} & \underline{0.757}   & 0.231  & 0.339 & {0.411} & {0.296} \\
        $R_{\text{transform}}$ &
        \textbf{0.741} & \underline{0.740}  & 0.644  & 0.571 & {0.646} & {0.258} \\ 
        \hline
        $F_{\text{transform}}$ &  
        \textbf{0.762} & \underline{0.748} & 0.340 & 0.426 & {0.502} & {0.275}\\
    \hline
    \end{tabular}
    }
    \vspace{-4mm}
\end{table}

\begin{table*}[t]
\iftoggle{full}
{
}
{
    \vspace{-22mm}
}
    \centering
    \caption{Join quality comparison.}
    \vspace{-4mm}
    \label{tab:join-quality}
    \resizebox{0.98\linewidth}{!}{
    \begin{tabular}{|c|cc|cc|cccccccc|c|}
    \hline
        \multirow{2}{*}{\textbf{Method}} & \multicolumn{2}{c|}{\textbf{\sys}} & \multicolumn{2}{c|}{\textbf{Join-only}} &\multicolumn{8}{c|}{\textbf{Transform + Join}} & {\textbf{End-to-end}} \\
        \cline{2-14}
        & 
        \textbf{\sysa-P} &
        \textbf{\sysa-O} &
        \textbf{BI-j} &
        \textbf{GPT-j} &
        \textbf{SOP+BI-j} &
        \textbf{MOP+BI-j} &
        \textbf{GPT-t+BI-j} &
        {\textbf{FT-GPT-t+BI-j}} &
        \textbf{SOP+GPT-j} &
        \textbf{MOP+GPT-j} &
        \textbf{GPT-t+GPT-j} &
        {\textbf{FT-GPT-t+GPT-j}} &
        {\textbf{FT-GPT}} \\
    \hline
        $P_{\text{join}}$ &
        \underline{0.806} & \underline{0.806}  & 0.737  & 0.274 & 0.767 & 0.769 & {0.762} & {\textbf{0.853}} & 0.222 & 0.244 & {0.221} & {0.270} & {0.518} \\
        $R_{\text{join}}$ &
        \textbf{0.734} & \underline{0.714} & 0.671  & 0.350 & 0.631 & 0.600 & {0.626} & {0.588} & 0.271 & 0.299 & {0.225} & {0.311} & {0.470}\\ \hline
        $F_{\text{join}}$ &  
        \textbf{0.769} & \underline{0.758} & 0.702 & 0.307 & 0.705 & 0.674  & {0.687} & {0.696} & 0.244 & 0.269 & {0.223} & {0.289} & {0.493} \\
    \hline
    \end{tabular}
    }
\end{table*}

\subsubsection{Quality Comparison: Transformation} \mbox{}\\
Table~\ref{tab:transformation-quality} shows a comparison of transformation quality between \sys (\sysa) and the existing transformation methods.  
Overall, both \sysa-P and \sysa-O significantly outperform existing transformation baselines\footnote{Note that we omit  ``transformation+join'' baselines here, since their transformation-predictions are identical to those of transformation-only methods (for in these baselines we always invoke transformation-predictions before join-predictions).}, correctly predicting more than $70\%$ transformations, with \sysa-P performing noticeably better than \sysa-O, as it uses more precise estimates of probabilities via iterative graph computation.

Recall that our task of transformation prediction has a large search space (with millions of possible operator and parameter combinations), where a random guess has an exceedingly small chance (less than 1\%) of being correct, such that GPT-t based on GPT-4 also produces low precision. 

Specialized algorithms like SOP and MOP are more competitive, but are still substantially less accurate than \sysa, which holistically optimize transformations/joins in a principled manner.

We experimented with 6 prompting variations to find the best prompt strategies for GPT (in Table~\ref{tab:gpt-variations}), including using few-shot vs. zero-shot, using COT~\cite{wei2022chain} vs. no-COT, and the use of RAG~\cite{lewis2020retrieval} vs. no-RAG. Few-shot with COT turns out to be the best, where few-shot examples provide demonstrations, with COT further enhancing structured reasoning. 

\begin{table}
    \centering
    \caption{{GPT-t results using 6 prompting variations.}}
    \vspace{-4mm}
    \label{tab:gpt-variations}
    \resizebox{0.9\linewidth}{!}{
    \begin{tabular}{|c|c|c|c|c|} 
    \hline
    \textbf{Example selection} &
        \textbf{Chain of thought} &
        $P_{\text{transform}}$ &
        $R_{\text{transform}}$ &
        $F_{\text{transform}}$ \\
    \hline
        \multirow{2}{*}{Zero-shot} & Without COT & 0.254 & 0.626 & 0.361 \\ \cline{2-5}
        & With COT & 0.083 & 0.554 & 0.145 \\ \hline

        \multirow{2}{*}{Few-shot} & Without COT & 0.161 & 0.821 & 0.269 \\ \cline{2-5}
        & With COT & 0.411 & 0.646 & 0.502 \\ \hline

        \multirow{2}{*}{RAG} & Without COT & 0.198 & 0.756 & 0.313 \\ \cline{2-5}
        & With COT & 0.129 & 0.619 & 0.213 \\ \hline

    \hline
    \end{tabular}
    }
\end{table}

\begin{figure*}
    \centering
    \includegraphics[width=0.75\textwidth]{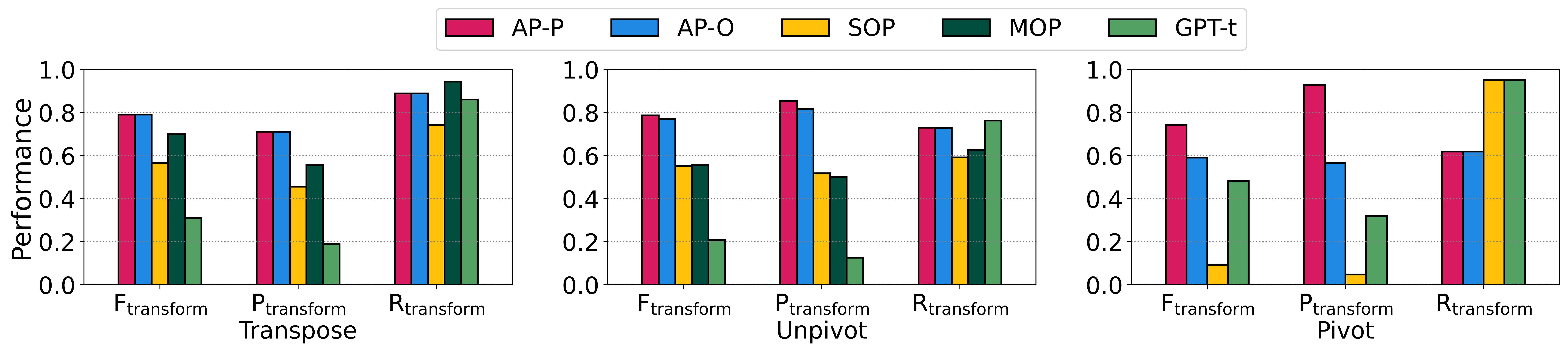}
    \vspace{-4mm}
    \caption{{Transformation quality breakdown by operator.}}
    \label{fig:transformation-breakdown}
\end{figure*}

\begin{figure*}
    \centering
    \begin{subfigure}[t]{0.24\textwidth}
        \centering
        \includegraphics[height=1.2in]{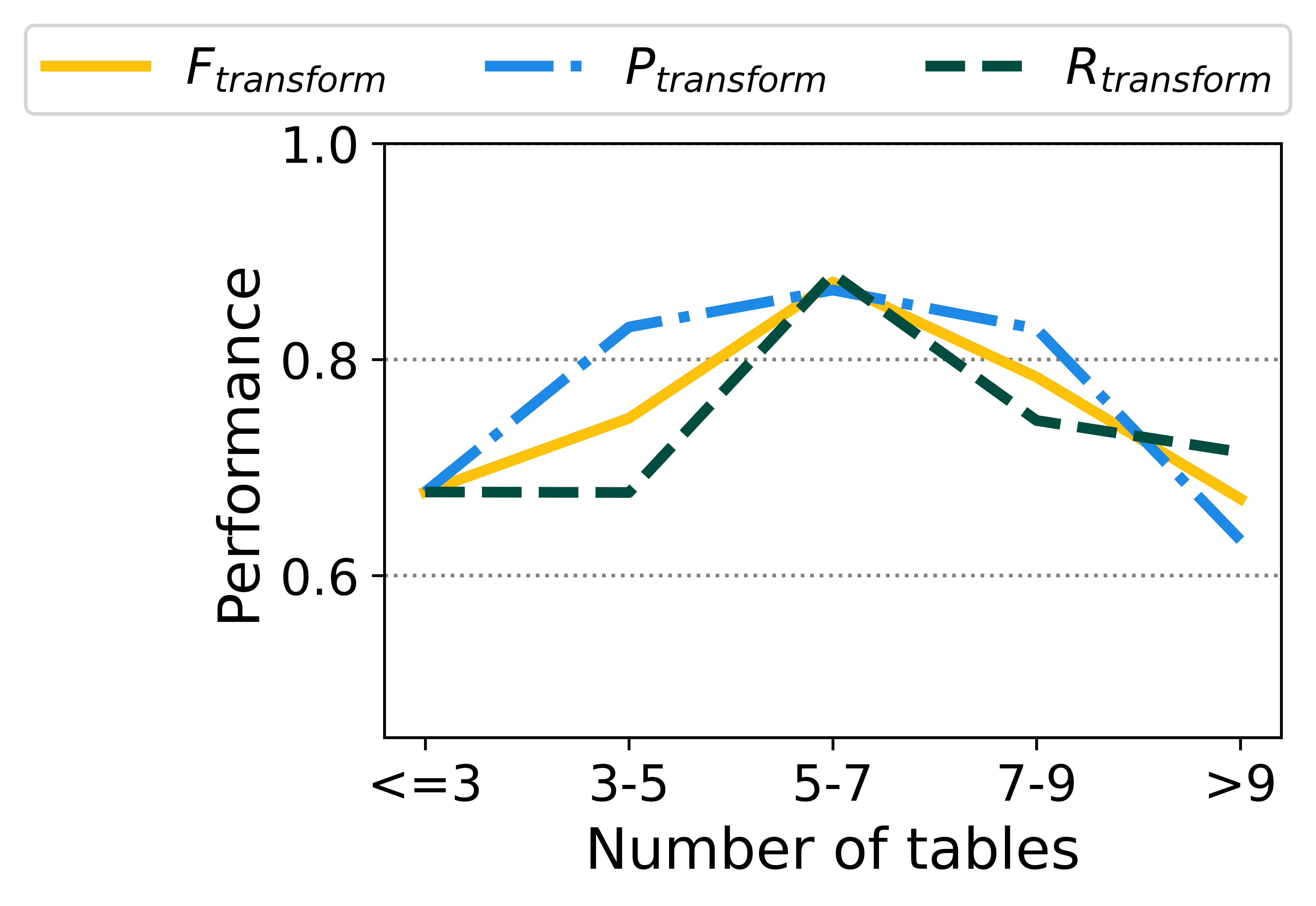}
        \vspace{-6mm}
        \caption{Transformation quality.}
        \label{fig:sensitity0}
    \end{subfigure}%
    \begin{subfigure}[t]{0.24\textwidth}
        \centering
        \includegraphics[height=1.2in]{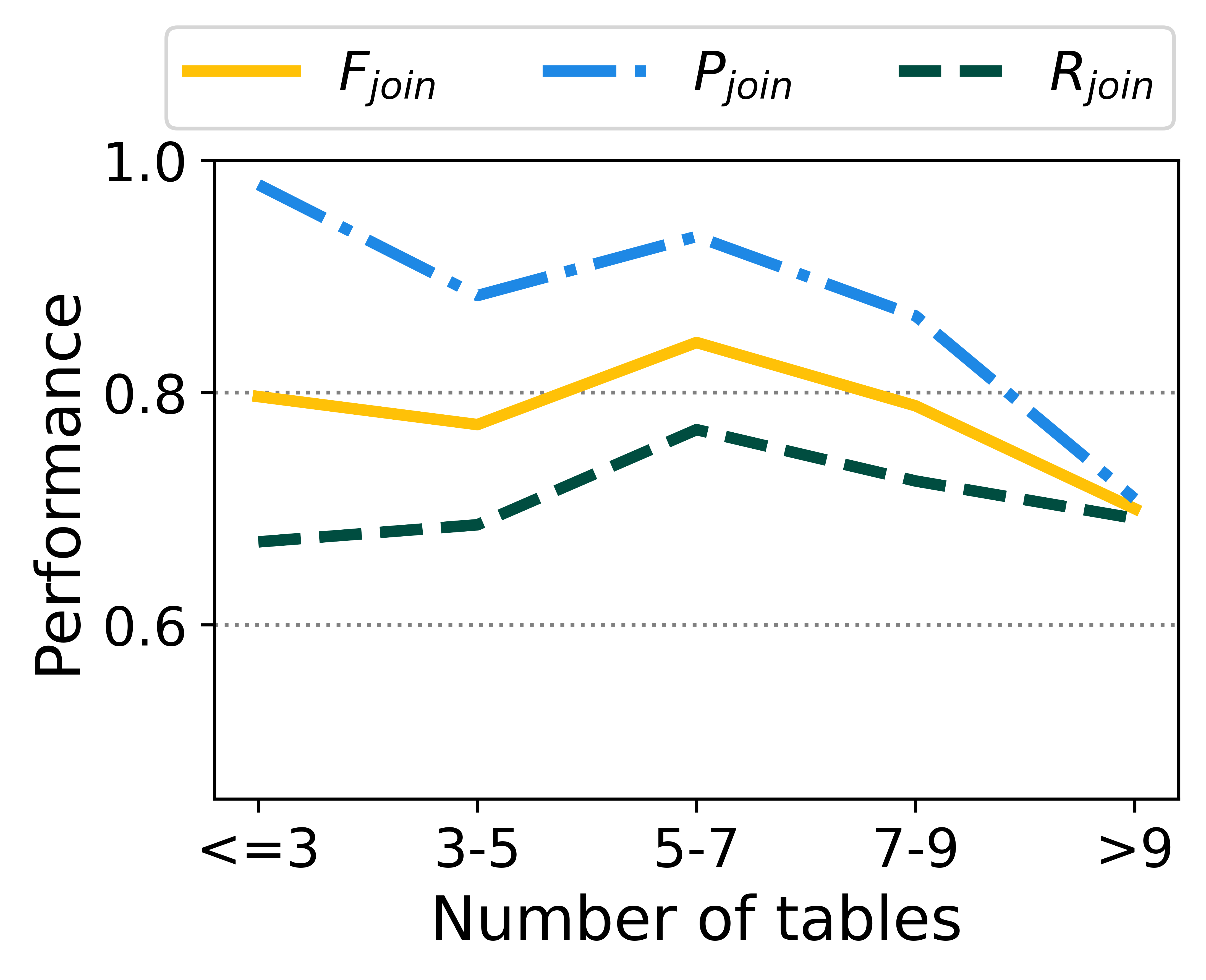}
        \vspace{-2mm}
        \caption{Join quality.}
        \label{fig:sensitity1}
    \end{subfigure}%
    \begin{subfigure}[t]{0.24\textwidth}
        \centering
        \includegraphics[height=1.2in]{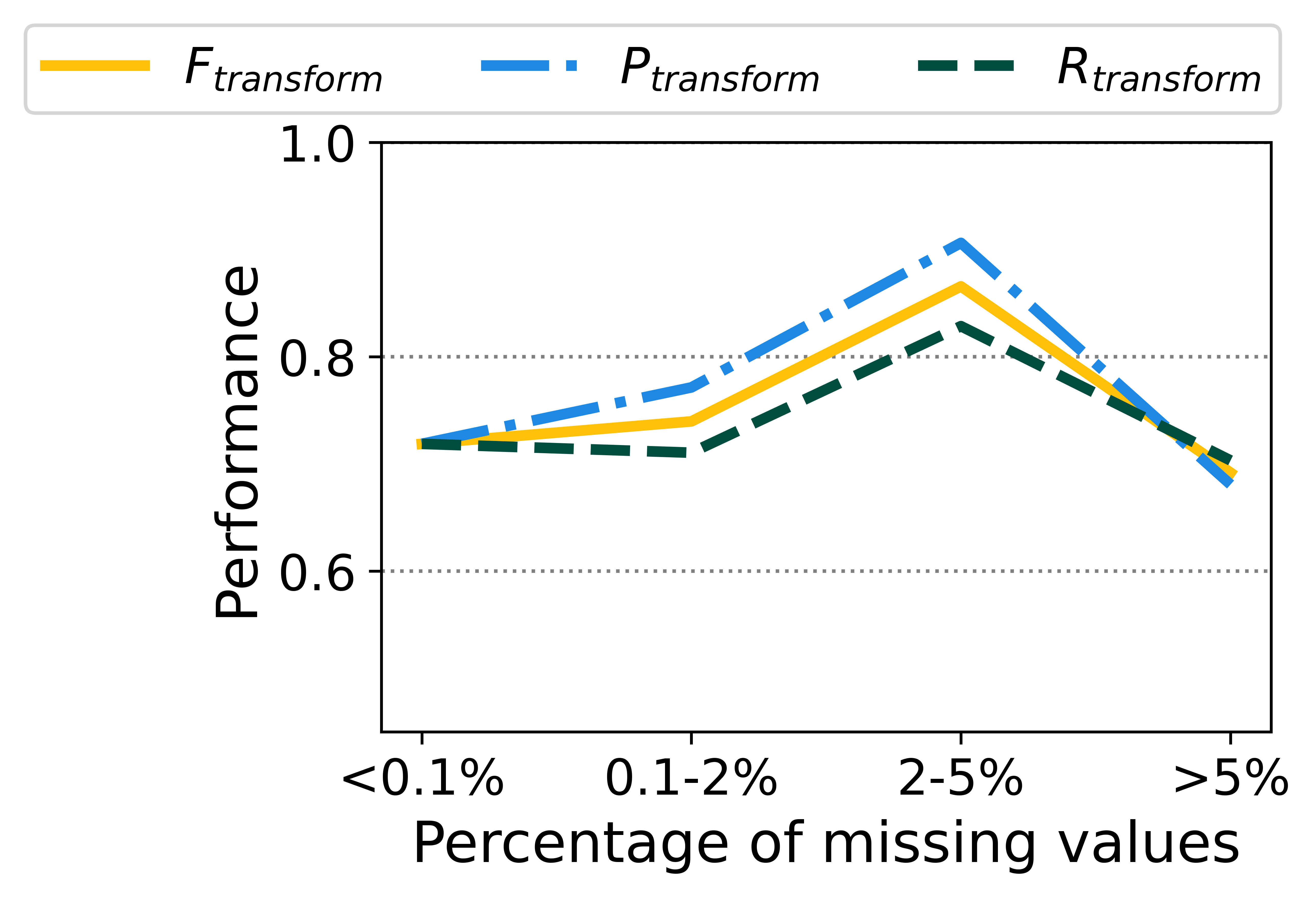}
        \vspace{-6mm}
        \caption{Transformation quality.}
        \label{fig:missing_cell_transform_sensitity}
    \end{subfigure}%
    \begin{subfigure}[t]{0.24\textwidth}
        \centering
        \includegraphics[height=1.2in]{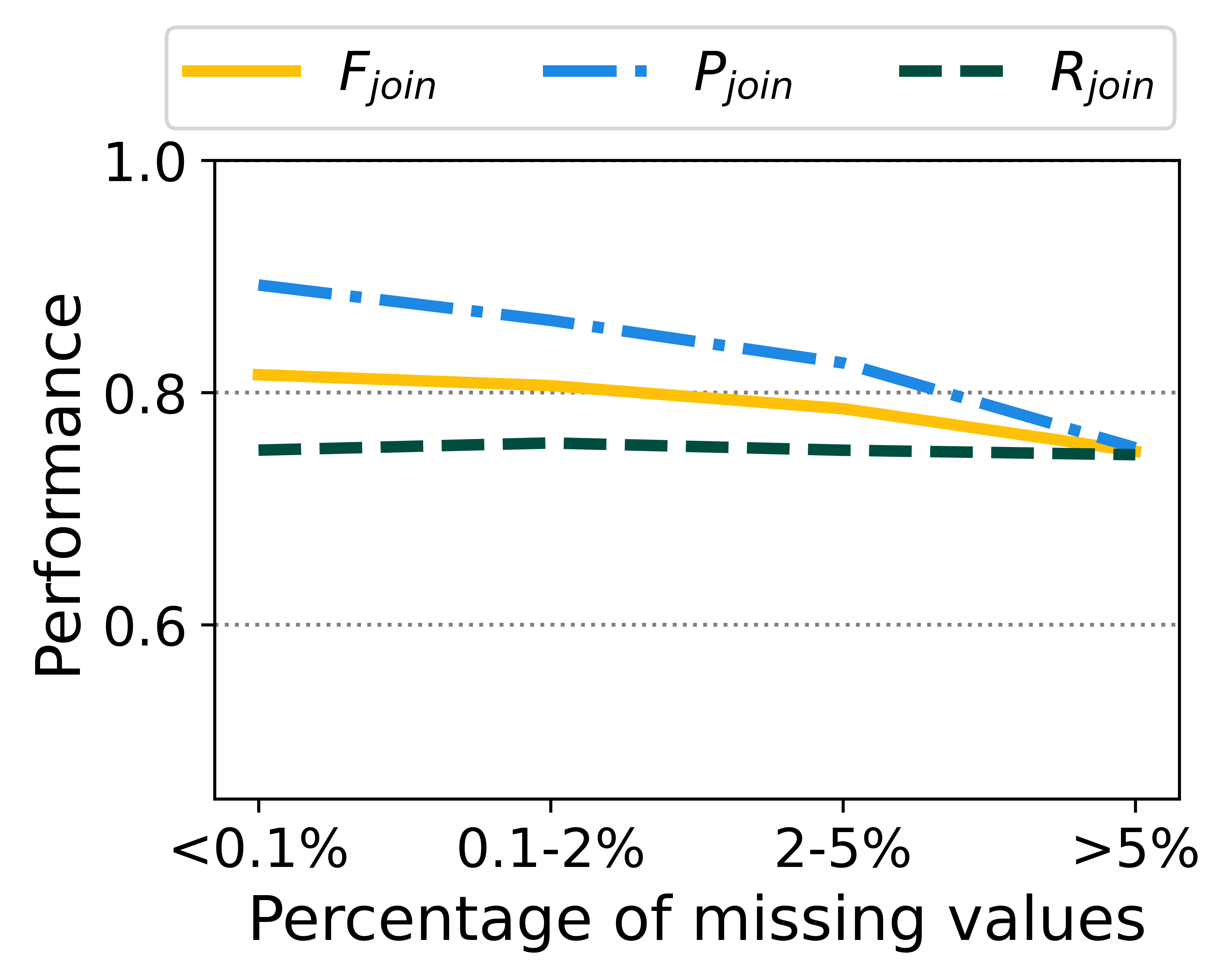}
        \vspace{-2mm}
        \caption{Join quality.}
        \label{fig:missing_cell_join_sensitity}
    \end{subfigure}
    \vspace{-4mm}
    \caption{{Sensitivity analysis.  (a, b): Varying \# of tables per BI project. (c, d): Varying \% of missing values per BI project.}}
    \vspace{-2mm}
\end{figure*}

Figure~\ref{fig:transformation-breakdown} shows the quality comparison broken down by operator, which confirms the benefit of using \sysa across operators.


\subsubsection{Quality Comparison: Join}\mbox{}\\
Table~\ref{tab:join-quality} shows a comparison of the join quality, where we compare \sysa-P and \sysa-O, against both join-only baselines (BI-j and GPT-j), and transform+join baselines (6 methods). 
Overall, both \sysa-P and \sysa-O  considerably outperform all baselines, showing the benefit of our holistic graph optimizations. 

Among join-only baselines, the specialized BI-j is substantially better than GPT-j, which is not surprising as it directly leverages numeric features tailored to the join problem (e.g., column overlap in Jaccard similarity and containment). 

The transformation + join baselines (e.g., SOP+BI-j and MOP+BI-j) outperform join-only methods (BI-j) in join quality, underlining the need to predict transformations together with joins.  However, these baselines are still inferior to \sysa methods, as simply invoking transformation-predictions followed by join-predictions is still sub-optimal to principled graph optimizations.  



\subsubsection{Sensitivity Analysis}\mbox{}\\
We perform sensitivity analysis to understand how the performance of \sys would change with different parameters.

\underline{Sensitivity to the number of tables.} BI projects come with varying numbers of input tables, often reflecting their inherent complexity for data preparation. Table~\ref{tab:benchmark-stats} shows that on average, each BI project has $4.8$ tables, which can go up to 13 at the $95\textsuperscript{th}$ percentile. 

Figure~\ref{fig:sensitity0} shows the transformation prediction quality, on BI projects with different numbers of input tables. As the number of tables increases, the transformation quality improves initially (likely because having more tables in a project provides more global signals for transformation predictions), which then decreases on challenging projects with more than $9$ tables. 
Figure~\ref{fig:sensitity1} shows a similar analysis for join quality, which reveals a more consistent trend as we vary the number of input tables.

\iftoggle{full}
{

    \begin{figure*}
        \centering
        \begin{subfigure}[t]{0.24\textwidth}
            \centering
            \includegraphics[height=1.2in]{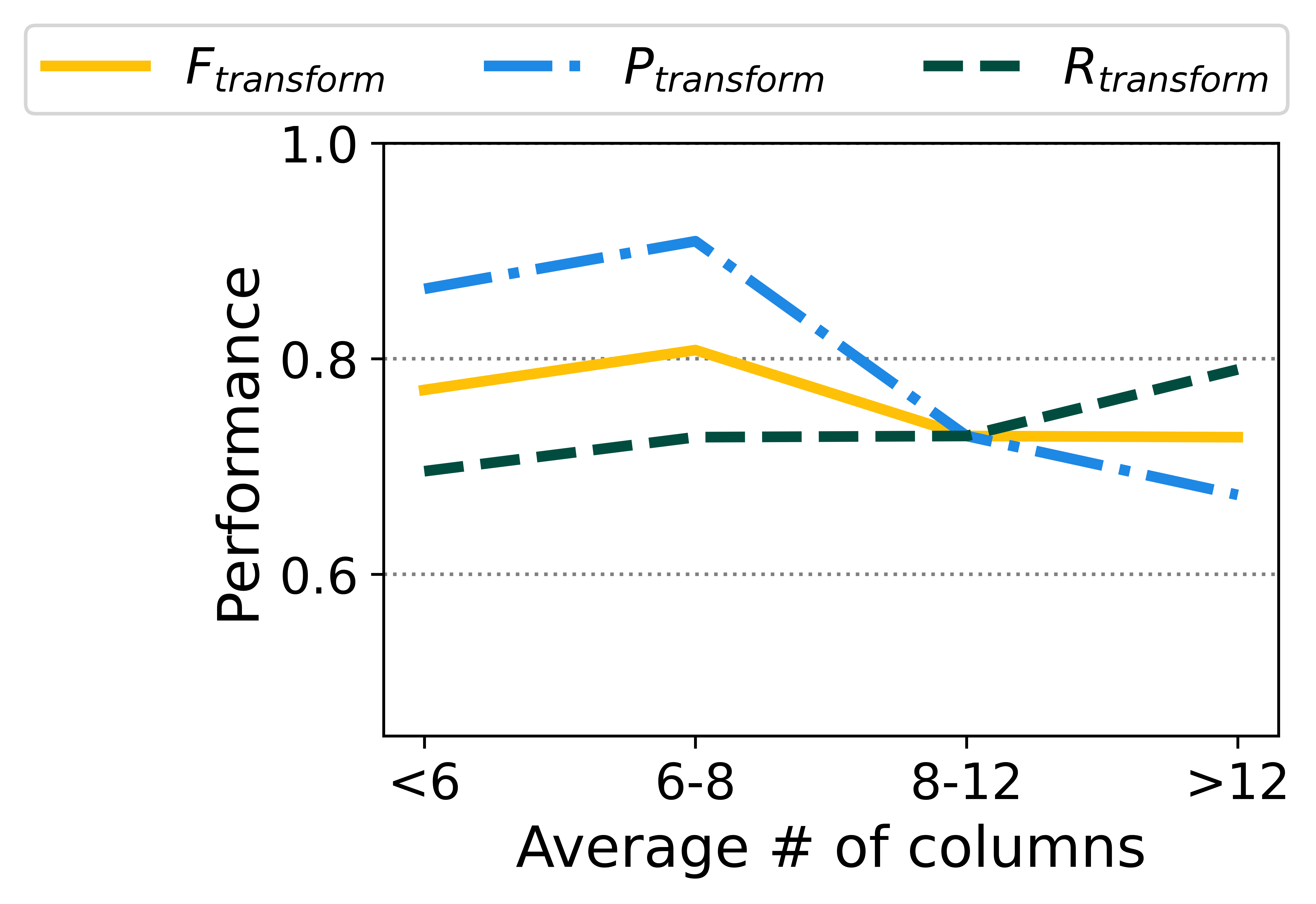}
            \vspace{-6mm}
            \caption{Transformation quality.}
            \label{fig:avg_cols_transform_sensitity}
        \end{subfigure}%
        \begin{subfigure}[t]{0.24\textwidth}
            \centering
            \includegraphics[height=1.2in]{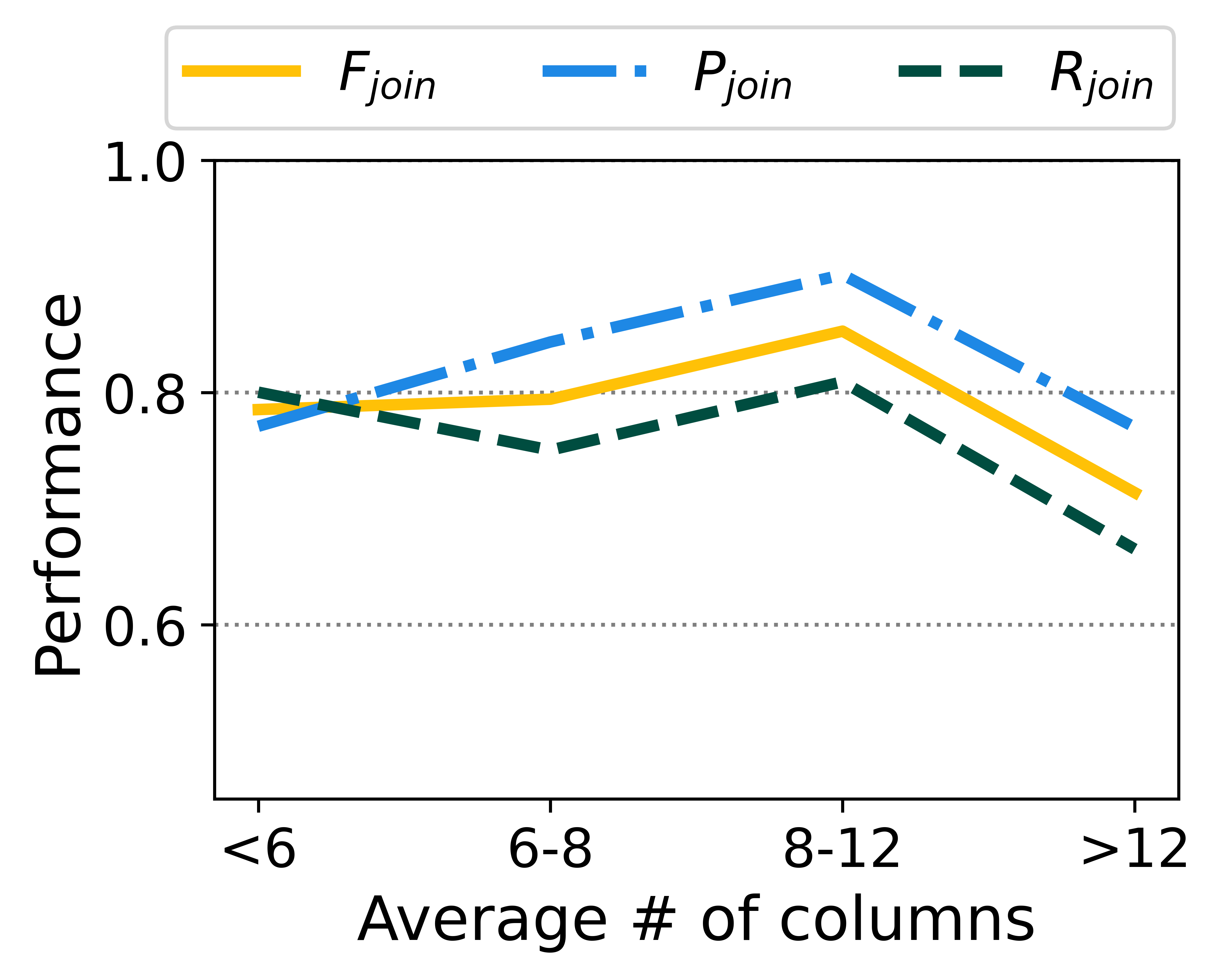}
            \vspace{-2mm}
            \caption{Join quality.}
            \label{fig:avg_cols_join_sensitity}
        \end{subfigure}%
        \begin{subfigure}[t]{0.24\textwidth}
            \centering
            \includegraphics[height=1.2in]{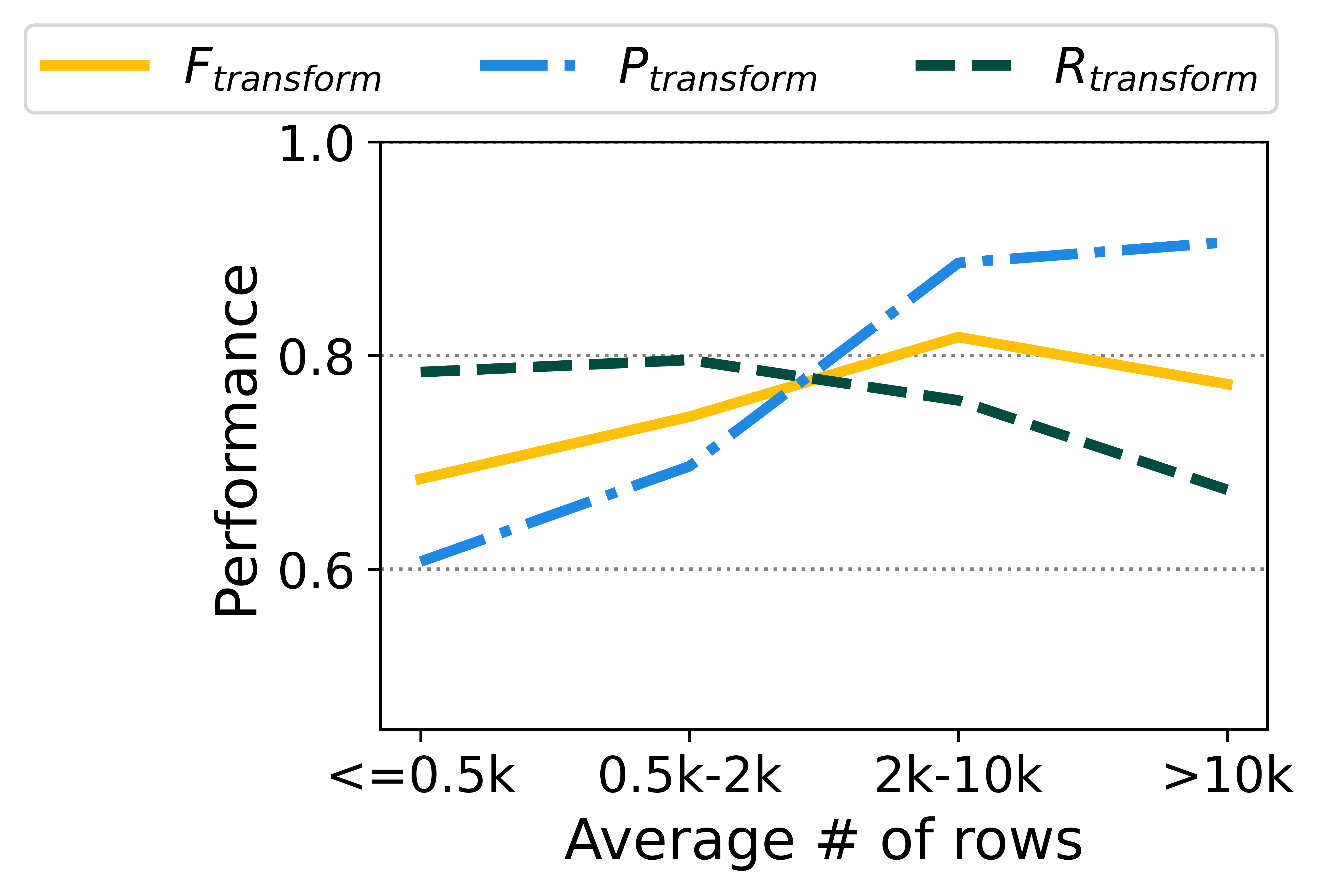}
            \vspace{-6mm}
            \caption{Transformation quality.}
            \label{fig:avg_rows_transform_sensitity}
        \end{subfigure}%
        \begin{subfigure}[t]{0.24\textwidth}
            \centering
            \includegraphics[height=1.2in]{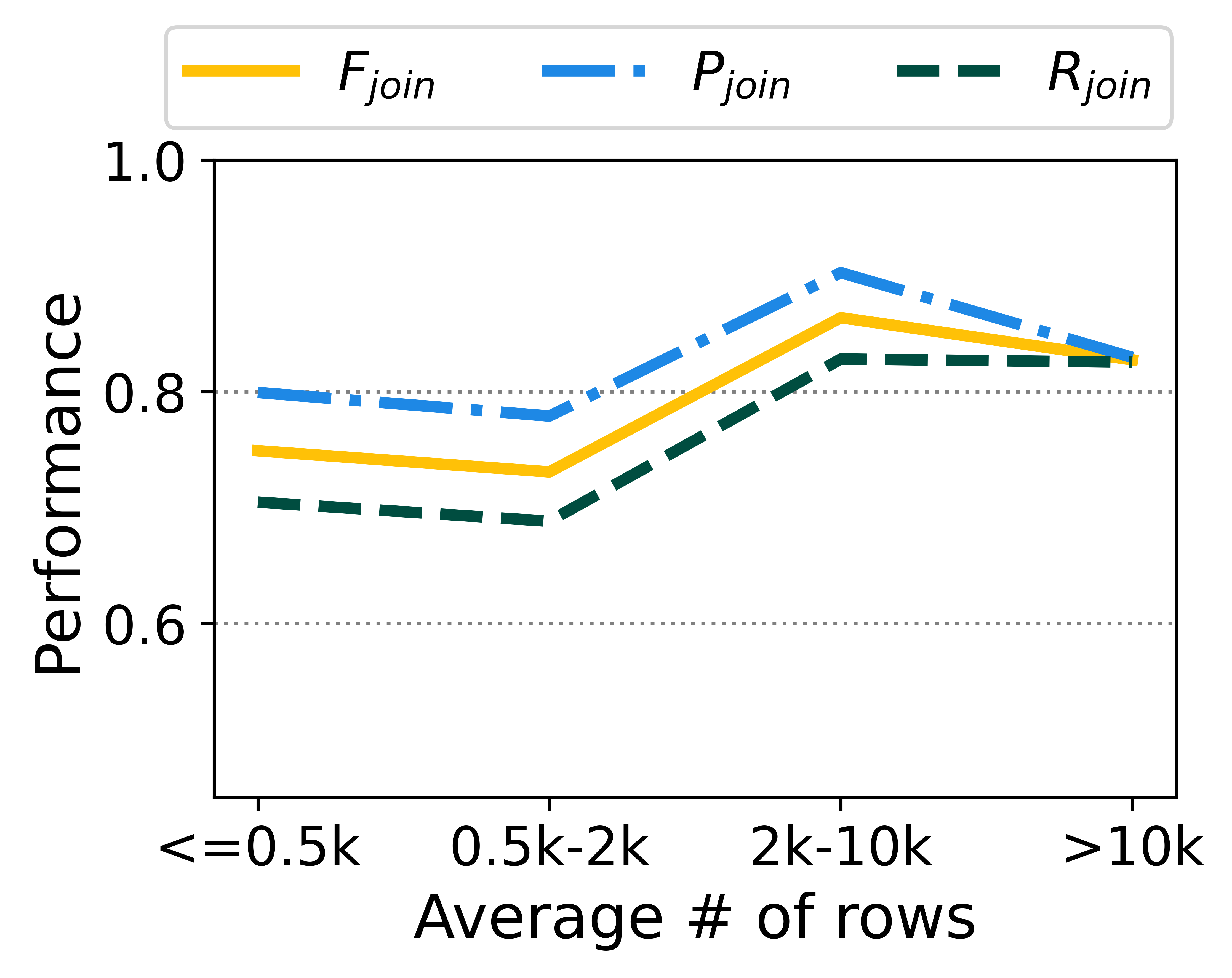}
            \vspace{-2mm}
            \caption{Join quality.}
            \label{fig:avg_rows_join_sensitity}
        \end{subfigure}
        \vspace{-4mm}
        \caption{{Sensitivity analysis.  (a, b): Varying average \# of columns per BI project. (c, d): Varying average \# of rows per BI project.}}
        \vspace{-2mm}
    \end{figure*}

    \underline{Sensitivity to the number of columns.} In addition, we bucketize all BI-projects based on the average number of columns per table in each project, which is another indicator of project complexity, and 
    analyze the corresponding quality of predicted transformations and joins,  in Figure~\ref{fig:avg_cols_transform_sensitity} and Figure~\ref{fig:avg_cols_join_sensitity}, respectively. As can be seen from the figures, the prediction quality stays reasonably stable as we vary the number of columns in projects, showing the robustness of our method on complex BI projects.

    \underline{Sensitivity to the number of rows.} We further bucketize BI projects by the average number of rows per table, and analyze the corresponding result quality for transformation and join. 
    In Figure~\ref{fig:avg_rows_join_sensitity} for join quality, we can see that with larger tables (e.g., with over 2K or 10K rows), the join quality (precision/recall/F-1) all goes up, which is perhaps not surprising as it is likely more reliable to predict joins on larger tables than on smaller tables with only tens of rows. In terms of transformation quality in Figure~\ref{fig:avg_rows_transform_sensitity}, we observe that while precision increases with larger tables, recall dips slightly, leading to a stable overall F1-score as the number of rows increases.
}

\underline{Sensitivity to the amount of missing values.} Another factor that can influence prediction quality is the amount of missing/empty cells in tables, so we group BI projects based on the average percentage of missing cells in tables, and analyze the corresponding prediction quality. In Figure~\ref{fig:missing_cell_join_sensitity}, we see that the join quality slips slightly as the fraction of missing values increases, while in Figure~\ref{fig:missing_cell_transform_sensitity}, we observe that the transformation quality can go up slightly with a moderate amount of missing values, which decreases when the amount of missing values increases.



\iftoggle{full}
{

    \begin{table}
        \centering
        \caption{{Sensitivity of transformation quality to domains.}}
        \vspace{-4mm}
        \label{tab:sensitivity-transformation-domain}
        \resizebox{0.9\linewidth}{!}{
        \begin{tabular}{|c|c|ccccc|} 
        \hline
        \textbf{} &
            \textbf{Full (\sysa-P)} &
            \textbf{Finance} &
            \textbf{HR} &
            \textbf{Marketing} &
            \textbf{Sales} &
            \textbf{Other} \\
        \hline
            $P_{\text{transform}}$ &
            \textbf{0.785}  & 0.789  & 0.857 & 0.733 & 0.795 & 0.687  \\
            $R_{\text{transform}}$ &
            \textbf{0.741}   & 0.800  & 0.923 & 0.786 & 0.775 & 0.633 \\ 
            \hline
            $F_{\text{transform}}$ &  
            \textbf{0.762}  & 0.795 & 0.889 & 0.759 & 0.786 & 0.659 \\
        \hline
        \end{tabular}
        }
    \end{table}

    \begin{table}
        \centering
        \caption{{Sensitivity of join quality to varying domain.}}
        \vspace{-4mm}
        \label{tab:sensitivity-join-domain}
        \resizebox{0.85\linewidth}{!}{
        \begin{tabular}{|c|c|ccccc|} 
        \hline
        \textbf{} &
            \textbf{Full (\sysa-P)} &
            \textbf{Finance} &
            \textbf{HR} &
            \textbf{Marketing} &
            \textbf{Sales} &
            \textbf{Other} \\
        \hline
            $P_{\text{join}}$ &
            \textbf{0.806}  & 0.866  & 0.805 & 0.795 & 0.926 & 0.662  \\
            $R_{\text{join}}$ &
            \textbf{0.734}   & 0.778  & 0.858 & 0.746 & 0.798 & 0.652 \\ 
            \hline
            $F_{\text{join}}$ &  
            \textbf{0.769}  & 0.819 & 0.831 & 0.770 & 0.857 & 0.657 \\
        \hline
        \end{tabular}
        }
    \end{table}
    
    \underline{Sensitivity to the domains of BI projects.}
    We further categorize the underlying ``domain'' of data present in different BI projects, and classify them into topics with the help of GPT (based on samples of table data). The top 4 categories are Sales ($38.2\%$), Finance ($23.6\%$), Marketing ($4.4\%$), and HR ($2.9\%$), and we group the rest into Other ($30.9\%$).
    We then analyze the prediction quality across BI projects in different domains, to see whether quality may be sensitive to the domains.  Table~\ref{tab:sensitivity-transformation-domain} and Table~\ref{tab:sensitivity-join-domain} show the transformation-quality and join-quality in different domains. We can see that while prediction quality are reasonably consistent across different domains, results in the Finance and HR domains tend to be higher (perhaps because these tend to involve simple schemas and analysis), while Other tend to be lower (perhaps because these schemas are more varied and more challenging to optimize).

}

\underline{Sensitivity to the depth of search tree $m$.}
Recall that in Section~\ref{sec:graph-rep}, we use a parameter $m$ to control the depth of the search graph, which is a hyper-parameter that we use to determine the length and therefore the complexity of the transformation sequences that we search for in the algorithm. In our main experiments this $m$ is always fixed at 2, in Table~\ref{tab:sensitivity-level} we vary the search depth $m$ from 1 to 3, and report the quality of our predictions. As can be seen from the table, our join and transformation quality do not change significantly to a different setting of $m$, which is desirable.

\begin{small}
    \begin{table}
    \centering
    \caption{Sensitivity of quality to varying $m$ (tree depth).}
    \vspace{-4mm}
    \label{tab:sensitivity-level}
    \resizebox{0.8\linewidth}{!}{
    \begin{tabular}{|c|ccc||c|ccc|} 
    \hline
        \textbf{Depth} &
        \textbf{1} &
        \textbf{2} &
        \textbf{3} &
        \textbf{Depth} &
        \textbf{1} &
        \textbf{2} &
        \textbf{3} \\
    \hline
        $P_{\text{transform}}$ &
        0.788 & 0.785  & 0.787  & 
        $P_{\text{join}}$ & 
        0.805 & 0.806 & 0.804  \\
        $R_{\text{transform}}$ &
        0.742 & 0.741  & 0.734  & 
        $R_{\text{join}}$ &
        0.714 & 0.734 & 0.712 \\ 
        \hline
        $F_{\text{transform}}$ &  
        0.765 & 0.762  & 0.760 &  
        $F_{\text{join}}$ & 
        0.756 & 0.769 & 0.756 \\
    \hline
    \end{tabular}
    }
    \end{table}
\end{small}

\subsubsection{Ablation Studies}\mbox{}\\
We perform ablation studies to understand the importance of various components, which are reported in Table~\ref{tab:abaltion-transform} and~\ref{tab:abaltion-join}.

\ignore{
\begin{table}
    \centering
    
    \resizebox{\linewidth}{!}{
    \begin{tabular}{@{\extracolsep{4pt}}ccccccc@{}} 
    \toprule
    \multirow{2}{*}{\textbf{Method}} & \multicolumn{3}{c}{\textbf{Table transformation quality}}&\multicolumn{3}{c}{\textbf{Join quality}} \\
        \cline{2-4} \cline{5-7}
        & 
        \textbf{Full} &
        \textbf{No model} &
        \textbf{No calibrate} 
        &
        \textbf{Full} &
        \textbf{No model} &
        \textbf{No calibrate} \\
    \midrule
        $P_{\text{transform}}$ &
        0.769 & 0.146 & 0.736 & 0.797 & 0.742 & 0.795 \\
        $R_{\text{transform}}$ &
        0.734 & 0.697 & 0.717 & 0.704 & 0.413 & 0.701 \\
        $F_{\text{transform}}$ &  
        0.751 & 0.242 & 0.727 & 0.747 & 0.531 & 0.745 \\
    \bottomrule
    \end{tabular}
    }
    
    \caption{Ablation studies for \sys-Precise. \eugenie{Put this into a figure instead?}}
    \label{tab:ablation}
    \vspace{-6mm}
\end{table}
}

\begin{small}
\begin{table}
    \centering
    \caption{Ablation  studies: Transformation quality.}
    \vspace{-4mm}
    \label{tab:abaltion-transform}
    \resizebox{0.98\linewidth}{!}{
    \begin{tabular}{|c|c|ccc|} 
    \hline
    \textbf{} &
        \textbf{Full (\sysa-P)} &
        \textbf{No holistic graph search}&
        \textbf{No classifier} &
        \textbf{No calibration}  \\
    \hline
        $P_{\text{transform}}$ &
        \textbf{0.785}    & 0.645  & 0.146 & 0.736 \\
        $R_{\text{transform}}$ &
        \textbf{0.741}  & 0.707  & 0.697 & 0.717  \\ 
        \hline
        $F_{\text{transform}}$ &  
        \textbf{0.762}  & 0.674 & 0.242 & 0.727 \\
    \hline
    \end{tabular}
    }
    \vspace{-4mm}
\end{table}

\begin{table}
    \centering
    \caption{Ablation studies: Join quality.}
    \vspace{-4mm}
    \label{tab:abaltion-join}
    \resizebox{0.98\linewidth}{!}{
    \begin{tabular}{|c|c|ccc|} 
    \hline
    \textbf{} &
        \textbf{Full (\sysa-P)} &
        \textbf{No holistic graph search} &
        \textbf{No classifier} &
        \textbf{No calibration} \\
    \hline
        $P_{\text{join}}$ &
        \textbf{0.806}  & 0.801  & 0.746 & 0.798   \\
        $R_{\text{join}}$ &
        \textbf{0.734}   & 0.692  & 0.416 & 0.703 \\ 
        \hline
        $F_{\text{join}}$ &  
        \textbf{0.769}  & 0.742 & 0.534 & 0.748 \\
    \hline
    \end{tabular}
    }
    \vspace{-4mm}
\end{table}
\end{small}

\ignore{
\begin{figure}
    \centering
    \includegraphics[width=0.96 \linewidth]{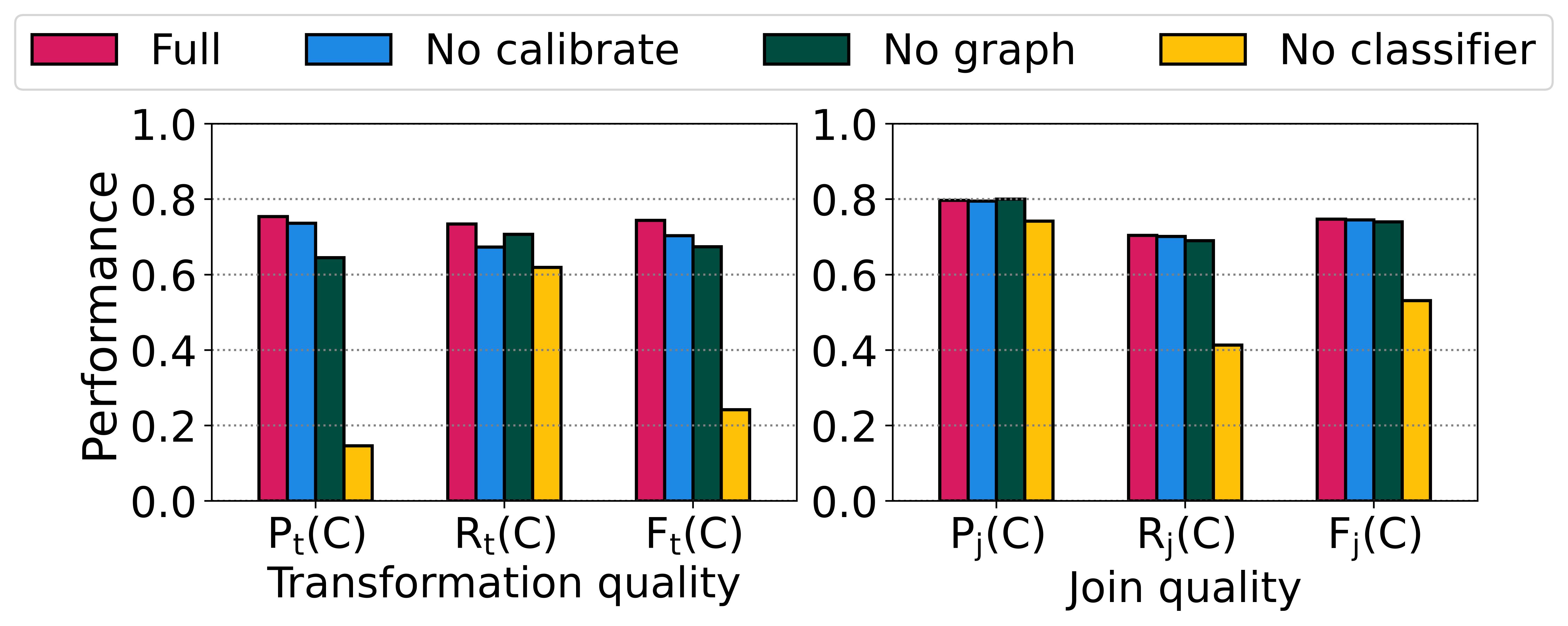}
    \vspace{-5mm}
    \caption{Ablation studies for \sys-Precise.}
    \vspace{-2mm}
    \label{fig:ablation}
\end{figure}
}

\underline{No holistic graph search.} To show the benefit of our graph-based optimization (Section~\ref{sec:steiner-tree}), we replace our principled graph-search algorithm with a greedy heuristic, which greedily picks candidates based on the highest transformation/join probabilities. We observe that both transformation and join quality drop noticeably. 

\underline{No classifiers.} To understand the importance of our classification models, we replace our $M^+_T$ models (Section~\ref{sec:offline-training}) using $M_T$ models without global-level features and calibrated scores, leading to a substantial decrease in prediction quality.

\underline{No calibration.} To study the benefit of calibrating probability scores from classifiers (Section~\ref{sec:offline-training}), we remove the calibration step, and observe a drop in result quality, which is less significant than previous ablations. 

\vspace{-2mm}
\subsubsection{Efficiency Comparison}\mbox{}\\
We perform an efficiency comparison to understand the search and end-to-end latency of the system.

\begin{figure}
    \centering
    \includegraphics[width=0.6 \linewidth]{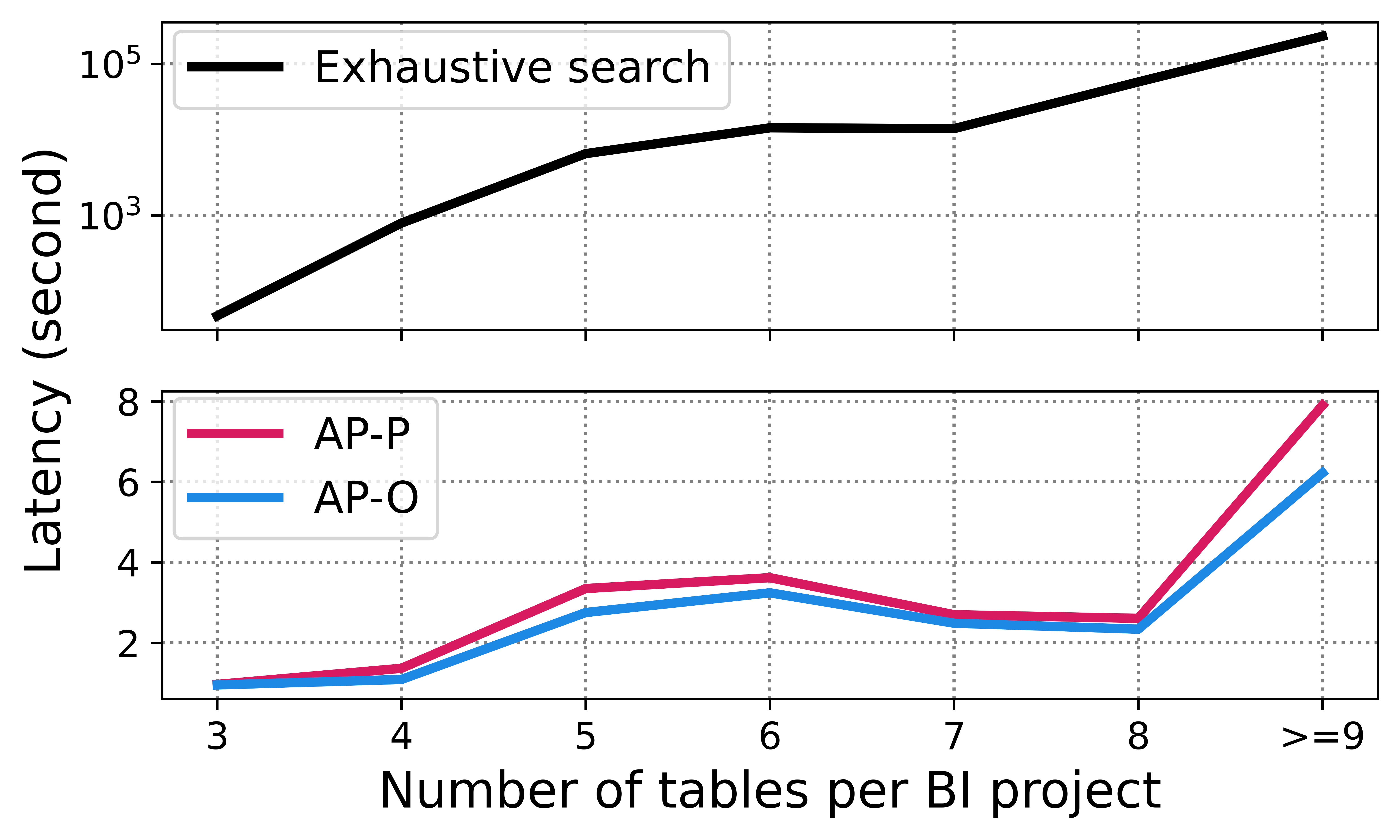}
    \vspace{-5mm}
    \caption{Comparison of average search latency. 
    }
\iftoggle{full}
{
}
{
    \vspace{-3mm}
}
    \label{fig:runtime}
\end{figure}

\underline{Search latency.} Figure~\ref{fig:runtime} shows the average latency of our graph algorithm based on Steiner-tree, and exhaustive search, bucketized by the number of input tables in each BI project (x-axis).

\ignore{
\begin{table}
    \centering
    \resizebox{\linewidth}{!}{
    \begin{tabular}{@{\extracolsep{4pt}}ccccccccc@{}}
    \toprule
        \multirow{2}{*}{\textbf{Method}} & & & \multicolumn{3}{c}{\textbf{ABI}}&\multicolumn{3}{c}{\textbf{GPT-j}} \\
        \cline{4-6} \cline{7-9}
        & 
        \textbf{\sysa-P} &
        \textbf{\sysa-O} &
        \textbf{AS} &
        \textbf{AT} &
        \textbf{GPT-tt} 
        &
        \textbf{AS} &
        \textbf{AT} &
        \textbf{GPT-tt}\\
    \midrule
        50\textsuperscript{th} \%tile &
        0.43 & 0.43   & 0.75  & 0.23 & 0.85 & 0.62 & 0.19 & 0.48 \\
        90\textsuperscript{th} \%tile &
        1.36 & 1.35  & 9.72  & 0.50 & 1.95 & 9.45 & 0.49 & 1.21 \\
        95\textsuperscript{th} \%tile &
        2.09 & 2.08  & 17.15  & 0.63 & 2.97 & 17.35 & 0.69 & 1.68 \\
        Average &  
        0.81 & 0.80 & 3.73 & 0.41 & 1.11 & 3.62 & 0.49 & 0.74 \\
    \bottomrule
    \end{tabular}
    }
    \caption{End-to-end latency per BI project (minute).}
    \vspace{-6mm}
    \label{tab:latency-comparison}
\end{table}
}

The average latency of our search is less than 4 seconds with less than 8 tables in a BI project, which grows to 8 seconds on large BI projects with over 9 tables. In comparison, exhaustive search takes over an hour on BI projects with 5 tables (since the search space grows exponentially in the number of input tables), showing the importance of our principled graph search. 

\iftoggle{full}
{
    \underline{End-to-end latency.} Figure~\ref{fig:runtime-end2end} shows the end-to-end latency comparison between \sysa and select baselines (to avoid clutter in the figure). The overall end-to-end latency includes components such as (1) predicting transformations; (2) predicting joins; and (3) performing an overall search-based optimization (Algorithm~\ref{alg:solve-using-steiner} and the main focus of this work, whose latency is reported in Figure~\ref{fig:runtime}). 
    
    As can be seen, among all methods that score competitively, GPT-based methods are the slowest, given the number of inference calls it requires (e.g., for transformation alone, it requires one inference call per table and per transformation type, in order to produce the best prediction results). Standalone methods, such as AP and SOP/MOP, have similar latency numbers. Note that within AP methods, search-based optimization (the focus of this work in Algorithm~\ref{alg:solve-using-steiner}) only accounts for a small fraction of overall latency, while performing transformation and join predictions on a large search graph (with many tables and derived versions of candidate tables) is substantially more costly than our core search algorithm. 
    

    \begin{figure}
        \centering
        \includegraphics[width=0.8 \linewidth]{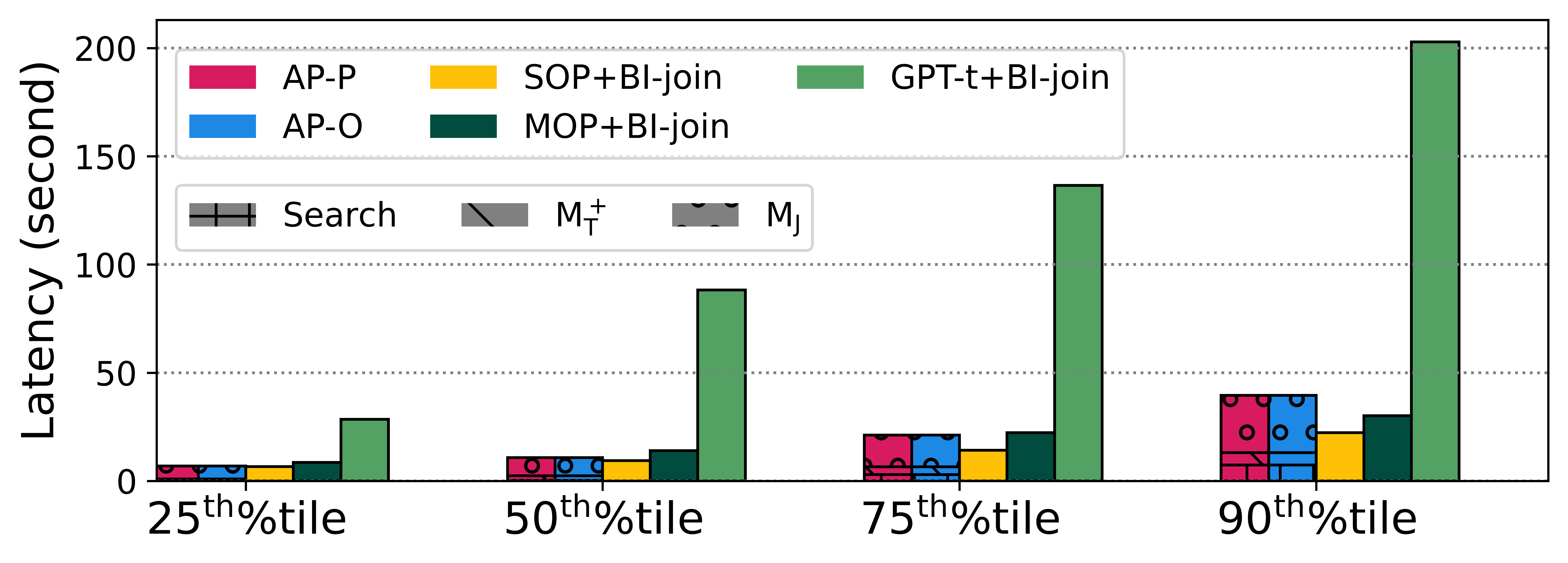}
        \vspace{-5mm}
        \caption{{End-to-end latency.}}
        \vspace{-4mm}
        \label{fig:runtime-end2end}
    \end{figure}
    
}
{
    Additional latency analysis, such as end-to-end latency, can be found in~\cite{full} in the interest of space.
}

\section{Conclusions}

Observing the need to predict both transformations and joins holistically in real BI projects, we propose a new BI-prep problem that has been overlooked thus far, and develop new graph algorithms for holistic optimizations.
Future directions include exploring holistic predictions in scenarios beyond BI (e.g., in ML and ETL workflows), and further improving predictions on large and complex BI projects.


\clearpage

\bibliographystyle{ACM-Reference-Format}
\bibliography{Auto-Prep}
\clearpage

\iftoggle{full}
{
    \begin{appendices}

\section{Transformation Models} \label{ap:local-features}


We build ML-based classification models to predict transformation $O$, based on the characteristics of input tables $T$, with the predicted probability $P(O|T)$. 

Our transformation classification models are constructed following prior work on transformation predictions~\cite{yan2020auto-suggest, li2023autotables}.  In addition, we enhance the transformation models $M_T$ from prior work, with additional features to train our models $M^+_T$ for each operator, where many of features are based on signals from across multiple tables in the same BI project (thereby moving away from the one-table-at-a-time paradigm of prior work).

\subsection{Transpose Classification Model} \label{sec:transpose-classifier} 


We enhance the Transpose classifier with the following global cross-table features:


\underline{Column-row-ratio} is a local-level feature that indicates how wide/long a table is (e.g., Table~\ref{tab:example-country} has a column-row-ratio of $5/3=1.67$). Wide tables are more likely to be transposed.

\underline{Column-Header-overlap} is a global-level feature captures the header overlap between $T_i$ and other tables in $\mathcal{T}$, and use the maximum value (e.g., Table~\ref{tab:example-country} has a header-overlap of $\text{max}([0.2, 0.2, 0])=0.2$). A low header-overlap indicates a possible misplace of data, meaning that the headers are supposed to be column values instead.

\underline{Value-domain-overlap} is a global-level feature that captures how much value domain overlap there is, between table $T_i$ and other tables $T_j \in \mathcal{T} \setminus \{T_i\}$. A high overlap indicates the domain values are commonly seen in other tables, and thus the probability of transpose is lower. For each column in $T_i$, its column-level overlap is the maximum overlap of its value domain with other columns' in $T_j \in \mathcal{T}\setminus\{T_i\}$. The max-column-domain-overlap for $T_i$ is the mean over the column-level overlaps. Given $T_4$, the column-level overlap for the \codeq{Code} column is $\text{max}([0, 0, 0,0,0,0,0,0,0,0,0])=0$, and for the \codeq{POL} column is $\text{max}([0.33, 0, 0,0,0,0,0,0,0, 0,0])=0.33$, so following the steps, the max-column-domain-overlap for $T_4$ is $\text{mean}([0, 0.33, 0.33,0.33,0.33])=0.26$.

\underline{Headers-value-overlap} is a global feature that indicates the possibility of the column headers being in column domains. A high overlap indicates a possible misplace of data, meaning that the headers are supposed to be in column domains instead. Given $T_i$, the max-headers-as-column-domains-overlap is the maximum of the overlap between $C_i$ and each column domains of $T_j \in \mathcal{T}\setminus\{T_i\}$. $T_4$ has a value of $\text{max}([0.8, 0, 0, 0, 0, 0, 0, 0.2, 0, 0])=0.8$.


\subsection{Unpivot Classification Model} 

For Unpivot, in addition to using all features listed in Section~\ref{sec:transpose-classifier} for Transpose, we also include the following local features to train the classification model:

\underline{Unpivot-columns-sparsity} captures how the sparsity of the unpivot columns $p_k$, where 0 is the least sparse. $p_k$ with higher sparsity is more likely to be the desired columns to unpivot.

\underline{Number-of-numeric-column-names} indicates the number of column names. Columns with a numerical column name are more likely to be in the unpivot parameter. We extract this feature for both $T_i$ and $O_{\code{unpivot}k}(T_i)$. For example, Table~\ref{tab:example-fertility} has $3$ numerical column names, while the desired $O_{\code{unpivot}k}(T_i)$ Table~\ref{tab:example-fertility-after} has $0$, which is the transformed table after applying $O_{\code{unpivot}1}(T_i)$, where $p_1=$ \{``2010'', ``2011'', ``2012''\}. A less desired parameter for unpivot on Table~\ref{tab:example-fertility} is $p_2=$ \{``2010'', ``2011''\}, which would lead to $1$ numerical column name in $O_{\code{unpivot}k}(T_i)$. 

\underline{Number-datetime-column-names} is the number of column names that are in types of date and time, which can often indicate a need to perform Unpivot. 

\underline{Number-of-float-column-names} is the number of column names that are float numbers, which would suggest the need to Unpivot.







\subsection{String-Transform Classification Model} 
The classification model developed for String-transform~\cite{zhu2017auto} takes two tables ($T_i$, $T_j$) as input. Given table $T_i$, we iterate through each $T_j \in \mathcal{T}/\{T_i\}$ to extract the following features to train models for String-transform:

\underline{Key-side-join-percentage} is an output of $M_T$ for String-transform. It indicates the join participation rate, defined as the fraction of records in the target table that participate in the join~\cite{zhu2017auto}. A higher percentage is more desired. 

\underline{Foreign-key-side-join-percentage} is an output of $M_T$ for String-transform. Similar to key-side-join-percentage, it is the join participation rate on the foreign key side. A higher percentage is more desired. 

\underline{Type} is an output of $M_T$ for String-transform. The three types of string operators are \code{split}, \code{concatenate}, \code{substring}. 

\underline{Delim} is an output of $M_T$ for String-transform. Our training data includes $12$ different string delimiters such as \code{``,''}, \code{``|''}, and \code{``.''}. 

\underline{Number-of-numerical-columns} is the number of numerical columns in the target table. Depending on Type, this feature has different indications. For example, if a split is done on a floating number column, it is likely to be a false positive. 

\underline{Number-of-datetime-columns} is the number of date-time columns in the target table, included for the same reason as the above.


We applied the Python SMOTE~\cite{smote} library in training and observed a significant improvement in model performance.

\section{Proof of theorem~\ref{theorem:np-complete}} \label{ap:theorem1}

First we show the MPBP problem is in NP. Given a solution $G' \subseteq G$, we verify that $G'$ is a tree that connects all the root nodes $R$. We then sum the weights of all edges in $G'$ and verify maximality. These steps can be done in polynomial time. 

To show the hardness of MPBP problem, we reduce from the Exact Cover by 3-Sets (X3C) problem~\cite{karp2010reducibility}. Given a finite set $X$ with $|X| = 3q$ and a collection $C$ of 3-element subsets of $X$, X3C looks for an exact cover for $X$, that is, a subcollection $C' \subseteq C$ such that every element of $X$ occurs in exactly one member of $C'$.

Given an instance of X3C, defined by the set $X=\{x_1, \dots, x_{3q}\}$ and a collection of 3-element sets $C=\{C_1, \dots, C_n\}$, we build an MPBP instance specifying the graph $G=(V,E)$ and the set of root tables $R$. We first define the vertices $V$ as:

\begin{gather}
    V(G) = \{v\} \cup \{c_1, \dots, c_n\} \cup \{x_1, \dots, x_{3q}\} \cup \{x_1', \dots, x_{3q}'\} 
\end{gather}

We construct a new node $v$ and a node for each $C_i \in C$. For each $x_j \in X$, we construct two nodes $x_j$ and $x_j'$. 

We then construct the set of transformation edges as:
\begin{gather}
    E_t(G) = \{vc_1, \dots, vc_n\} \cup (\bigcup_{x_j \in X} \{x_j'x_j\} )
\end{gather}

Next we construct the set of join edges as:
\begin{gather}
    E_\text{join}(G) = ( \bigcup_{x_j \in C_i} \{c_ix_j'\} )
\end{gather}

We set all transformation edge weights to be unit weight, and all join weights to be $0.5$.

Finally, we set root tables $R \in V$ as:
\begin{gather}
    R = \{v, x_1, \dots, x_{3q}\}
\end{gather}

As we can see, this construction can easily be done in polynomial time. Now we show by this construction, $C'$ is an exact cover of X3C iff $G'$ is a solution to the MPBP instance. 

If there is an exact cover $C'$ for the X3C problem, $C'$ uses exactly $q$ subsets $C'=\{C_1,\dots,C_q\}$, without loss of generality. Then the tree consists of transformation edges $\{vc_1, \dots, vc_q\} \cup (\bigcup_{x_j \in C_i, 1\leq i \leq q} \{x_j'x_j\} )$ and join edges $( \bigcup_{x_j \in C_i, 1\leq i \leq q} \{c_ix_j'\} )$ is a maximal Steiner tree solving the MPBP problem with $3q = |R|-1$ join edges. Therefore, if there is an exact cover $C'$, then there is a maximal Steiner tree using exactly $(|R|-1)$ join edges.

If there is a maximal Steiner tree $G'$ that uses exactly $(|R|-1)$ join edges, $G'$ touches all the terminal nodes in $R=\{v, x_1, \dots, x_{3q}\}$ by definition. In addition, the degree of $c_i$ nodes is $4$ by construction, $1$ transformation edge from $v$ and $3$ join edges connecting $x'_j$ nodes. Hence, $G'$ contains exactly $q$ $c_i$ nodes. We conclude that, without loss of generality, these nodes are $\{c_1, \dots, c_q\}$, so then the $C'=\{C_1, \dots, C_q\}$ is the subcollection such that every element of $X$ occurs in exactly one member of $C'$.

\section{Proof of Proposition~\ref{prop:solution-is-steiner}} \label{ap:proof-prop-solution-is-steiner}

We show why a valid solution to MPBP-G, written as $E = E_T \cup E_J$ in Definition~\ref{def:valid-solution},  must be a valid Steiner tree for  the given terminals $R = \{v(T_i) | i \in [n] \}$.

First, we show that edges in $E = E_T \cup E_J$ must be connected, forming a single connected component that also connect $R$. Observe that for a valid solution to MPBP-G in Definition~\ref{def:valid-solution}, $E = E_T \cup E_J$,  $E_T$ would contain unique paths that connect each root $v(T_i)$, with a leaf-vertex $v(L_i)$, in each transformation-tree $G(T_i)$, where the union of the leaf vertices $v(L_i)$ is written as $\mathcal{L} = \{v(L_i) | i \in [n] \}$. 
On the other hand, $E_J$ form a spanning tree that connects all vertices in $\mathcal{L}$ (Definition~\ref{def:valid-solution}). Because $E_J$ connect $\mathcal{L}$ in a single connected component, and each $v(L_i)$ is connected to $v(T_i)$ via a unique path, we conclude that $E = E_T \cup E_J$ forms a single connected component that includes $R = \{v(T_i) | i \in [n] \}$, therefore $R$ must be connected by $E = E_T \cup E_J$.

Next, we show that edges in $E = E_T \cup E_J$ must form a tree. Recall that with $n$ transformation-trees (one for each input table), each with $m$ levels. The unique paths in $E_T$ contain exactly $n \cdot m$ edges, while the spanning tree in $E_J$  should contain exactly $n-1$ edges (to connects $n$ vertices), for a total of $n \cdot m + (n-1)$ edges. In terms of vertices, the sub-graph induced by $E = E_T \cup E_J$ has exactly $(m+1) \cdot n = n\cdot m + n$ vertices for on transformation paths (join-edges only connect vertices at the end of transformation paths, and do not bring in new vertices). Given that the sub-graph induced by $E = E_T \cup E_J$  has $n \cdot m + (n-1)$ edges to connect $n\cdot m + n$ vertices, we know that this sub-graph must be a tree.

Combining the fact that $E = E_T \cup E_J$  is a tree that connects all of $R$, we conclude that it is a valid Steiner tree with respect to $R$.


\section{Proof of Proposition~\ref{prop:solution-edge-count}} \label{ap:proof-prop-solution-edge-cnt}

We show any valid solution to MPBP-G in Definition~\ref{def:valid-solution} has exactly $(m \cdot n)$ transformation-edges and $(n-1)$ join edges.

Recall that valid solutions to MPBP-G requires a unique transformation path (with exactly $m$ edges) on each transformation tree $G(T_i)$, which translates to exactly $(m \cdot n)$ transformation-edges, since we have $n$ transformation trees, each of which is of depth $m$. 

Furthermore, we know that any valid solution to MPBP-G also contains a spanning tree with join-edges (Definition~\ref{def:valid-solution}), to connect $n$ vertices (representing transformed tables). Since the spanning tree connects $n$ vertices, it will have exactly $(n-1)$ join edges.

Combining, we show that any valid solution to MPBP-G will ahve exactly $(m \cdot n)$ transformation-edges and $(n-1)$ join edges.

\section{Proof of Proposition~\ref{prop:steiner-edge-count}} \label{ap:proof-prop-steiner-edge-count}

We first show the first half of Proposition~\ref{prop:steiner-edge-count}, that a valid Steiner tree on the search graph $G(\mathcal{T})$ of MPBP-G that connects all terminals $R = \{v(T_i) | i \in [n] \}$, has at least $(m \cdot n) + (n-1)$ edges.

Observe that any valid Steiner tree that can connect all terminals in $R = \{v(T_i) | i \in [n] \}$, require at least one path of length $m$ from the root $v(T_i)$ to a leaf vertex $v(L_i)$ in each transformation tree $G(T_i)$, which translates into $(m \cdot n)$ edges. Furthermore, given $n$ connected components (each of which is a sub-graph of a transformation-tree $G(T_i)$, induced by the transformation-path),  at least $(n-1)$ join edges is required. Combining, that translates into at least $(m \cdot n) + (n-1)$ total edges.

We now show the second part of Proposition~\ref{prop:steiner-edge-count}, which states that any valid Steiner tree with exactly $(m\cdot n) + (n-1)$ edges is a valid solution to MPBP-G. 
Given the construction of the search graph $G(\mathcal{T})$, which consists of $n$ disjoint transformation-trees $G(T_i)$, that are connected by join-edges at the leaf-levels of $G(T_i)$, we know that a valid Steiner tree on $G(\mathcal{T})$ that connects $R$ must have at least $n-1$ join edges (the minimum number of edges to connect $n$ disjoint transformation-trees). 

Suppose there are more than $n-1$ join edges in a valid Steiner tree with  $(m\cdot n) + (n-1)$ edges, then the number of transformation-edges will be less than $(m\cdot n)$, which however is impossible since $(m\cdot n)$ is the minimum number of transformation edges required to connect $R$ with leaf-level vertices of all transformation trees  $G(T_i)$ (otherwise it is no longer a valid tree). Therefore we have exactly $n-1$ join edges, and given a total of $(m\cdot n) + (n-1)$ edges in the valid Steiner tree, that leaves us with $(m\cdot n)$ transformation edges. 

Since it is a valid Steiner tree and therefore all of $R$ is connected, each $v(T_i) \in R$ should connect with a leaf vertex $v(L_i) \in G(T_i)$, requiring exactly a total of $(m\cdot n)$ transformation edges, forming exactly $n$ transformation paths from the root of each transformation-tree, to a leaf vertex $v(L_i) \in G(T_i)$, satisfying the requirement on $E_T$ in Definition~\ref{def:valid-solution}. 

Because the valid Steiner tree further connects leaf vertices in $\{v(L_i) | i \in [n] \}$ in a connected component, with $n-1$ join edges, we know this must form a spanning tree ($n-1$ edges to connect $n$ vertices), therefore satisfying the requirement on $E_J$  in Definition~\ref{def:valid-solution}.

Combining, we know that a valid Steiner tree with exactly $(m\cdot n) + (n-1)$ edges is also a valid solution to MPBP-G.

\section{Proof of Theorem~\ref{theorem:steiner-tree-solve}} \label{ap:proof-thm-steiner-tree-solve}

We first prove that Algorithm~\ref{alg:solve-using-steiner} produces a valid solution to MPBP-G, in time polynomial to the input problem size.

\underline{Algorithm~\ref{alg:solve-using-steiner} produces a valid solution to MPBP-G.} To see why the  Algorithm~\ref{alg:solve-using-steiner} must return a valid solution to MPBP-G, first notice that $S_b$ is a valid solution to MPBP-G (therefore also a valid Steiner tree per Proposition~\ref{prop:solution-is-steiner}). 
Furthermore, $S_s$ must be a valid Steiner tree by the virtue of Kou's algorithm~\cite{kou1981fast}, and from Proposition~\ref{prop:steiner-edge-count}, we know that $S_s$ must have at least $(m\cdot n) + (n-1)$ edges for being a valid Steiner tree.

Suppose  $S_s$ has exactly $(m\cdot n) + (n-1)$ edges in $|S_s|$, then by Proposition~\ref{prop:steiner-edge-count} it must be a valid solution to MPBP-G. Now since both $S_s$ and $S_b$ are valid solutions to MBPB-G, the better of the two returned by Algorithm~\ref{alg:solve-using-steiner} must also be a valid solution to MBPB-G. 

Alternatively, if $S_s$ has more than $(m\cdot n) + (n-1)$ edges  in $|S_s|$, then  its sum of edge-weight $W_s$ is at least $2p \cdot (m\cdot n + n)$ from the penalty weight alone, which is already greater than that of the solution $S_b$, which will have exactly $(m\cdot n) + (n-1)$ edges (Proposition~\ref{prop:solution-edge-count}), whose total edge-weight $W_b$ is therefore no greater than $2p \cdot (m\cdot n + (n-1)) + p$. This forces $S_b$, a valid solution to MBPB-G, be picked over $S_s$. 

Combining, we know that Algorithm~\ref{alg:solve-using-steiner} will always produce a valid solution to MPBP-G. 

\underline{Algorithm~\ref{alg:solve-using-steiner} computes a solution in polynomial time.}
 Furthermore, we show that the time complexity of  Algorithm~\ref{alg:solve-using-steiner} is $O(n^3 \cdot m^2)$, because it can be verified that the most expensive step in Algorithm~\ref{alg:solve-using-steiner} is Line~\ref{ln:solve_mst}, which invokes MST that has with a complexity of $O(|R||V|^2)$, where $|R|$ is the size of the terminal nodes, and $|V|$ is the total number of vertices in the graph~\cite{kou1981fast}. In our case with $n$ terminals and $n\cdot m$ vertices in $G(\mathcal{T})$, it follows that our overall complexity is $O(n^3 \cdot m^2)$, which is polynomial in the size of the input problem. This completes the first half of our proof of Theorem~\ref{theorem:steiner-tree-solve}.

\underline{Algorithm~\ref{alg:solve-using-steiner} has an approximation ratio of 
 $(2-(2/n))$.} We will now prove that Algorithm~\ref{alg:solve-using-steiner} has an approximation ratio of 
 $(2-(2/n))$, where $n$ is the number of input tables, to complete the second half of Theorem~\ref{theorem:steiner-tree-solve}.

First, consider the case when $S_s$ is picked over $S_b$ and returned as the solution in Algorithm~\ref{alg:solve-using-steiner}, 
$S_s$ must also be a valid Steiner tree of $G(\mathcal{T})$, since it is produced by the MST algorithm from Kou et al.~\cite{kou1981fast}. Let $S^*$ be the optimal solution to MPBP-G, which is different from $S_s$ (otherwise we found the optimal solution and is done). We know from Proposition~\ref{prop:solution-is-steiner}, that $S^*$ being a valid solution to MPBP-G, must also be a valid Steiner tree of $G(\mathcal{T})$ with respect to terminals $R$. 

Recall that Kou's algorithm has an approximation ratio of $2 - \frac{2}{t}$, where $t$ is the number of leaves in the optimal Steiner tree~\cite{kou1981fast}. 
Since Kou's algorithm produces a valid Steiner tree solution $S_s$ instead of the valid Steiner tree $S^*$ that has a better cost, we know that the cost of $S_s$ is within a factor $2 - \frac{2}{t}$ of the cost of $S^*$, given its approximation guarantee. Since the our construction of $G(\mathcal{T})$ ensures that the optimal Steiner tree always has  all root nodes of all transformation-trees as its leaves, we know $t=n$ where $n$ is the total number of input tables, ensuring that the cost of our solution $S_s$ is within a factor $2 - \frac{2}{n}$ of the cost of $S^*$, which is the optimal solution to MPBP-G.

Next consider the case when $S_b$ is picked over $S_s$ and returned as the solution in Algorithm~\ref{alg:solve-using-steiner}. 
Let $S^*$ be the optimal solution to MPBP-G, with cost $W^*$, which is different from the solution $S_b$ (otherwise we found the optimal solution). Similar to above, we know from Proposition~\ref{prop:solution-is-steiner}, that $S^*$ being a valid solution to MPBP-G, must also be a valid Steiner tree of $G(\mathcal{T})$ with respect to terminals $R$, yet Kou's algorithm returns a different valid Steiner tree $S_s$ with cost $W_s$, which given the $2-\frac{2}{n}$ factor approximation ratio, ensures that $W_s  \leq (2-\frac{2}{n}) W^*$. Furthermore, we know that $W_b < W_s$ since $S_b$ is picked over $S_s$, which together with the previous inequality, leads to  $W_b < W_s  \leq (2-\frac{2}{n}) W^*$, ensuring that the cost of solution $S_b$ that we return is also within a factor of $2-\frac{2}{n}$ from the optimal solution $S^*$ (since $W_b < (2-\frac{2}{n}) W^*$). 

Combining both cases in which $S_s$ and $S_b$ are returned as solutions in Algorithm~\ref{alg:solve-using-steiner}, where both are within a factor of $2-\frac{2}{n}$ from the optimal solution $S^*$, we conclude that Algorithm~\ref{alg:solve-using-steiner} produces a solution to MPBP-G with an approximation ratio of 
 $(2-(2/n))$.
 \end{appendices}
}

\end{document}